\pdfoutput=1
\documentclass[PAPER, atlasdraft=false, UKenglish, cernpreprint, texmf, orcidlogo]{atlasdoc}
\usepackage{atlaspackage}
\usepackage{atlasbiblatex}
\usepackage{multirow}

\usepackage{atlasphysics}
\graphicspath{{logos/}{figures/}}

\usepackage{ANA-HIGG-2020-20-PAPER-defs}

\AtlasTitle{Measurements of $WH$ and $ZH$ production with Higgs boson decays into bottom  quarks and direct constraints on the charm Yukawa coupling in $13\,\mathrm{TeV}$ $pp$ collisions with the ATLAS detector}

\AtlasAbstract{%

A study of the Higgs boson decaying into bottom quarks ($H \to b\bar{b}$) and charm quarks ($H \to c\bar{c}$) is performed, in the associated production channel of the Higgs boson with a $W$ or $Z$ boson, using $140\,\mathrm{fb}^{-1}$ of proton--proton collision data at $\sqrt{s} = 13\,\mathrm{TeV}$ collected by the ATLAS detector.
The individual production of $WH$ and $ZH$ with $H \to b\bar{b}$ is established with observed (expected) significances of $5.3$ ($5.5$) and $4.9$ ($5.6$) standard deviations, respectively. Differential cross-section measurements of the gauge boson transverse momentum within the simplified template cross-section framework are performed in a total of 13 kinematical fiducial regions.

The search for the $H \to c\bar{c}$ decay yields an observed
(expected) upper limit at 95\% confidence
level of $11.5$ ($10.6$) times the Standard Model
prediction. The results are also used to set constraints on the charm
coupling modifier, resulting in  $|\kappa_c| < 4.2$ at 95\% confidence
level. Combining the $H \to b\bar{b}$ and $H \to c\bar{c}$
measurements constrains the absolute value of the ratio of Higgs-charm and Higgs-bottom coupling modifiers ($|\kappa_c/\kappa_b|$) to be less than
$3.6$ at  95\% confidence level.

}

\AtlasRefCode{HIGG-2020-20}
\PreprintIdNumber{CERN-EP-2024-237}

\AtlasJournalRef{\JHEP 04 (2025) 075}
\AtlasDOI{10.1007/JHEP04(2025)075}

\hypersetup{pdftitle={ATLAS document},pdfauthor={The ATLAS Collaboration}}

\begin{document}

\maketitle

\tableofcontents

%
%

%

\section{Introduction}
\label{sec:intro}

The discovery of a heavy scalar particle  with a mass of about
\SI{125}{\GeV} by the ATLAS and
CMS collaborations~\cite{HIGG-2012-27,CMS-HIG-12-028} at the Large Hadron Collider (LHC) provided experimental confirmation of the
Brout-Englert-Higgs (BEH)
mechanism~\cite{Englert:1964et, Higgs:1964ia, Higgs:1964pj,
Guralnik:1964eu,Higgs:1966ev,Kibble:1967sv}, which spontaneously
breaks electroweak (EW) gauge symmetry and generates mass terms for
the $W$ and $Z$ gauge bosons. The observation of Higgs boson decays into
$\tau$-leptons~\cite{CMS-HIG-16-043, HIGG-2017-07},
bottom-quarks~\cite{HIGG-2018-04,CMS-HIG-18-016} and associated
production with top-quarks~\cite{CMS-HIG-17-035,HIGG-2018-13} provided
strong evidence to support the BEH mechanism in the Standard Model
(SM) where the fermion masses are generated via Yukawa interactions.
All measured couplings are so far consistent with the
SM~\cite{HIGG-2021-23,CMS-HIG-22-001}.

Since the decay of the Higgs boson into $b$-quarks has the largest
branching fraction ($\mathcal{B}$) with a SM expectation of $58.2\% \pm 0.7\%$ for
a mass of $\SI{125}{\GeV}$~\cite{deFlorian:2016spz}, the $b$-quark Yukawa coupling is of particular interest. It
is the most sensitive decay channel to study some of the rarer Higgs boson  processes,
such as associated production with a $W$ or $Z$
boson ($VH$)~\cite{HIGG-2018-51,CMS-HIG-20-001}, $WH$ production via vector boson fusion~\cite{ATLAS:2024vxc}, Higgs boson
production at high transverse
momentum~\cite{CMS-HIG-19-003,HIGG-2018-52,HIGG-2021-08,HIGG-2021-11} and Higgs boson pair
production, in which the most sensitive channels have one or both Higgs
bosons decaying into $b$-quarks~\cite{CMS-HIG-19-018,CMS-HIG-20-005,CMS-HIG-20-010, HDBS-2018-40,HDBS-2019-29, HDBS-2021-10}.
In the SM the $b$-quark Yukawa coupling has the largest impact on the Higgs boson's width and thus measuring it is  crucial to set constraints on
phenomena beyond the SM~\cite{HIGG-2021-23,CMS-HIG-22-001}.
Furthermore, this decay mode has allowed differential measurements of the $VH$ cross-section
in kinematic fiducial volumes defined in the simplified template cross-section (STXS) framework~\cite{deFlorian:2016spz}.

Measurement of the decay of the Higgs boson into $c$-quarks is much more
challenging than its decay into $b$-quarks due to the smaller expected
$\mathcal{B}$ (2.9\% for a Higgs boson mass of $\SI{125}{\GeV}$~\cite{deFlorian:2016spz}),
the increased difficulty in identifying $c$-hadrons, due to their
smaller lifetimes and masses, and the larger background from vector boson plus $c$-quark production.
Phenomena  beyond the SM could significantly enhance the
coupling of the Higgs boson to the charm-quark, and in turn the $H \to c\bar{c}$
branching fraction~\cite{PhysRevD.89.033014,PhysRevD.92.033016,Botella:2016krk,PhysRevD.98.055001,GHOSH2016504,Egana-Ugrinovic:2018znw,Egana-Ugrinovic:2019dqu}.
Searches in the $VH$ channel~\cite{HIGG-2021-12,CMS-HIG-21-008} offer the best direct constraints on the Higgs-charm coupling.
This paper reports on a combined study of \hbb and \hcc in the $VH$
production mode, using $\SI{140}{\ifb}$ of proton--proton collision data recorded with the ATLAS detector in the years 2015-2018
at a centre-of-mass energy of
\SI{13}{\TeV}. The associated $VH$ production
mode is used as it provides a clean signature with the vector boson decaying
into charged leptons or neutrinos and allows the separation of the \hbb and \hcc
signals from the large multijet background. In the analysis presented in this paper, the $W$ boson is searched for
in its decays into an electron,  muon or $\tau$-lepton ($\ell$) in addition to
a neutrino and the $Z$ boson
in its decays into electron, muon or neutrino pairs. The channels are labelled
as 0-,1- or 2-lepton corresponding to the number of reconstructed charged leptons %
in the final state.  The vector boson must have a minimum transverse
momentum (\pTV) of \SI{75}{\GeV} in the 1- and 2-lepton channels and
\SI{150}{\GeV} in the 0-lepton channel.

The Higgs boson is reconstructed either from two small-radius (small-$R$) jets or from a single large-radius (large-$R$) jet,
whose definitions are given in \Sect{\ref{sec:objects}}.
The \hbb and \hcc sensitive regions are defined according to flavour
tagging information of the jets.  The \hcc search exclusively uses
small-$R$ jets.
To reconstruct the Higgs boson decay products, the \hbb analysis uses small-$R$ jets for $\pTV<\SI{400}{\GeV}$ (resolved
regime) and large-R jets for $\pTV>\SI{400}{\GeV}$ (boosted regime).
This approach offers good sensitivity across
the \pTV range.  Events are further classified
according %
to the transverse momentum of the vector boson and the number of jets.
Multivariate discriminants, built from variables that describe the
kinematic properties, jet flavour and missing transverse momentum content of the
selected events, are used to maximise the analysis sensitivity in all signal
regions. The output distributions are used as inputs to  binned
maximum-likelihood fits, which allows the yields and kinematic properties of both
the signal and the background processes to be estimated.  The measurements  are validated using the diboson process (WZ and ZZ production),
where the multivariate discriminant is modified to
extract the $VZ$, $Z \rightarrow b\bar{b}$  and $Z \rightarrow c\bar{c}$ diboson processes.  In
addition to overall signal yields, differential STXS measurements of
the $VH$ cross-section in the \hbb channel are presented.  The
results are also interpreted in the $\kappa$-framework~\cite{Heinemeyer:2013tqa, deFlorian:2016spz} to constrain the
coupling modifiers of the Higgs boson to $b$- and $c$-quarks, $\kappa_b$
and $\kappa_c$. In the following, the sensitive region targeting the
\hbb decay is referred to as the \hbbCAT and the sensitive region
targeting the \hcc decay is referred to as the \hccCAT.

This paper updates previous results presented in Refs.~\cite{HIGG-2018-51,HIGG-2018-52,HIGG-2021-12}, which used the same data sample
as the present analysis, with several
improvements:  better reconstruction and calibration of leptons and jets, an improved flavour tagging algorithm that combines
$b$- and $c$-jet identification and has  a more precise calibration,
extended acceptance for the $WH$ process to events with $\pTV<\SI{150}{\GeV}$,
re-optimisation of multivariate discriminants and first application to the  \hbb boosted regime and \hcc search,
updated theoretical predictions, improved definition of signal and control regions
and increased granularity of STXS measurements both at high transverse momentum and as a function of jet multiplicity.

%
%
%

%
%


%
%

\section{ATLAS detector}
\label{sec:detector}

\newcommand{\AtlasCoordFootnote}{%
ATLAS uses a right-handed coordinate system with its origin at the nominal interaction point (IP)
in the centre of the detector and the \(z\)-axis along the beam pipe.
The \(x\)-axis points from the IP to the centre of the LHC ring,
and the \(y\)-axis points upwards.
Polar coordinates \((r,\phi)\) are used in the transverse plane,
\(\phi\) being the azimuthal angle around the \(z\)-axis.
The pseudorapidity is defined in terms of the polar angle \(\theta\) as \(\eta = -\ln \tan(\theta/2)\) and is equal to the
rapidity $y =\frac{1}{2} \ln \left( \frac{E+p_z}{E-p_z} \right) $ in the relativistic limit.
Angular distance is measured in units of \(\Delta R \equiv \sqrt{(\Delta y)^{2} + (\Delta\phi)^{2}}\).

}

The ATLAS detector~\cite{PERF-2007-01} at the LHC covers nearly the entire solid angle around the collision point.\footnote{\AtlasCoordFootnote}
It consists of an inner tracking detector surrounded by a thin superconducting solenoid, electromagnetic and hadron calorimeters,
and a muon spectrometer incorporating three large superconducting air-core toroidal magnets.

The inner-detector system (ID) is immersed in a \qty{2}{\tesla} axial magnetic field
and provides charged-particle tracking in the range of \(|\eta| < 2.5\).
The high-granularity silicon pixel detector covers the vertex region and typically provides four measurements per track,
the first hit normally being in the insertable B-layer (IBL) installed before Run~2~\cite{ATLAS-TDR-19,PIX-2018-001}.
It is followed by the SemiConductor Tracker (SCT), which usually provides eight measurements per track.
These silicon detectors are complemented by the transition radiation tracker (TRT),
which enables radially extended track reconstruction up to \(|\eta| = 2.0\).
The TRT also provides electron identification information
based on the fraction of hits (typically 30 in total) above a higher energy-deposit threshold corresponding to transition radiation.

The calorimeter system covers the pseudorapidity range \(|\eta| < 4.9\).
Within the region \(|\eta|< 3.2\), electromagnetic calorimetry is provided by barrel and
endcap high-granularity lead/liquid-argon (LAr) calorimeters,
with an additional thin LAr presampler covering \(|\eta| < 1.8\)
to correct for energy loss in material upstream of the calorimeters.
Hadron calorimetry is provided by the steel/scintillator-tile calorimeter,
segmented into three barrel structures within \(|\eta| < 1.7\), and two copper/LAr hadron endcap calorimeters.
The solid angle coverage is completed with forward copper/LAr and tungsten/LAr calorimeter modules
optimised for electromagnetic and hadronic energy measurements respectively.

The muon spectrometer (MS) comprises separate trigger and
high-precision tracking chambers measuring the deflection of muons in a magnetic field generated by the superconducting air-core toroidal magnets.
The field integral of the toroids ranges between \num{2.0} and \qty{6.0}{\tesla\metre}
across most of the detector.
Three layers of precision chambers, each consisting of layers of monitored drift tubes, cover the region \(|\eta| < 2.7\),
complemented by cathode-strip chambers in the forward region, where the background is highest.
The muon trigger system covers the range \(|\eta| < 2.4\) with resistive-plate chambers in the barrel, and thin-gap chambers in the endcap regions.

The luminosity is measured mainly by the LUCID--2~\cite{LUCID2} detector that records Cherenkov light produced by the quartz windows of photomultipliers located close to the beam pipe.

Events are selected by the first-level trigger system implemented in custom hardware,
followed by selections made by algorithms implemented in software in the high-level trigger~\cite{TRIG-2016-01}.
The first-level trigger accepts events from the \qty{40}{\MHz} bunch crossings at a rate below \qty{100}{\kHz},
which the high-level trigger further reduces to record events to disk at about \qty{1}{\kHz}.

A software suite~\cite{SOFT-2022-02} is used in data simulation, in the reconstruction
and analysis of real and simulated data, in detector operations, and in the trigger and data acquisition
systems of the experiment.


%
%

\section{Data and simulated event samples}
\label{sec:data}

The data  were collected using triggers that require an electron, or a muon or large missing
transverse momentum in the event.
Events are selected for analysis only if they are of good quality and
if all the relevant detector components are known to be in good
operating condition~\cite{DAPR-2018-01}.
The recorded events contain an average of 34 inelastic $pp$ collisions
per bunch-crossing (\pileup).

Monte Carlo (MC) simulated events are used to
model most of the backgrounds from SM processes and the \vhbbcc signal
processes.  A summary of all the generators used for the simulation of
the signal and background processes is given in \Tab{\ref{tab:samples}}.
The \Vjets and $qq \to VV$ diboson processes are now modelled using the \SHERPA[2.2.11] generator,
using the configuration described in \Refn{\cite{PMGR-2021-01}}. These are
multijet merged samples with next-to-leading-order (NLO) accuracy in the matrix element calculation.
Virtual EW loop-terms are included at NLO accuracy for the \Vjets and diboson processes.
Furthermore, the default treatment of interference between the \ttbar and $Wt$ processes now uses the
diagram subtraction scheme~\cite{Frixione:2007vw} instead of the previously used diagram removal scheme~\cite{Frixione:2008yi},
as it was found to give a better description of the data.
Samples produced with alternative generators are used to estimate systematic
uncertainties in the event modelling, as described in \Sect{\ref{sec:background}}.
All simulated processes are normalised using higher-order
theoretical cross-section predictions and are
generated at least to  NLO accuracy, except for
the \ggZH~and \ggVV~processes, which are generated at leading-order (LO).
For samples containing top-quarks or Higgs bosons, the \EVTGEN[1.6.0] programme~\cite{Lange:2001uf}
is used to model the decays of $b$- and $c$-hadrons.

\begin{table}[hbt!]
\begin{center}{\fontsize{9}{10.8}\selectfont
\caption{Generators used for the simulation of the signal and background
processes. If not specified, the order of the cross-section calculation refers to the expansion in the strong coupling constant ($\alphas$).
The acronyms ME, PS and UE stand for matrix element, parton shower and underlying event,  respectively.
$(\dagger)$ The NNLO(QCD)+NLO(EW) cross-section calculation for the $pp \to ZH$ process already includes the $gg\to ZH$ contribution.
The $qq\to ZH$  process is normalised using the cross-section for the $pp \to ZH$ process,
after subtracting the $gg\to ZH$ contribution. An additional scale factor is applied to
the $qq \rightarrow VH$ processes as a function of the transverse momentum of the vector
boson, up to $\pTV = \SI{1}{\TeV}$, to account for electroweak (EW) corrections at NLO\@. This makes use of the $VH$ differential cross-section
computed with \textsc{Hawk}~\cite{Denner:2011id,Denner:2014cla}. Contributions from photon-quark processes are
also included for $pp \to WH$~\cite{deFavereaudeJeneret:2009db}. ($\ddagger$) For the diboson samples the cross-sections
are calculated by the Monte Carlo generator at NLO accuracy in QCD.
\protect\label{tab:samples}}
\begin{tabular}{llllll}
\toprule
\hsp Process & ME generator & ME PDF &PS and & UE  & Cross-section \\
& & & Hadronisation & tune & order\\
\midrule
\multicolumn{6}{l}{Signal, mass set to \SI{125}{\GeV} and $b\bar{b}$ branching fraction to 58\%} \\
\midrule
\hsp $qq\to VH$  &\POWHEGBOX[v2]~\cite{Alioli:2010xd} + &
\PDFforLHC[15NLO] & \PYTHIA[8.245]  & AZNLO & NNLO(QCD)$^{(\dagger)}$+  \\
&   \textsc{GoSam}~\cite{Cullen:2011ac}+  & \cite{Butterworth:2015oua} & \cite{Sjostrand:2007gs} & \cite{STDM-2012-23} &NLO(EW)~\cite{Ciccolini:2003jy,Brein:2003wg,Ferrera:2011bk,Brein:2011vx,Ferrera:2013yga,Ferrera:2014lca,Campbell:2016jau} \\
&    \textsc{MiNLO}~\cite{Hamilton:2012np,Luisoni:2013cuh} &  & &   \\

\hsp $gg  \to ZH$ &\POWHEGBOX[v2] & \PDFforLHC[15NLO] &\PYTHIA[8.245] & AZNLO & NLO+  \\
&  & & & &NLL~\cite{Altenkamp:2012sx,Hespel:2015zea,ggzhnll,Harlander:2013mla,Brein:2012ne}\\
\midrule
\multicolumn{6}{l}{Top quark, mass set to \SI{172.5}{\GeV}}  \\
\midrule
\hsp $t\bar{t}$ &\POWHEGBOX[v2]~\cite{Frixione:2007nw} &  \NNPDF[3.0NLO] &\PYTHIA[8.230] &A14~\cite{ATL-PHYS-PUB-2014-021}& NNLO+NNLL~\cite{Czakon:2011xx} \\
\hsp $s$-chan.\ single top&\POWHEGBOX[v2]~\cite{Alioli:2009je} & \NNPDF[3.0NLO] &\PYTHIA[8.230] &A14 & NLO~\cite{PhysRevD.81.054028} \\
\hsp $t$-chan.\ single top&\POWHEGBOX[v2]~\cite{Alioli:2009je} & \NNPDF[3.0NLO] &\PYTHIA[8.230] &A14& NNLO~\cite{Campbell:2020fhf} \\
\hsp $Wt$ &\POWHEGBOX[v2]~\cite{Re:2010bp} & \NNPDF[3.0NLO] &\PYTHIA[8.230] &A14& Approx. NNLO+NNLL~\cite{PhysRevD.82.054018} \\
\midrule
\multicolumn{6}{l}{Vector boson + jets} \\
\midrule
\hsp \Vjets & \SHERPA[2.2.11] & \NNPDF[3.0NNLO] &  \SHERPA[2.2.11] & Default & NNLO \\
& \cite{Bothmann:2019yzt,Cascioli:2011va,Gleisberg:2008fv} &  &  \cite{Schumann:2007mg,Hoeche:2012yf} & \cite{Catani:2009sm}  &   \\
\midrule
\multicolumn{6}{l}{Diboson} \\
\midrule
\hsp $qq \to VV$ & \SHERPA[2.2.11] & \NNPDF[3.0NNLO] &  \SHERPA[2.2.11] & Default & NLO$^{(\ddagger)}$  \\
\hsp $gg \to VV$ & \SHERPA[2.2.2] & \NNPDF[3.0NNLO] &  \SHERPA[2.2.2] & Default & NLO$^{(\ddagger)}$  \\
\bottomrule
\end{tabular}}
\end{center}
\end{table}


All samples of simulated events are processed with  the ATLAS detector
simulation~\cite{SOFT-2010-01} based on \GEANT~\cite{Agostinelli:2002hh}.
The effects of multiple interactions in the same or nearby bunch crossings
are  modelled by overlaying minimum-bias events, simulated using the soft QCD processes
of \PYTHIA[8.186] with the A3~\cite{ATL-PHYS-PUB-2016-017} set of tuned parameters (tune) and
\NNPDF[2.3LO]~\cite{Ball:2012cx}  parton distribution functions (PDFs).


%
%

\section{Event reconstruction}
\label{sec:objects}
Tracks measured in the inner detector are used to reconstruct interaction vertices~\cite{ATL-PHYS-PUB-2015-026}, of which the one with the
highest sum of squared transverse momenta of associated tracks is selected as the primary vertex of the hard interaction.

Electron candidates are reconstructed by matching ID
tracks to clusters of energy deposited in the electromagnetic calorimeter.
Electrons must have $\pt>\SI{7}{\GeV}$ and $|\eta|<2.47$. The associated
track must have $|d_0|/\sigma_{d_0}<5$ and $|z_0|\sin \theta<\SI{0.5}{\milli\metre}$,
where $d_0$ ($z_0$) is the transverse (longitudinal) impact parameter
relative to the primary vertex and $\sigma_{d_0}$ is the uncertainty
in $d_0$. %
Candidates are identified with a likelihood method and must
satisfy the `loose' identification criteria and `loose' isolation
requirements as described in \Refn{\cite{EGAM-2018-01}}.
The electron energy is calibrated as described in
\Refn{\cite{EGAM-2018-01}}.  The electron reconstruction, identification and trigger efficiencies in
the simulation are corrected using comparisons with
data~\cite{EGAM-2018-01}.

Muon candidates are required to be reconstructed within the acceptance of the MS
$\abseta<2.7$ and to have $\pt>\SI{7}{\GeV}$. They are further required to have $|d_0|/\sigma_{d_0}<3$, and
$|z_0|\sin \theta<\SI{0.5}{\milli\metre}$.
Muons are selected using a `loose' quality criterion as described in
\Refn{\cite{MUON-2018-03}} and a loose track isolation requirement. The muon momentum is calibrated as described in
\Refn{\cite{MUON-2022-01}}.
The muon reconstruction, identification and trigger efficiencies in the simulation are corrected using comparisons with
data~\cite{MUON-2018-03}.

Hadronically decaying $\tau$-lepton candidates (\tauhad) are reconstructed~\cite{ATL-PHYS-PUB-2015-045} in the range of
$|\eta|<2.5$ and $\pt>\SI{20}{\GeV}$ and identified using a tagger based on a recurrent neural network that uses  tracking and calorimeter information. The  `loose'
working point as described in \Refn{\cite{ATL-PHYS-PUB-2019-033}} is used. No
attempt is made  to distinguish between leptonically
decaying $\tau$-leptons and electrons or muons.

The lepton selections as described above are used to remove overlaps between physics
objects (see below) and classify the events into 0-, 1- and 2-lepton
channels, but more stringent requirements are made in the 1-lepton
channel to suppress background from multijet processes: the quality
is raised to `tight' for electrons and `medium' for muons and
`HighPtCaloOnly' isolation is used for both lepton flavours~\cite{EGAM-2018-01,MUON-2018-03}.

Small-$R$ jets\footnote{Unless explicitly stated the term `jet' refers
to small-$R$ jets.} are formed using objects from a particle-flow (PFlow)
algorithm~\cite{PERF-2015-09}, which combines energy deposits in the calorimeter with
inner detector tracks. The PFlow objects are
combined into jets using the anti-$k_t$
algorithm~\cite{Cacciari:2008gp,Fastjet} with a radius parameter $R=0.4$.
These jets are then initially calibrated  to the particle level by applying a jet energy scale derived
from simulation with in situ corrections based on the methodology derived in \Refn{\cite{JETM-2018-05}}.
Central jets, i.e.\ those with $|\eta|<2.5$, are required to have $\pt>\SI{20}{\GeV}$
while forward jets ($2.5<|\eta|<4.5$) must satisfy the $\pt>\SI{30}{\GeV}$
requirement. A jet-vertex-tagging
technique using a multivariate likelihood~\cite{PERF-2014-03, ATL-PHYS-PUB-2019-026} is
applied to central (forward) jets with  $\pt<\SI{60}{\GeV}$ (\SI{120}{\GeV}) to suppress jets
that are not associated with the event's primary vertex.

Simulated jets are labelled as $b$, $c$ or $\tauhad$ by matching a $b$- or
$c$-hadron or $\tau$-lepton having $\pT \geq \SI{5}{\GeV}$ within a cone of
$\Delta R = 0.3$~\cite{FTAG-2019-07} around the jet axis; the matching procedure prioritises $b$-hadrons over $c$-hadrons and  $\tau$-leptons, and $c$-hadrons over  $\tau$-leptons.
Jets without a valid match are  labelled as light ($l$) jets.
The DL1r flavour tagger~\cite{FTAG-2019-07}  computes for each jet the probability of containing a $b$-hadron, a $c$-hadron or being an $l$-jet ($p_b$, $p_c$, $p_l$);
these probabilities are combined to define dedicated one-dimensional discriminants %
optimised to select $b$-jets (\DLoner
\footnote{\DLoner=$\ln{\left(p_b / ( f_c\cdot p_c+ (1-f_c) \cdot p_l  ) \right)}$  with $f_c=0.018$.}) and $c$-jets (\DLonerc
\footnote{\DLonerc=$\ln{\left(p_c / ( f_b\cdot p_b+ (1-f_b) \cdot p_l  ) \right)}$  with $f_b=0.3$.}).

\begin{figure}
\begin{center}
\includegraphics[width=0.6\textwidth]{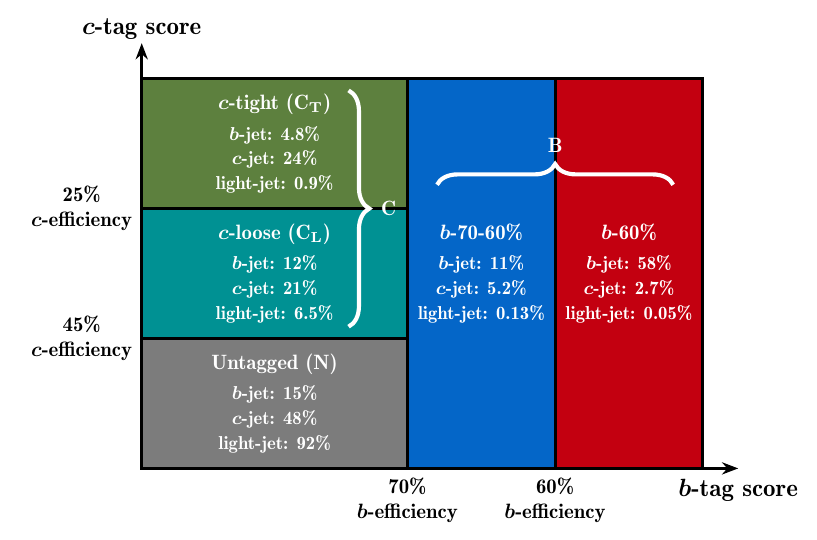}\\
\caption{A schematic of the flavour tagging regions as used in the resolved regime. The efficiencies for the various jet flavours in the various regions are extracted from a simulated $t\bar{t}$ sample following the
procedure detailed in Ref.~\cite{FTAG-2019-07}.}
\label{fig:ftagregions}
\end{center}
\end{figure}
Central jets are then classified as either $b$-tagged, $c$-tagged or non-tagged in a scheme that is specifically designed for this analysis
and is illustrated in \Fig{\ref{fig:ftagregions}}.
Any jet that satisfies the 70\% $b$-tagging operating point of the \DLoner
discriminant ($b$-70\%)~\cite{FTAG-2019-07} is classified as $b$-tagged. Jets that do not fall
into this category are classified as $c$-tagged if they satisfy the a \DLonerc
discriminant working point designed to have a 45\% efficiency for $c$-jets.
All other jets are classified as non-tagged.
Jets that are $c$-tagged are further subdivided into
loose and tight categories.
In the $t\bar{t}$ simulation the fraction of $b$, $c$, $l$, and $\tauhad$ jets that satisfy the $b$-tagging selection is 69\%,
7.9\%, 0.18\%, and 2.2\% respectively. The corresponding fractions that satisfy the exclusive loose $c$-tagging
selection are 11\%, 21\%, 6.5\%, and 19\% respectively, while those that satisfy the tight are 4.8\%, 24\%, 0.9\%, and 20\% respectively.
A tighter 60\% $b$-tagging working point is also used in the multi-variate
analysis (MVA) as described in \Sect{\ref{sec:mva}}.
The tagging efficiencies are measured with a data-driven method following the approaches described
in Refs.~\cite{FTAG-2018-01,FTAG-2020-08,FTAG-2019-02} and the results are used to correct the simulation.

Large-$R$ jets %
are formed from topological clusters of energy depositions~\cite{PERF-2014-07}
clustered with the anti-$k_t$ algorithm with  $R=1.0$.
They are trimmed~\cite{Krohn:2009th,JETM-2018-06} to mitigate the effects of
\pileup and soft radiation.  The jet mass is computed using
tracking and calorimeter information %
and calibrated following the technique described  in Ref.~\cite{JETM-2018-02}.
Such jets are required to have $\pT > \SI{250}{\GeV}$, $|\eta|< 2.0$ and a mass larger than \SI{50}{\GeV}.
Track-jets with $p_T>\SI{10}{\GeV}$ are used to identify the \hbb decay within the
large-$R$ jet~\cite{HIGG-2018-52}.  Track-jets are built with the
anti-$k_T$ algorithm with a variable radius (VR) \pT-dependent parameter,
from tracks reconstructed in the ID~\cite{Krohn:2009zg,ATL-PHYS-PUB-2017-010,PERF-2015-03}.
Track-jets are ghost associated~\cite{Cacciari:2008gn} to large-$R$ jets.
In a similar manner to small-$R$ jets, the DL1r algorithm is used to $b$-tag
track-jets. For the 85\% working point chosen for the definition of the Higgs boson candidate,
the fraction of $b$-, $c$- and $l$-track-jets that satisfy the $b$-tag selection in the $t\bar{t}$
simulation is 85\%, 34\% and 2\% respectively.
Tighter working points at 77\%, 70\% and 65\% are used in the MVA.
The tagging efficiencies are measured in a similar way to small-$R$ jets.

To maximise the statistical power of the available MC samples, the tagging requirement is not applied to the \Vjets and $s$- and $t$-channel single-top-quark samples,
or to the $t\bar{t}$ and $Wt$ samples in the boosted regime. Instead, events are weighted by the probabilities for each jet to fall into a given tagging category.
The probabilities are extracted with a neutral network approach~\cite{ATL-PHYS-PUB-2022-041}, which considers the flavour label and the kinematic properties of every jet in the event.
In the \hccCAT\ all jets are considered in the calculation of the event weight, while only light and $c$-jets are used in the \hbbCAT.
For $b$-jets in the \hbbCAT, the tagging requirement is applied.
This neural network approach improves the simple jet-based parameterisation used in \Refn{\cite{HIGG-2021-12}} since it is capable of directly capturing the dependency of the tagging efficiency on nearby jet activity.

Further corrections are applied to the four-momentum of the large-$R$ jet or pair of small-$R$ jets used to build the
Higgs boson candidate to account for semileptonic decays of $b$- and
$c$-hadrons and the different energy response of $b$-jets and $c$-jets compared with
$l$-jets. Muons, before imposing the isolation requirement, which are within a \pT dependent $\Delta R$ cone\footnote{
The cone size is optimised as a function of the reconstructed muon \pT as $\Delta R = \min{(0.4,\,0.04+10/(\pT\,[\GeV]))}$
} of the (track-) jet axis, are added to the jet  four-momentum
for both jet types~\cite{HIGG-2016-29,HIGG-2018-52,PERF-2017-04} if the jets are $b$- or $c$-tagged.
An additional MC-based correction is applied for small-$R$ $b$-tagged
jets to correct to the energy to the truth level using a similar method to
\Refn{\cite{HIGG-2016-29}}; a separate correction is applied to jets
with and without an associated muon.
In the 2-lepton channel, additional techniques are employed to improve the resolution on the reconstructed Higgs boson candidate mass ($m_H$).
A per-event kinematic likelihood fit~\cite{HIGG-2016-29}, primarily exploiting the excellent energy and momentum resolution
of electrons and muons, is adopted for the boosted regime and events in the resolved regime with less than four jets.
Finally in the resolved regime, if an additional central jet is sufficiently close  %
to both of the two jets forming the Higgs boson candidate, the four-momentum of such a jet is added to the Higgs boson four-momentum
\HCand,  to compute its invariant mass  and other related quantities.  The proximity criteria for the sum of the  $\Delta R$  values between the
jet and each of the two Higgs boson candidate jets, is optimised
as a function of the reconstructed vector-boson $\pt$ and varies between 2.85 at $\ptv=\SI{75}{\GeV}$ and 0.91 at $\ptv=\SI{400}{\GeV}$.
This procedure is referred to as final state radiation  (FSR) recovery and aims at minimising the effect of hard QCD radiation of heavy quarks in the Higgs boson candidate reconstruction.
The additional jet employed in this approach is not included in the jet multiplicity calculation.
The combined effect of all these dedicated techniques improve the Higgs boson mass resolution by up to 12\% in the 0-lepton and 1-lepton channels and by up to 40\% in the 2-lepton channel.

An overlap removal procedure is applied to avoid any double-counting
between electrons, muons, $\tauhad$ candidates and jets.
In the boosted regime, any small-$R$ jet within a cone of $\Delta R = 1.0$ of the
large-$R$ jet axis is removed. %

The missing transverse momentum, \metV, is reconstructed in each event as the
negative vector sum of the transverse momenta of electrons, muons,
$\tauhad$s, jets, and a `soft-term'.
The soft-term is calculated as the vectorial sum of the \pt of tracks
matched to the primary vertex but not associated with a reconstructed
lepton or jet~\cite{JETM-2020-03}. The magnitude of \metV~is referred
to as \met.
The four-momentum of the vector-boson ($\vec{V}$) is equivalent to \metV\ in the 0-lepton channel. In the
1-lepton channel it is reconstructed as vector sum of the lepton and
the neutrino where the transverse component of the neutrino momentum
is identified with $\metV$ and the longitudinal component is
obtained by applying a $W$ mass constraint to the lepton-neutrino
system. In the 2-lepton channel $\vec{V}$ is reconstructed
from the two-lepton system.
The transverse momentum of $\vec{V}$ is denoted by \pTV.


%
%

\section{Event selection and categorisation}
\label{sec:event}

\subsection{Event selection}

The events are classified into three channels depending on the number of charged leptons.
0-lepton events contain no electrons, muons or $\tauhad$ candidates.
Events in the 1-lepton channel contain exactly one electron, muon or \tauhad candidate.
The electron or muon must fulfil the stricter lepton selection described in \Sect{\ref{sec:objects}}.
Events in the 2-lepton channel have exactly two electrons or two muons. %
Since the signal-to-background ratio increases with \pTV, the analysis focuses on the  high \pTV phase space:
$\pTV  \geq \SI{150}{\GeV}$ in the 0-lepton channel  and $ \pTV  \geq \SI{75}{\GeV}$ in the  1- and 2-lepton channels.
Jet cleaning criteria are used to identify jets arising from non-collision backgrounds or noise in the calorimeters, and events containing such jets are removed, using the `tight cleaning' defined in \Refn{\cite{ATLAS-CONF-2015-029}}.

Events are triggered either by the single electron, single muon or \met~triggers depending on the channel and  \pTV.
Events with one or two electrons are collected with the single electron triggers~\cite{TRIG-2018-05}; for these events at least one electron
must have an offline $\pT$ greater than \SI{27}{\GeV} to ensure a high trigger efficiency.
The single muon triggers~\cite{TRIG-2018-01} together with an offline muon requirement of $\pT>\SI{25}{\GeV}$ are used for the 1-lepton
muon sub-channel for $\pTV<\SI{150}{\GeV}$ and the 2-lepton muon sub-channel for $\pTV<\SI{250}{\GeV}$.
The $\met$ triggers~\cite{TRIG-2019-01} are used for the 0-lepton channel, the 1-lepton $\tau$ sub-channel
and at higher values of \pTV in the 1- and 2-lepton muon sub-channels, where it becomes efficient because muons do not deposit much energy in the
calorimeters.  A requirement on the scalar sum of the transverse
momenta of the jets, $\htV>120$ (150) \GeV for 2 ($\ge 3$) jet events, removes a region where the $\met$ trigger efficiency depends mildly on the number of jets in the event, corresponding to less than 1\% of the phase space.

This section describes the main selection criteria for the signal regions (SRs). Most of the selections are also valid for the control
regions (CRs), with any differences discussed in \Sect{\ref{sec:cat}}.

\begin{figure}[tb!]
\centering
\includegraphics[width=0.6\textwidth]{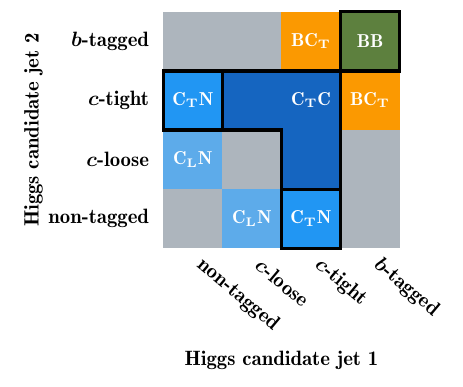}
\caption[]{A representation of the analysis regions in the resolved regime as defined by the tagging requirements
for the Higgs boson jet candidates. The bold outline signifies the categories used in the SRs.
The green regions are used in the \hbbCAT, the blue regions are used in the \hccCAT,
and the orange regions are used in the Top CRs, as defined in the text. Events in the grey categories are not used in the analysis.}
\label{fig:cat2}
\end{figure}

In the resolved \hbbCAT reconstruction, events must
contain exactly two $b$-tagged jets (BB regions).
These two jets are labelled  $j_1$ and
$j_2$ and define the Higgs boson candidate along with a possible FSR jet (see \Sect{\ref{sec:objects}}). Other jets in the event with $p_T>\SI{30}{\GeV}$ are ordered in \pt\ and labelled $j_i$
with $i>2$. They are referred to in the following as `additional jets'.
Events in the \hccCAT must contain at least one $c$-tagged jet and
no $b$-tagged jets. The events are further separated into tight-tight ($\ctct$),
tight-loose ($\ctcl$) and  tight-non-tagged ($\ctn$) categories. Loose-non-tagged ($\cln$) events are used to define dedicated CRs (see \Sect{\ref{sec:cat}}). The
Higgs boson candidate jets $j_1$ and $j_2$ are the two $c$-tagged
jets with the highest $\pT$ or, for the $\ctn$ and $\cln$ categories, the $c$-tagged jet and the
leading non-tagged central jet.  In the SRs, the
$\ctct$ and $\ctcl$ categories are combined into a $\ctc$ category
and the flavour tagging scores of the jets are used in the boosted-decision-tree (BDT) training
to further improve the signal-to-background separation as explained in \Sect{\ref{sec:mva}}.
A further category, in which the event contains a $b$-tagged jet and a tight $c$-tagged jet ($\bct$), is used to define a CR as described below.
A representation of the various
analysis regions as a function of the tagging requirements for the
Higgs boson jet candidates is shown in \Fig{\ref{fig:cat2}}.

The resolved regimes require $m_H>\SI{50}{\GeV}$ and the angular
separation of the two Higgs boson candidate jets $\drjj <\pi$ to remove events with back-to-back
jets that are far away from the signal regions; the leading
jet of the pair is further required to have $\pT>\SI{45}{\GeV}$.

In the boosted Hbb regime  the Higgs boson is reconstructed from the
large-$R$ jet with the highest $\pT$.  This jet must have mass greater than $\SI{50}{\GeV}$
and at least two matched track-jets. Exactly two of the three leading
matched track-jets ordered in $\pT$ must be $b$-tagged and are
labelled $j_1$ and $j_2$ with the remaining track-jet labelled $j_3$ if present.

A dedicated event selection is then applied in each lepton channel with the aim to profit from the particular event topology to reduce key background contributions;
the selection is optimised independently for the resolved and boosted regimes.

\textbf{0-lepton channel:} High \met in multijet events typically
arises from mismeasured jets in the calorimeters; such events are
efficiently removed by requirements on: the minimum azimuthal difference between \metV and any
jet min $[\Delta \phi (\metV, \vec{j_i})]>20^\circ$ ($30^\circ$) for 2 jet
($\ge 3$ jet and boosted) events; the azimuthal difference between \metV\ and
\HCand,  $\Delta \phi (\metV, \HCand)>120^\circ$; and the
azimuthal  difference between the two jets forming the Higgs boson candidate in the resolved
category $\Delta \phi (\vec{j_1}, \vec{j_2})<140^\circ$.
Events with greater than two (one) additional jets
are rejected in the
Hbb (Hcc) resolved regime to reduce the top-quark background.

\textbf{1-lepton channel:} In the electron sub-channel an additional
selection of \met > 30 (50) \GeV is applied in the resolved (boosted)
regime to reduce the background from multijet production;
for similar reasons, the $W$ boson transverse mass,\footnote{$m_{\mathrm{T}}^W = \sqrt{2 \pt^{\ell} \pt^{\nu} (1 -
\cos(\phi^{\ell} - \phi^{\nu}))}$, where  $\pt^{\ell}$ and $\phi^{\ell}$ are the transverse momentum and azimuthal angle of the charged lepton and
the corresponding quantities for the neutrino, $\pt^{\nu}$ and $\phi^{\nu}$, are reconstructed
from \metV.} \mtw, is requested to
be larger than \SI{20}{\GeV} in events with $\pTV < \SI{150}{\GeV}$.
Multijet background  in the $\tau$ sub-channel is suppressed using the same azimuthal
requirements as in the 0-lepton channel, in addition to $\mtw>\SI{10}{\GeV}$.
Events with greater than one additional jet %
in the resolved
regime are rejected to reduce top-quark background.

\textbf{2-lepton channel:} The invariant mass of the lepton pair
($m_{\ell\ell}$) must be close to the $Z$ boson mass ($\SI{81}{\GeV}<
m_{\ell\ell} < \SI{101}{\GeV}$ for the resolved regime  and $\SI{66}{\GeV}<
m_{\ell\ell} < \SI{116}{\GeV}$ for the boosted regime). In di-muon events, the
two muons are required to have opposite-sign charge. This requirement is not used
in the electron sub-channel, where the charge misidentification rate
is not negligible. %

\subsection{Event categorisation}
\label{sec:cat}
\begin{figure}[htpb!]
\centering
\rotatebox[origin=c]{90}{\includegraphics[width=1.3\textwidth]{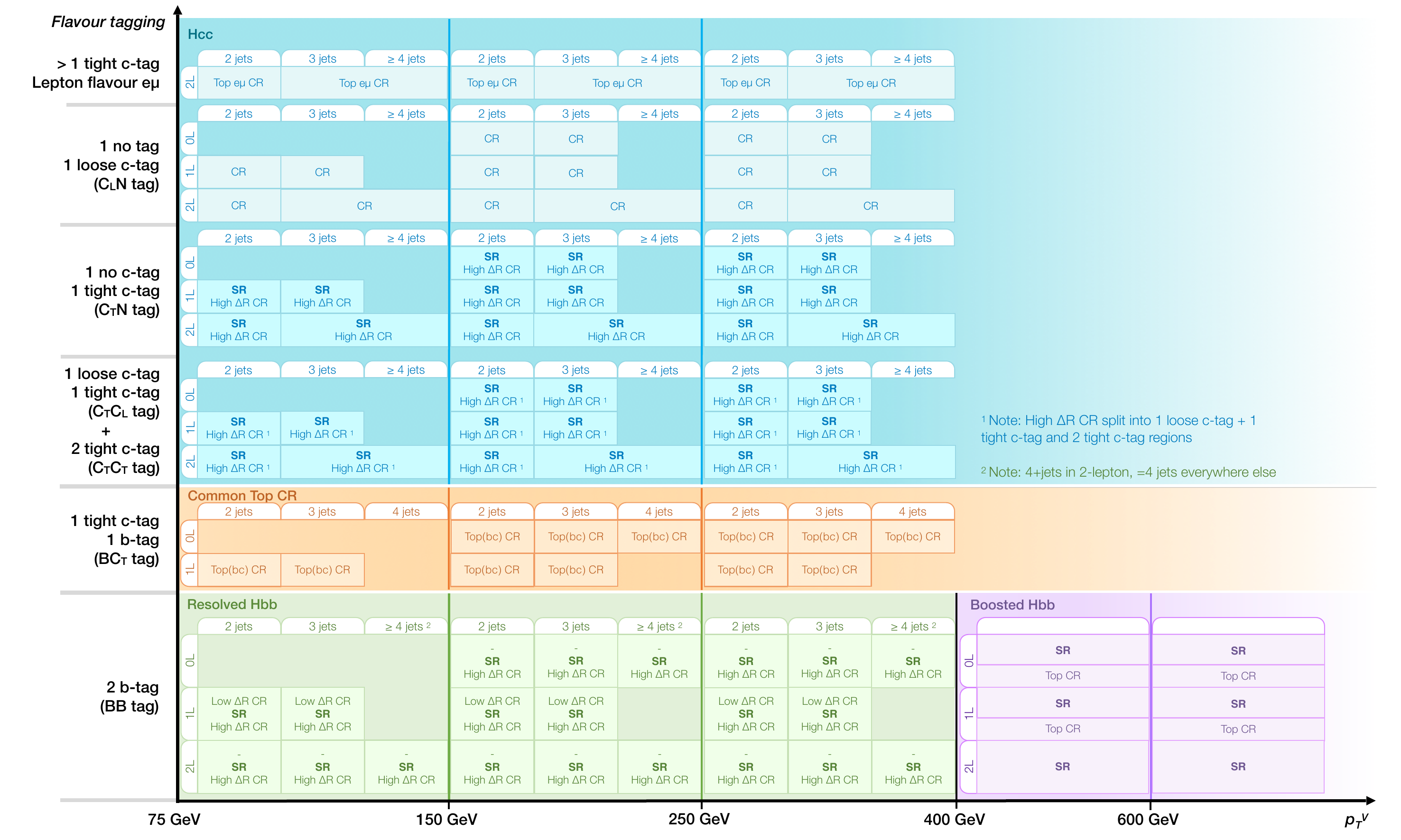}}
\caption[]{A schematic view of all the signal and control regions used in the analysis.
The green region is used in the \hbbCAT, the blue regions are used in the \hccCAT,
and the orange regions are used in the Top CRs, as defined in the text.}
\label{fig:categories}
\end{figure}

A graphical summary of the analysis regions is given in \Fig{\ref{fig:categories}} and their composition is discussed in detail in \Sect{\ref{sec:background}}.

Events are categorised according to the number of $b$- and $c$-tagged jets,
the value of \pTV   and, in the resolved regime, the number of additional jets.
Regions at large \pTV and those with no additional jets have higher signal purity.
The \hccCAT\ is split into regions of \pTV with boundaries
75, 150 and \SI{250}{\GeV}. The \hbbCAT\ has additional \pTV boundaries at
400 and \SI{600}{\GeV} where the boosted reconstruction is used.
In the 2-lepton resolved regime, which can have
larger jet multiplicities (defined as two more than the number of additional jets)  than the other channels, regions with at least four
(three) jets are combined in the Hbb (Hcc) category.

Events are divided into the SRs, which contain most of the signal
events, and several CRs, which are defined by changing exactly one
selection criterion of the SR, such that they are orthogonal and
extrapolations to the SR are small.  The CRs are designed to control
the contribution of specific backgrounds and allow theoretical uncertainties in the background predictions
to be reduced.

\textbf{Signal region} events are defined with different criteria for the resolved and boosted regimes.
The resolved SRs are defined using a continuous selection on \drjj, as a
function of \ptv following a procedure similar to that described in \Refn{\cite{HIGG-2021-12}}, but
reoptimised for the present analysis.  An upper requirement on \drjj\
that depends also on jet multiplicity is designed to
retain 95\% (85\%) of the signal in the SRs with exactly two  (at least
three) jet SRs. In the 1-lepton channel of the \hbbCAT, a lower
limit on  \drjj\ is also applied to retain 90\% of the diboson
background in the SRs. This requirement ensures little signal loss, while retaining a large fraction of the diboson events that are used  to validate the analysis (see \Sect{\ref{sec:stat}}).
In the 0-and 1-lepton channels in the \hbbCAT\ the top-quark background is further reduced by rejecting the event
if any additional jet passes the tight c-tag criterion.%
In the boosted regime, the events are required to have no $b$-tagged track-jet outside the large-$R$ jet
in order to reduce the top-quark contribution.

The \hcc~contamination in the \hbbCAT\ is negligible, while the expected \hbb~yield in the $\ctc$ ($\ctn$) SRs of the
\hcc~category is approximately 1.5 (1.9) times that of the expected \hcc~process. %
Due to the different tagging requirement, the Hbb and Hcc categories are not orthogonal for events
with $\pTV>\SI{400}{\GeV}$. The \vhcc signal (data) events selected by the boosted Hbb regime represents at most 1.9\% (0.4\%) of
the total amount of \vhcc signal (data) events selected in the \hccCAT. This overlap has a negligible impact on the results.

\textbf{$\hdr$ control regions} are defined by
considering events with \drjj\ above the upper boundary (see above).
Such CRs are designed to have a similar composition of \Vjets backgrounds to the SR and  also serve to constrain the top-quark background.
In the Hcc category $\ctct$ events are considered in separate regions from $\ctcl$
events in order to better constrain different \Vjets flavour
components as discussed in \Sect{~\ref{sec:background}}.

\textbf{BB $\ldr$ control regions} are formed %
from 1-lepton Hbb resolved events that are below the lower boundary in
\drjj. The CRs are dominated by \ttbar\ and $W$+jets backgrounds  with an enhanced contribution of
$W$+jets  with respect to the SRs and $\hdr$ CRs.

\textbf{$\cln$ $V$+light  control regions} mimic the Hcc SRs  but consist of events
in which one of the Higgs boson candidate jets is a loose $c$-tagged jet and
the other is non-tagged. These CRs are used to constrain the backgrounds where a $W$ or $Z$ boson is
associated with light jets. This background is primarily relevant in the Hcc category.
These regions replace the previously 0 $c$-tag region used in \Refn{\cite{HIGG-2021-12}}.
\textbf{$\topbc$ control regions } in the 0- and 1-lepton channels
are defined with at least one tight $c$-tagged jet and at least one
$b$-tagged jet and are used to constrain the top-quark background in the resolved regime.
This CR targets the top-quark process where a $b$-jet and a $c$-jet from
the same hadronically decaying top-quark are considered as a Higgs boson candidate.

\textbf{$\topboost$ control regions} in the 0- and 1-lepton channels are defined by using the same selection on the large-$R$ jet on the boosted Hbb categories,
but requiring the presence of a $b$-tagged track-jet in the rest of the event. This CR enhances the fraction of $\ttbar$ events
by effectively targeting events with two $b$-tagged jets with large angular separation.

\textbf{$\topem$ control regions} are defined for the resolved regime by using the same selection as the SR but replacing the
same-flavour lepton selection with a selection containing an electron and a muon
in order to significantly increase the fraction of top-quark events.
In the
Hbb category, the $e\mu$  CRs  directly  provides the background
prediction for the top-quark background in a
data-driven way,
as it is detailed in \Sect{~\ref{sec:background}}.
In the Hcc category this data-driven method is not used due to low
statistical precision. Instead, the $e\mu$ CRs are used to provide a constraint
on the normalisation of the MC-based predictions. The CRs are not
split according to jet tagging and therefore events must have at least one tight $c$-tagged jet.

\subsection{Categorisation for the simplified template cross-section measurement}
In the \hbbCAT, cross-section measurements are conducted in an extended version of the $VH$, $V\to$ leptons stage-1.2 STXS region scheme~\cite{Badger:2016bpw,Berger:2019wnu}. %
In this scheme, \qqZH~and \ggZH~are treated as a single $ZH$ process, since there is currently not enough sensitivity to distinguish
between them.
The expected signal distributions and acceptance times efficiencies for each STXS region are estimated from the simulated signal samples by selecting events using information from the generator's `truth' record,
in particular the truth $\pTV$, denoted by \pTVt,~and the number of truth-level jets with $\pT>\SI{30}{\GeV}$ and $|\eta|<4.5$ (additional truth jets, $\Njet$) reconstructed from stable truth particles after excluding the Higgs boson decay products. The truth templates are categorised as a function of \pTVt following the same \pTV boundaries used in the definition of the SRs. Only for the $ZH$ process with  $\pTVt <\SI{400}{\GeV}$, a further separation is performed to
distinguish between events with zero or at least one additional truth jet. The choice is driven by the larger  top-quark background in the 1-lepton channel in events with one additional jet. A set of measurements is also made without this split in jet multiplicity.

\begin{figure}[tb!]
\centering
\includegraphics[width=1.0\textwidth]{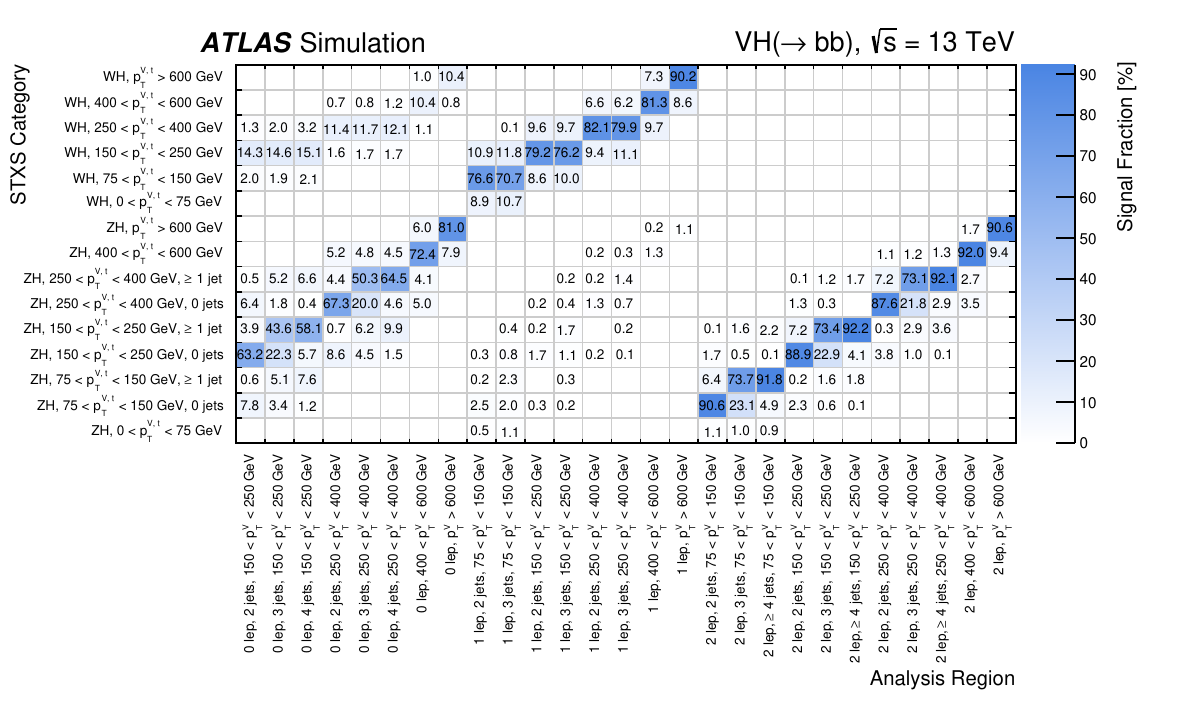}
\caption[]{The predicted fraction of  \vhbb\ signal events passing all selection criteria (in percent) in every reconstructed-event category ($x$-axis)
from each STXS category ($y$-axis). In the y-axis label, "jet" refers to additional truth jets, $\Njet$, as specified in the text. Entries with signal fractions below 0.1\% are not shown.}
\label{fig:STXScat}
\end{figure}
The fraction of events from each STXS category satisfying the
requirements of the SRs of the \hbbCAT\ is shown in
\Fig{\ref{fig:STXScat}}.  The correlation between `truth' and
reconstructed categories is improved with respect to the previous
iteration of the analysis thanks to the dedicated treatment of the
$\tau$ sub-channel, which reduces the $WH$ contribution in the
0-lepton channel, and the increase  in the  $\pT$ requirement  from 20 to $\SI{30}{\GeV}$
for the additional reconstructed
central jets to align the
definition with the targeted STXS observable.

%
%
%
%
%
%
%
%
%
%
%
%
%
%
%
%

%
%
%
%
%
%
%
%
%
%
%
%
%
%
%
%


%
%

\section{Multivariate discriminants}
\label{sec:mva}

A multivariate approach  is used to maximise the sensitivity of the
analysis.  Several %
BDTs are used in the
various signal regions of the analysis.  A nominal set, referred to as
BDT$_{VH}$, is designed to discriminate the $VH$ signal from the
background processes. A second set, BDT$_{VZ}$, which
aims to separate the $VZ$ diboson process from the $VH$ signal and other
background processes, is used to measure the $WZ$ and $ZZ$ processes,
which serve as a validation of the  analysis.  Finally, in the
1-lepton channel \ldr CR for the resolved \hbbCAT, additional
BDTs are trained to separate the $V$+jets events from the other background,
which is dominated by top-quark production, called BDT$_{\ldr\,\mathrm{CR}}$.

The BDTs are generally trained separately for each SR but to reduce
complexity, resolved SRs with $\ge 3$ jets, or with $\ptv<\SI{150}{\GeV}$ or
with $\ptv>\SI{150}{\GeV}$ are trained together.
The same training procedures as those detailed in \Refn{\cite{HIGG-2018-51}} are used.

The BDT input variables are listed in \Tab{\ref{tbl:MVAVars}} and
briefly discussed below.  Different sets of variables are used for the
0-, 1- and 2-lepton channels in resolved and boosted regimes, but the \hccCAT\
and \hbbCAT\ resolved channels use the same variables. The BDT$_{\ldr\,\mathrm{CR}}$
uses the same variables as the 1-lepton resolved \hbbCAT. Variables are only
included in the training if they improve the overall sensitivity. The distributions
of all input variables of the BDTs are compared between data and simulation, and
good agreement is found within the uncertainties.

The reconstructed Higgs boson mass $m_H$ is the most discriminating variable,
since no  backgrounds peaks at the same mass value. The
transverse momenta of the tagged jets or track-jets $\pT^{j_1}$ and
$\pT^{j_2}$ are used together with the scalar sum of additional jets with
$p_T>\SI{20}{\GeV}$ in the resolved channel, $\sum\limits\pT^{j_i}$ with $i> 2$, or the
third track-jet $\pT^{j_3}$ in the boosted regime. Flavour tagging
information for $j_1$ and $j_2$ is taken into account by using the
binned output of the \DLoner tagging algorithm
$\mathrm{bin}_{\DLoner}(j_{1})$ and
$\mathrm{bin}_{\DLoner}(j_{2})$, with bin boundaries as listed
in \Sect{\ref{sec:objects}}. All channels use $\ptv$, which is equivalent
to \met in the 0-lepton channel.  In the resolved channel, the three jet
system $m_{{j_1j_2j_3}}$ is built considering the Higgs boson candidate jets and the highest \pT additional jet not retained for the FSR recovery procedure.

The 1-lepton channel uses $\met$ as
an additional variable. In the 2-lepton resolved channel, where the signal does
not have significant \met, the variable $\met/\sqrt{\st}$ is used,
where $\st$ is the scalar sum of $H_T$ and the $\pt$ of the two
leptons.

Angular variables include $\Delta R(\vec{j_1},\vec{j_2})$ and the minimum
separation between either of the Higgs boson candidate jets and any other
jet, $\min[\Delta R(\vec{j_i}, \vec{j_1} \text{~or~} \vec{j_2})]$
with $i>2$, and the
azimuthal and rapidity difference between the reconstructed Higgs boson and
vector boson vectors, $|\Delta \phi(\vec{V},\vec{\HCand})|$ and
$|\Delta y(\vec{V},\vec{\HCand})|$ respectively.

Variables that are  used in all lepton channels of the boosted regime are the total
number of track-jets in the Higgs boson candidate jet, the number of additional
small-$R$ jets in the event and `colour ring', which is a variable
designed to exploit the difference in colour-flow between gluon
splitting and a decay from QCD singlet states~\cite{Buckley:2020kdp}.
Two variables are only used in the 0-lepton resolved channel: the
pseudorapidity difference between the Higgs boson candidate jets $|\Delta
\eta(\vec{j_{1}},\vec{j_{2}})|$ and the sum of $\htV$ and $\met$.
Variables that are only used in the 1-lepton resolved channel are $\mtw$, and the
reconstructed top-quark  mass assuming the event contains a top-quark,
$\mtop$~\cite{HIGG-2018-51}. The transverse momentum of the lepton
$\pT^{\ell}$ and $(\pT^{\ell} - \met)/\ptv$, which is a proxy for the
$W$ boson's lepton $\pt$ asymmetry, are only used in the 1-lepton boosted channel.
The dilepton mass $m_{\ell\ell}$ is used in the 2-lepton resolved channel
and  $\cos{\theta^*(\vec{\ell^-},\vec{V})}$, which
is calculated using the lepton direction in the $Z$ boson rest
frame and the flight direction of the $Z$ boson in the laboratory
frame, is used both in  the 2-lepton resolved and boosted regimes.

\clearpage
%
%
%
\begin{table}[!htbp]
\caption{Variables used for the multivariate discriminant in each of the channels. The \checkmark\ symbol indicates the inclusion of a variable. The BDT$_{\ldr\,\mathrm{CR}}$ uses the same variables as the 1-lepton resolved \hbbCAT as described in the text.}
\centering
\setlength\aboverulesep{0pt}
\setlength\belowrulesep{0pt}
\resizebox{\linewidth}{!}{%
\begin{tabular}{c|ccc|ccc}
\multicolumn{1}{c|}{} & \multicolumn{3}{c|}{Resolved~\vhbbcc} & \multicolumn{3}{c}{Boosted~\vhbb}
\\\toprule
Variable & 0-lepton & 1-lepton & 2-lepton & 0-lepton & 1-lepton & 2-lepton
\\\midrule
$m_H$
& \checkmark & \checkmark & \checkmark
& $\checkmark$ & $\checkmark$  & $\checkmark$
\\\midrule
$m_{{j_1 j_2 j_3}}$
& $\checkmark$ & $\checkmark$ & $\checkmark$
& & &
\\\midrule
$\pT^{j_1}$
& $\checkmark$ & $\checkmark$ & $\checkmark$
& $\checkmark$ & $\checkmark$ & $\checkmark$
\\\midrule
$\pT^{j_2}$
& $\checkmark$ & $\checkmark$ & $\checkmark$
& $\checkmark$ & $\checkmark$ & $\checkmark$
\\\midrule
$\pT^{j_3}$
& & &
& $\checkmark$ & $\checkmark$ & $\checkmark$
\\\midrule
$\sum\limits\pT^{j_i}$, $i> 2$
& $\checkmark$ & $\checkmark$ & $\checkmark$
& & &
\\\midrule
$\mathrm{bin}_{\DLoner}(j_{1})$
& $\checkmark$ & $\checkmark$ & $\checkmark$
& $\checkmark$ & $\checkmark$ & $\checkmark$
\\\midrule
$\mathrm{bin}_{\DLoner}(j_{2})$
& $\checkmark$ & $\checkmark$ & $\checkmark$
& $\checkmark$ & $\checkmark$ & $\checkmark$
\\\midrule
\pTV
& $\equiv \met $ & $\checkmark$  & $\checkmark$
& $\equiv \met $ & $\checkmark$  & $\checkmark$
\\\midrule
\met
& $\checkmark$ & $\checkmark$ &
& $\checkmark$  & $\checkmark$ &
\\\midrule
$\met/\sqrt{\st}$
& & & $\checkmark$
& & &
\\\midrule
$|\Delta \phi(\vec{V},\vec{\HCand})|$
& $\checkmark$ & $\checkmark$ & $\checkmark$
& $\checkmark$ & $\checkmark$ & $\checkmark$
\\\midrule
$|\Delta y(\vec{V},\vec{\HCand})|$
& & $\checkmark$ & $\checkmark$
& & $\checkmark$ & $\checkmark$
\\\midrule
$\Delta R(\vec{j_1},\vec{j_2})$
& $\checkmark$ & $\checkmark$ & $\checkmark$
& $\checkmark$ & $\checkmark$ & $\checkmark$
\\\midrule
$\min[\Delta R(\vec{j_i}, \vec{j_1} \text{~or~}  \vec{j_2})]$, $i>2$
& $\checkmark$ & $\checkmark$ &
& & &
\\\midrule
$N(\text{track-jets in $J$})$
& & &
& $\checkmark$ & $\checkmark$ & $\checkmark$
\\\midrule
$N(\text{add. small-$R$ jets})$
& & &
& $\checkmark$ & $\checkmark$ & $\checkmark$
\\\midrule
colour ring
& & &
& $\checkmark$ & $\checkmark$ & $\checkmark$
\\\midrule
$|\Delta \eta(\vec{j_{1}},\vec{j_{2}})|$
& $\checkmark$ & &
& & &
\\\midrule
\htV+\met
& $\checkmark$ & &
& & &
\\\midrule
\mtw
& & $\checkmark$ &
& & &
\\\midrule
\mtop
& & $\checkmark$ &
& & &
\\\midrule
$\mathrm{min}[\Delta\phi(\vec{\ell}, \vec{j_1} \text{~or~} \vec{j_2})]$
& & $\checkmark$  &
& & &
\\\midrule
$\pT^{\ell}$
& & &
& & $\checkmark$ &
\\\midrule

$(\pT^{\ell} - \met)/\ptv$
& & &
& & $\checkmark$ &
\\\midrule
$m_{\ell\ell}$
& & & $\checkmark$
& & &
\\\midrule
$\cos{\theta^*(\vec{\ell^-},\vec{V})}$
& & & $\checkmark$
& & & $\checkmark$
\\\bottomrule
\end{tabular}
}
\label{tbl:MVAVars}
\end{table}

\clearpage


%
%

%
\section{Background predictions}
\label{sec:background}

The leading backgrounds affecting the analysis are \ttbar\ and \Vjets production.
Subleading contributions arise from single top-quark production, %
multijet production and diboson production ($VV$);
for the latter case, the production of a vector boson and a $Z$ boson ($VZ$) decaying into $b\bar{b}$ or $c\bar{c}$ represent a process with similar topology to the $VH$
signal and is used to validate the analysis methodology.

The simulated event samples, which are summarised in \Sect{\ref{sec:data}}, are used to model all background processes except for the \ttbar~background in the Hbb 2-lepton resolved channel
and the multijet background in the 1-lepton channel, which are both estimated by using data-driven techniques.
In the following, a brief overview of the composition of the leading backgrounds and the techniques used to estimate them is given.

\begin{figure}[tb!]
\centering
\subfloat[]{
\includegraphics[width=0.5\textwidth]{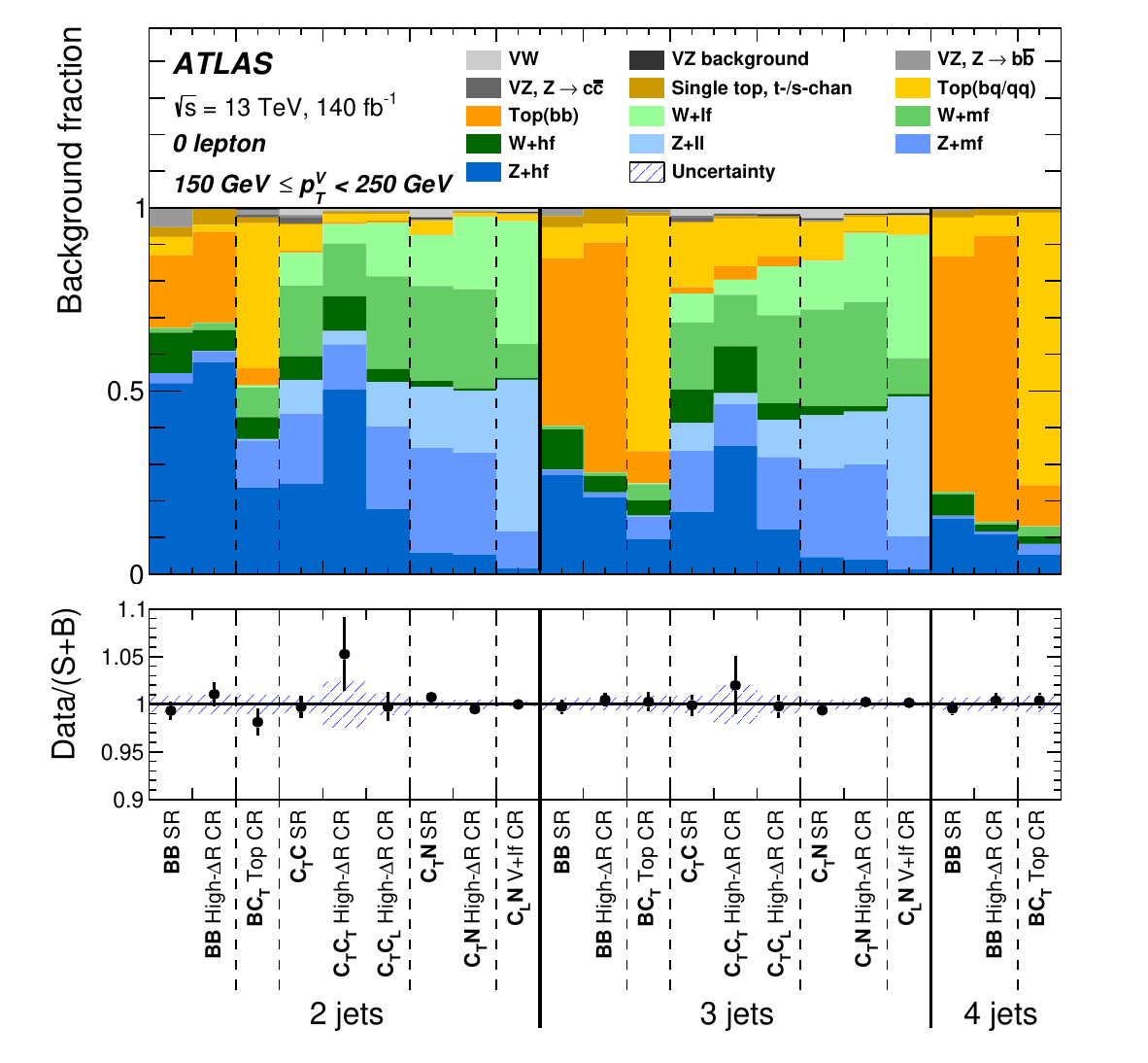}
\label{fig:BComp1a}
}
\subfloat[]{
\includegraphics[width=0.5\textwidth]{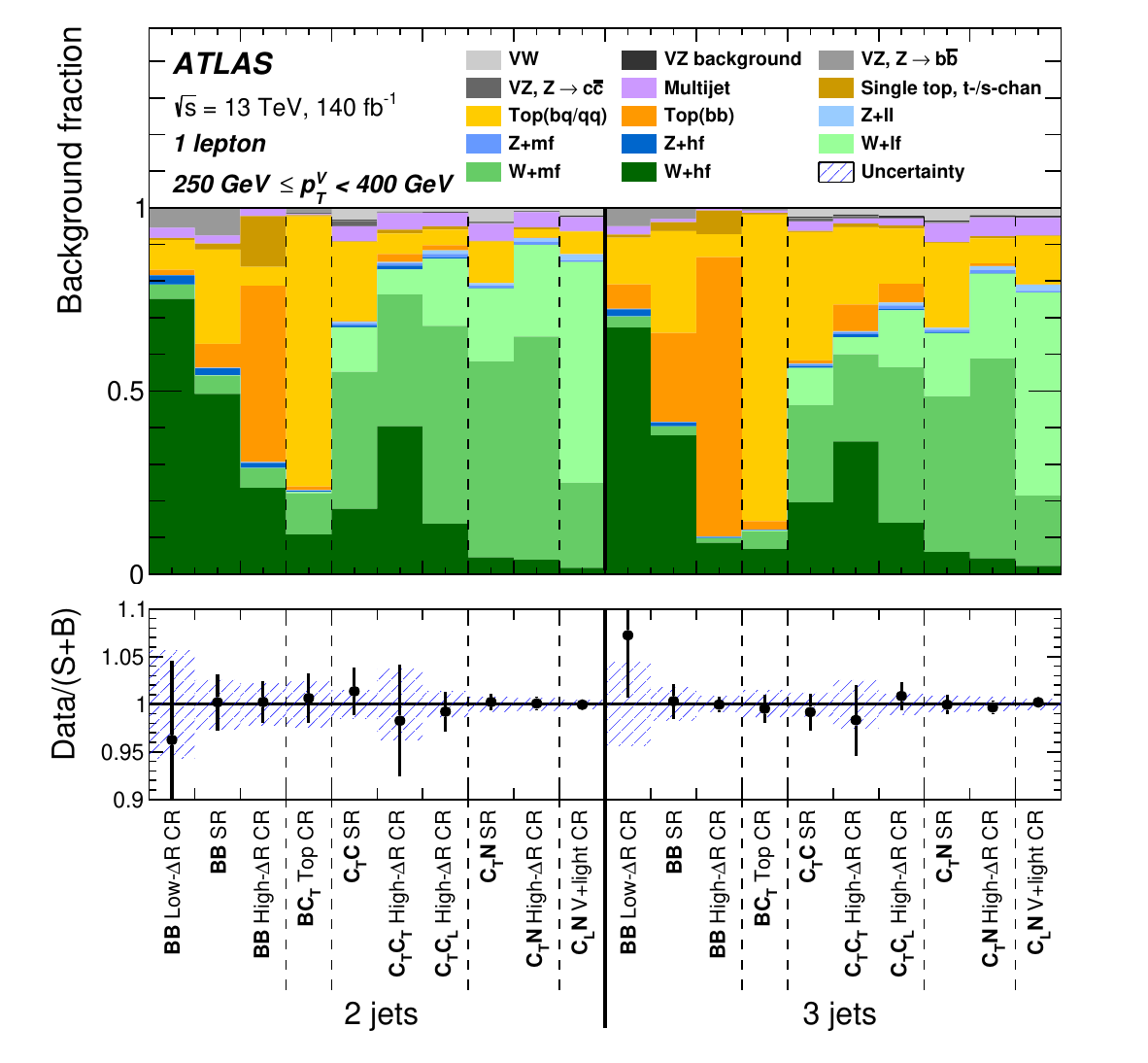}
\label{fig:BComp1b}
}
\caption[]{ The relative background composition in the signal and
control regions with \protect\subref{fig:BComp1a} $\SI{150}{\GeV} <\ptv<\SI{250}{\GeV}$ in
the 0-lepton channel and \protect\subref{fig:BComp1b}
$\SI{250}{\GeV}<\ptv<\SI{400}{\GeV}$ in the 1-lepton
channel. The ratio of the data to the total signal-plus-background (S+B) prediction is also
shown. The background predictions are adjusted with the results of   the $VH$ fit
to data described in \Sect{\ref{sec:stat}} and the uncertainty band
corresponds to the overall uncertainty in the background
predictions.}%
\label{fig:BComp1}
\end{figure}

\begin{figure}[tb!]
\centering
\subfloat[]{
\includegraphics[width=0.5\textwidth]{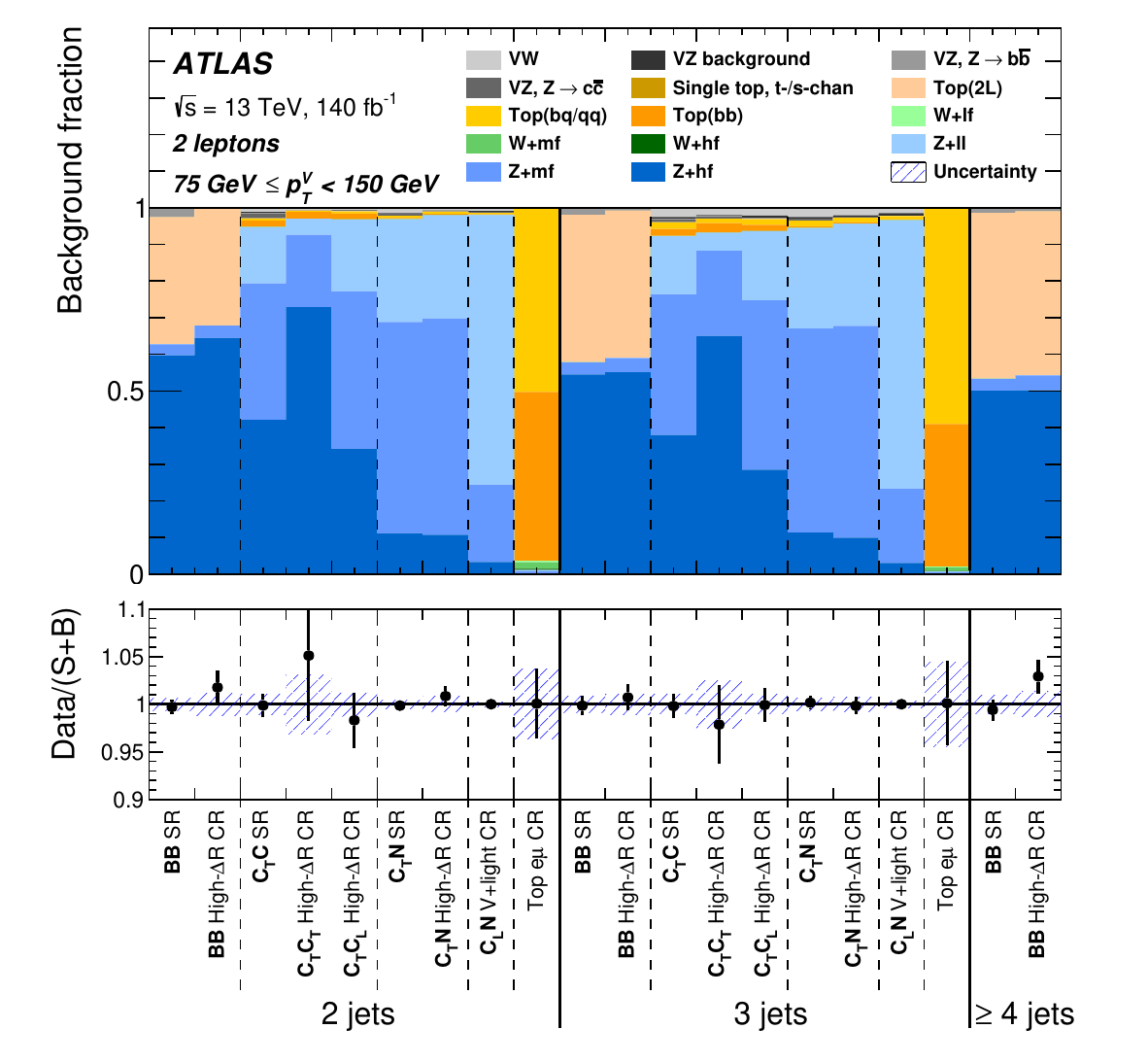}
\label{fig:BComp2a}
}
\subfloat[]{
\includegraphics[width=0.5\textwidth]{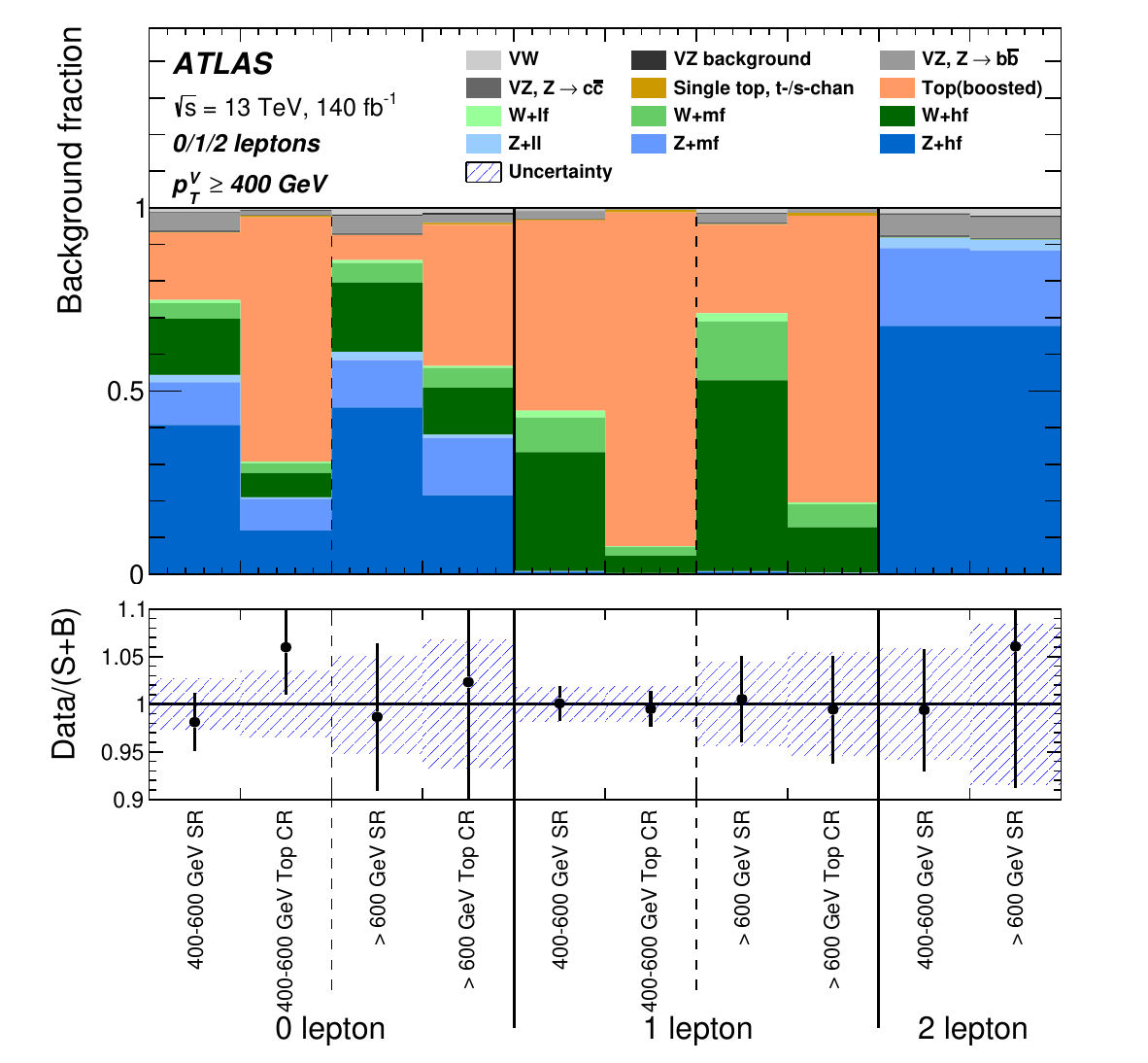}
\label{fig:BComp2b}
}
\caption[]{The relative background composition in the signal and control regions with
\protect\subref{fig:BComp2a} $\SI{75}{\GeV}<\ptv<\SI{150}{\GeV}$ in the 2-lepton channel and
\protect\subref{fig:BComp2b} all regions with $\ptv>\SI{400}{\GeV}$. The ratio of the data to the
total signal-plus-background (S+B) prediction is also
shown. The background predictions are adjusted with the results of   the $VH$ fit
to data described in \Sect{\ref{sec:stat}} and the uncertainty band
corresponds to the overall uncertainty in the background
predictions.}

\label{fig:BComp2}
\end{figure}

The detailed composition of background processes varies greatly across
signal and control regions. Examples of the background composition in
a representative $\ptv$ intervals for each of the three lepton channels  are shown in
\Figs{\ref{fig:BComp1}}{\ref{fig:BComp2}}.

\textbf{\Vjets:} \Vjets is a major background in all the SRs and its
relative importance increases with \ptv.  The 2-lepton channel
contribution is almost exclusively composed of \Zjets events, while
that of the 1-lepton channel is mostly \Wjet events. Both contributions are present in the
0-lepton channel with a predominance of \Zjet events. Simulated MC
events are categorised according to the flavours of $j_1$ and $j_2$:
\Vlf when they are both light-flavour; \Vhf for $cc$ and $bb$; mixed
flavour (\Vmf) for the remaining cases ($cl$, $bl$ and $bc$). Events from
the  $V{+}bb$ process  compose 85\% to 95\% of the \Vjets background in the resolved
Hbb regions; the fraction drops to 50\% in the boosted reconstruction
regions due to the relaxed flavour tagging
requirement.\footnote{Approximately 15\% of \Vjets events is $V{+}cc$ in
these regions.} %
In the Hcc category the flavour composition of \Vjets events is more
diverse and strongly depends on the tagging requirements. In
addition, the composition differs between \Wjets and \Zjets processes
due to $W{+}c$ production, which significantly enhances the fraction
of \Wmf events.  The \ctc\ SRs have a $V{+}cc$ fraction of 30\%--40\%
(15\%--20\%) for \Zjets (\Wjets); the values roughly double in the
\ctct\ \hdr\ CRs. Therefore, splitting the \hdr CR in \ctcl and \ctct events
provides better control of the \Vjets flavour composition.
More than 60\% of \Vjets events in the \ctn regions
are populated by \Vmf events; the composition is very consistent
between SRs and \hdr\  CRs.  Finally, \cln\ CRs are dominated by \Vlf
events.

The normalisations of the  \Vhf, \Vmf, and \Vlf components in the resolved regime
are free to float in the  fits to data independently in different
jet multiplicity (2-, or $\ge 3$-jets) and \ptv\ intervals.
For \Zhf there are additional normalisation parameters for the regions
with exactly three jets, to decouple the background normalisation between the 3-jet
and $\ge 4$-jet regions.
In the  boosted regime, a common normalisation is used for  the \Vhf and \Vmf components %
in the two  \pTV\ ranges.
The \Vhf and \Vmf components are constrained by the SRs, \ldr\ and  \hdr\ CRs, while
the \Vlf are constrained by the \cln regions.
In the 1-lepton channel of the resolved \hbbCAT, the \Whf normalisation constraints
are significantly improved with the introduction of a BDT for the $\ldr$ CRs (BDT$_{\ldr\,\mathrm{CR}}$) as described in \Sect{\ref{sec:mva}}. In comparison to a fit with only a single bin in the $\ldr$ CR, the uncertainty in the \Whf normalisation factor is reduced, in the $\SI{75}{\GeV}<\ptw<\SI{150}{\GeV}$ ($\SI{150}{\GeV}<\ptw<\SI{250}{\GeV}$) region by a factor of two (by 30\%).
In total,  42 free-floating normalisation factors are used in the  fits to achieve an adequate \Vjets background model.

\textbf{Top-quark background in 0- and 1-lepton channels:} Top-quark processes represent the leading background in several analysis regions of the 0-lepton and the 1-lepton channels;
its contribution is more relevant in the \hbbCAT than in the Hcc ones, which have a  veto on $b$-tagged jets.
The relative contribution strongly increases with jet multiplicity and decreases  as $\pTV$ increases.

The top-quark background primarily consists of \ttbar production and $Wt$ production, which are treated jointly
given the very similar final state topology and to minimise the impact of interference effects. The relative
fraction of $Wt$ with respect to the total top contribution varies from 5\% to 15\% with increasing \pTV in the \hbbCAT and from 15\% to 20\% in the \hccCAT,
depending on the tagging requirements.
Single-top $t$-channel background is most relevant in the \hbbCAT of the 1-lepton channel, reaching almost 10\% of the top-quark background in the
low \pTV SRs. The $s$-channel single-top process is negligible in all the analysis regions.
The contributions of \ttbar and $Wt$ arise primarily from events where one of the $W$ bosons decays
leptonically and the  other one hadronically; events enter the 0-lepton channel from cases where the lepton
is not reconstructed, fails to meet the identification criteria or is outside the detector acceptance.

In a similar way as the $V$+jets background, top-quark backgrounds are categorised
according to the truth flavour of $j_1$ and $j_2$.
The \topbb~process
primarily captures configurations where the Higgs boson candidate is built from the two $b$-jets from the top-quark decays.\footnote{The contribution of $\ttbar{+}\,\textit{hf}$ is suppressed by the upper requirement on the jet multiplicity.} The events are characterised by large values of $m_H$ and  $\drjj$ and represent the leading contribution in $\hdr$ CRs of the \hbbCAT, while in the \hccCAT~the relative contribution reaches at most 15\%.
The \topbq~contribution describes configurations  where the Higgs boson candidate is built from a $b$-jet and either a $c$- or light jet from the $W$ boson decay;  $m_H$ is therefore bound by the top-quark mass and the event  is more signal-like. The contribution increases with $\pTV$ and it represents up to 70\% of the relative fraction of the total top-quark background in the SRs of both Hbb and Hcc categories a well as the \topbc~CRs; in the latter, the usage of $m_{bc}$ as discriminating variable in the  fits further contributes to isolating the contribution of this critical background.
Finally, the \topqq~background represents events where the Higgs boson candidate jets are $c$- or light jets, and therefore it captures configurations where the Higgs boson candidate is primarily built from jets from the hadronically decaying $W$ boson;
its contribution is negligible in the \hbbCAT, while it reaches up to 40\% of the top-quark background for the loose tagging requirement regions in the \hccCAT.
In the boosted regime, the large-$R$ jet naturally captures the full hadronic decay of the top-quark. In \topbq~events, particularly  where one of the subjets is a $c$-jet, represent up to 85\% of the total top-quark background contribution. The value is consistent between SRs and the \topboost~CRs.

In the 0- and 1-lepton channels, separate free-floating normalisations are used for the \topbb component and the combined
\topbq and \topqq components, collectively referred to as the \topbqqq component. The normalisation factors are further
split depending on the jet multiplicity (2-, 3-, or 4-jet) and \ptv\ intervals. In total, there are 18 free-floating normalisation factors
for the top-quark background in the 0- and 1-lepton channels.

\textbf{Top-quark background in 2-lepton channels:} In the 2-lepton
channels the top-quark background comes mainly from di-leptonic
\ttbar\ and $Wt$ events.  In the Hbb resolved category samples are defined in the data following
a similar selection to the $\topem$ CR described in \Sect{\ref{sec:event}}. These samples have
the same selection as the SRs but require an electron and a muon rather than two same-flavour leptons
and are directly used as the background estimate. They provide top-quark samples with purity above 99\%
since most backgrounds and the signal are expected to contain same-flavour lepton pairs from a $Z$ boson decay.

In the Hcc category, where the
MC-based predictions are used, all components are correlated and there are five corresponding free-floating
normalisation factors in the fits for the \hccCAT, depending on \ptv and jet multiplicity.
Similarly for the boosted regime, since the top-quark background decreases rapidly with \ptv and is small in this region,
it is taken from MC simulation.

\textbf{Diboson:} Diboson production is comprised of the $WW$, $WZ$
and $ZZ$ processes and contributes  primarily to the
SRs.  In the Hbb resolved category, the $VZ$, \zbb\ process contributes approximately
to 90\% of the overall diboson events; the fraction
decreases to 60\% for $\ptv>\SI{400}{\GeV}$ due the  relaxed
tagging requirement in the boosted reconstruction.  In the \hccCAT,
the contribution from $VZ$, \zcc\ varies between
10\% and 40\% depending on the tagging requirements. The leading
process is $VW$, $W \to cs$.
The \zbb\  contamination in the Hcc category is always lower
than 5\% thanks to the excellent performance of the tagging algorithm,
while the contribution from other hadronic $Z$ boson decays is only
relevant in the \ctn~SRs.

\textbf{Multijet:}
The production of events with multiple jets, multijet (MJ), has a large cross-section and thus, despite not being a source of genuine missing transverse momentum or prompt leptons, can contribute a non-negligible amount of background in the analysis regions.
In the 1-lepton channel, the MJ background is
estimated in each analysis region by performing a template fit to the \mtw~distribution, which offers
discrimination between MJ and other background and signal processes. The fit is performed as a
preliminary step prior to the final analysis fit and it is used to extract overall normalisation of the MJ template
and the related uncertainties.
The shape of the multijet template is evaluated from data using a dedicated CR,
defined as the nominal event selection, except the strict lepton isolation requirement is inverted.
Contributions from other backgrounds in this CR are estimated with simulation and subtracted.
The MJ background was demonstrated to be negligible in both the 0- and 2-lepton channels and all boosted regions following the procedure described in Refs.~\cite{HIGG-2018-04,HIGG-2018-52,HIGG-2021-12}.

%
%
%
%
%
%
%
%
%
%
%
%
%
%
%
%
%
%
%
%
%
%
%
%
%
%
%
%
%
%
%
%
%
%
%
%
%
%
%
%
%
%
%
%
%
%
%
%
%
%
%
%
%
%
%
%
%
%
%
%
%
%
%
%
%


%
%

\section{Systematic uncertainties}
\label{sec:sys}

The sources of systematic uncertainty can be broadly divided into three groups: those of an experimental nature, those related to the theoretical modelling of the backgrounds and those associated with the Higgs boson signal simulation.
The estimation of the uncertainties builds upon the methodology outlined in Refs.~\cite{HIGG-2018-51,HIGG-2018-52,HIGG-2021-12} and is briefly summarised below.
The breakdown of the systematic uncertainties and their impact on the signal strength extraction is shown in \Sect{\ref{sec:res}}.

\subsection{Experimental uncertainties}
The dominant experimental uncertainties originate from the flavour-tagging performance, jet energy scale calibration and the modelling of the jet energy resolution.

The uncertainty in the flavour tagging correction factors (see \Sect{\ref{sec:objects}}) is decomposed into several independent and uncorrelated contributions. Additional simulation-based uncertainties are derived to cover the extrapolation to higher-\pT regimes outside the validity range of the calibrations;
following the procedure laid out in Refs.~\cite{FTAG-2018-01,FTAG-2020-08,FTAG-2019-02}.
Flavour-tagging uncertainties are treated as uncorrelated between small-$R$ and VR track-jets.

For small-$R$ jets, the uncertainties in the calibration of the energy scale and resolution,
derived in \Refn{\cite{JETM-2018-05}}; %
an additional uncertainty in the energy calibration of $b$- and $c$-jets is also included. %
Furthermore, an uncertainty is assigned to the efficiency of the jet-vertex-tagging requirement on jets~\cite{PERF-2014-03}.
For large-$R$ jets, the uncertainties in the energy and mass scales are based on a comparison of the ratio of calorimeter-based to track-based measurements in dijet data and simulation, as described in Ref.~\cite{JETM-2018-02}.

Uncertainties in the reconstruction, identification, isolation and trigger efficiencies of muons~\cite{MUON-2018-03}
and electrons~\cite{EGAM-2018-01} are considered, along with the uncertainty in their energy scale and resolution.
The uncertainties in the energy scale and resolution of the jets and leptons are propagated to the calculation of \met, which also has additional uncertainties from the modelling
of the underlying event and momentum scale, momentum resolution
and reconstruction efficiency of the tracks used to compute the soft-term~\autocite{JETM-2020-03}.
A dedicated uncertainty in the \met\ trigger selection efficiency, derived from the comparison of data and simulation
as explained in \Refns{\cite{HIGG-2018-51}}{\cite{HIGG-2021-12}}, is also included.

An uncertainty of 0.83\% in the total integrated luminosity~\cite{DAPR-2021-01}, obtained using the LUCID-2 detector~\cite{LUCID2} for the primary luminosity measurements and complemented by measurements using the inner detector and calorimeters, is
assigned to physics processes whose normalisations are taken from simulation. An uncertainty arising from the correction
of the \pileup distribution in simulation compared with that in data based on the measurement of the visible cross-section in minimum-bias events~\cite{STDM-2015-05}, is also considered.

\subsection{Background uncertainties}
\label{sec:bkgd}

Background modelling uncertainties for the simulated samples cover three areas: absolute normalisation uncertainties;
relative acceptance uncertainties that account for the differences in normalisation between samples with a common floating normalisation factor;
and shape uncertainties that account for uncertainties in the shapes of the different discriminants fit in the various regions, as listed in \Tab{\ref{tab:CRvars}}.

Shape uncertainties derived by comparing the nominal and alternative MC samples commonly suffer from statistical fluctuations, especially in the tails of the distributions.
In \Refn{\cite{HIGG-2018-51}}, this issue was mitigated with the BDT$_{S}$ reweighting technique, which is now further improved by the neural-based Calibrated Likelihood Ratio Estimator (CARL) algorithm~\cite{cranmer2016approximatinglikelihoodratioscalibrated}.
Utilising dense neural networks, the nominal and alternative (systematic variation) MC samples for a given
physics process are used in a classification problem to train an optimal classifier, using the same kinematic variables as the ones
used in the MVA  described in \Sect{\ref{sec:mva}}.
A transformation of the neural network output score is used to estimate the density ratio of the probability density functions for the alternative and nominal processes, as described in \Refn{\cite{rizvi2024learninglikelihoodratiosneural}}. The CARL methodology then reweights the MC simulation to match the probability distribution of the alternative process.
Due to the larger statistical sample size of the nominal dataset compared with the alternative, the
reweighting process yields a pseudo estimate of the alternative MC prediction with higher statistical precision, thereby providing
shape uncertainties that are less susceptible to statistical fluctuations.

\textbf{$V$+jets production:} The \Vjets normalisation uncertainties and relative acceptance uncertainties are estimated by varying the
renormalisation (\muR) and factorisation (\muF) scales by factors of 0.5 and 2.0, by varying the PDF, by varying the nominal scheme
of the inclusion of the virtual NLO EW corrections compared with the NLO QCD corrections (\textit{additive}) to alternative schemes (\textit{multiplicative} and \textit{exponentiated}).
These uncertainties are obtained by constructing an envelope from the alternative generator weights and using the LHAPDF prescription~\cite{Butterworth:2014efa} for the PDF uncertainty.
Furthermore, shape uncertainties are derived with the CARL technique by comparing the nominal \SHERPA[2.2.11] sample to an alternative
\Vjets sample produced with \MGNLOPY[8]~\cite{Alwall:2014hca} using FxFx merging~\cite{Frederix:2012ps}. The \MGNLOPY[8] sample is produced with up
to three additional partons at NLO accuracy and is described in more detail in \Refn{\cite{PMGR-2021-01}}.
All \Vjets modelling uncertainties are treated as uncorrelated between \Wjets and \Zjets processes.

Uncertainties are estimated for the relative normalisation of the heavy-flavour components
that constitute the \Vhf~and \Vmf~backgrounds. These uncertainties account for the relative difference
in the normalisation of the $bb$ to $cc$, $bl$ to $bc$, $bl$ to $cl$, and $bc$ to $cl$   components,
as a function  of \ptv and jet multiplicity. In the boosted regime, where \Vhf~and \Vmf~are floated coherently,
additional uncertainties are also considered in the relative ratio of these components.

Relative acceptance uncertainties for the \Wjets backgrounds are estimated for the ratio of the event yield in
the 0-lepton channel to that in the 1-lepton channel. For the \Zjets backgrounds, there is a relative acceptance
uncertainty in the ratio of the event yield in the 0-lepton channel to that in the 2-lepton channel. For both
\Wjets and \Zjets, relative acceptance uncertainties are estimated for the ratio of the event yield in the SR
to that in the \ldr\ and \hdr\ CRs. Furthermore, relative acceptance uncertainties are estimated for the ratio of the event yield
between $\SI{400}{\GeV} < \pTZ < \SI{600}{\GeV}$ and $\pTZ > \SI{600}{\GeV}$ regions where there is a common
\Zjets normalisation factor and for the 3-jet and 4-jet regions, which are covered by a common \Wjets normalisation factor
in the 0L \hbbCAT.

Additional shape uncertainties are estimated for the \Vjets backgrounds using the CARL technique by reweighting
the nominal \SHERPA[2.2.11] sample to the alternative \MGNLOPY[8] sample separately in two regions defined by
the $\Delta R(j_1, j_2) = 1.0$ threshold.\footnote{The $\Delta R(j_1, j_2) = 1.0$ value approximately corresponds to the
maximum kinematically allowed separation between two jets originating from gluon splitting with the ME+PS matching scale set
at $\SI{20}{\GeV}$, following the $m^2 \approx \pT^{j_1} \pT^{j_2} \Delta R(j_1, j_2)^2$ relation.} The corrections obtained in the \ldr\ CRs,
which are sensitive to the modelling of the parton shower and the matching between matrix element and parton shower, are not
necessarily applicable also to the \hdr\ regions.
Shape uncertainties are also derived for the \muR and \muF scale variations and variations in the scheme of the virtual NLO EW corrections.
Furthermore, dedicated \pTV shape uncertainties are derived in addition to the scale uncertainties by
comparing the \SHERPA[2.2.11] and  \SHERPA[2.2.1] MC samples separately for each \Vjets component.
These uncertainties are not covered by the \muR and \muF variations and provide the needed flexibility in the
likelihood fit to correct the \pTV spectra within a given \pTV bin.
The scale setting algorithms in \SHERPA[2.2.11] were updated for better computational performance, as explained in \Refn{\cite{PMGR-2021-01}}, which lead to 10\%-level differences in the \pTV shape between the two samples.
It is found that the central value of the \pTV shape modelled by \SHERPA[2.2.1] matches the data better in the
$\pTV < \SI{400}{\GeV}$ region while the \SHERPA[2.2.11] \pTV shape matches the data better in the $\pTV > \SI{400}{\GeV}$ region.
Most of these \pTV shape uncertainties are constrained in the CRs by directly fitting the \pTV distributions as summarised
in \Tab{\ref{tab:CRvars}}.

\textbf{$\ttbar$ and $Wt$ production:} Modelling systematic uncertainties are derived from comparisons between the nominal sample (\POWPY[8]) and alternative samples corresponding to matrix-element (\MGNLOPY[8]) and parton-shower (\POWHER[7]~\cite{Bellm:2015jjp}) generator variations.
The corresponding shape uncertainties are derived using CARL reweighting, similar to the \Vjets background. In addition, the impact of additional
radiation is assessed using (\MGNLOPY[8]) samples with modified parameter values for initial-state radiation (ISR) and  FSR.

Uncertainties in the relative composition of the two commonly normalised components, \topbq and \topqq, and the relative composition
between the \ttbar and $Wt$ processes, are estimated from the difference in their relative rates between the nominal
sample and the alternative matrix element and parton shower generator samples.  Furthermore, relative acceptance
uncertainties are estimated for the ratio of the top-quark event yields in the SRs
to that in the \ldr\ and \hdr\ CRs and for the 0-lepton to 1-lepton channel event yield ratios.
Lastly, dedicated acceptance plus normalisation uncertainties are evaluated from comparisons between the nominal
sample using the diagram subtraction scheme~\cite{Frixione:2007vw} and the diagram removal scheme~\cite{Frixione:2008yi}
to account for the interference between $Wt$ and \ttbar production.

The uncertainty in the top-quark background in the 2-lepton Hbb resolved channel is dominated
by the statistical uncertainty of the data control regions used to estimate it.

\textbf{Single top $t$-/$s$-channel:} The single top $t$-channel normalisation is constrained
by the theory prediction with a normalisation uncertainty applied, derived from the internal
MC renormalisation and factorisation scale and PDF variations. The acceptance uncertainties are estimated for the extrapolation between
the SRs and \ldr\ and \hdr\ CRs, between different \ptv regions, between different jet multiplicities, and between
the 0-lepton and 1-lepton channels. The shape uncertainties are derived using the CARL technique
from comparisons between the nominal sample (\POWPY[8]) and alternative samples corresponding
to matrix-element (\MGNLOPY[8]) and parton-shower (\POWHER[7]) generator variations, and from the impact
of additional radiation using \MGNLOPY[8] samples with modified parameter values for ISR and FSR.
For the single top $s$-channel, only a flat normalisation uncertainty is applied as its contribution to the total background is small.

\textbf{Diboson production:} The shape uncertainties, normalisation and relative acceptance uncertainties
are estimated similarly to the \Vjets processes:
by comparing the nominal \SHERPA[2.2.11] samples to the \SHERPA[2.2.1] samples, by comparing the nominal samples to
\POWPY[8] diboson samples (also used in previous results~\cite{HIGG-2018-51,HIGG-2018-52,HIGG-2021-12}),
by varying the renormalisation and factorisation scales, by varying
the PDF, and by varying the scheme of the virtual NLO EW corrections. The shape uncertainties from two-sample comparisons
are estimated by using the CARL technique. In addition to the different scale choice between
\SHERPA[2.2.11] and \SHERPA[2.2.1] samples, \SHERPA[2.2.11] samples have retuned heavy-flavour hadron
production rates and $b$-hadron fragmentation functions~\cite{PMGR-2021-01}, which lead to shape differences
between the \zbb\ and \zcc\ dijet invariant mass distributions.

\textbf{Multijet production:} The shape and normalisation uncertainties in the multijet background
in the 1-lepton channel are derived by following the procedure outlined in \Refn{\cite{HIGG-2018-04}}.
Two different uncertainty components are considered, those that alter the normalisation and those that alter the multijet template shape.

\subsection{Signal uncertainties}
\label{sec:sigmod}
The systematic uncertainties that affect the modelling of the signal %
are estimated with procedures that closely follow those outlined in Refs.~{\cite{HIGG-2018-50, ATL-PHYS-PUB-2018-035, Bendavid:2018nar, deFlorian:2016spz}}.
The systematic uncertainties in the calculations of the $VH$ production cross-sections and the $H\to\bbbar$, $\ccbar$  branching fraction\footnote{These systematic uncertainties are fully degenerate with the signal yield and do not affect the calculation of the significance relative to the background-only prediction and STXS cross-section measurement.} are assigned following the recommendations of the LHC Higgs Cross Section Working Group~\cite{Dittmaier:2011ti,Dittmaier:2012vm,Heinemeyer:2013tqa,Brein:2012ne}.

Uncertainties in acceptance and shape of the BDT output are estimated, as described in \Refn{\cite{HIGG-2018-04}}, from renormalisation and factorisation scale variations, PDF and $\alphas$ (PDF+$\alphas$)
uncertainties, from varying the parton shower and underlying event (PS/UE) models using \PYTHIA[8] internal variations and from comparisons with alternative parton-shower generator samples (\POWHER[7]).
The effects are treated as uncorrelated between $qq$-initiated and $gg$-initiated processes.

In addition, a systematic uncertainty from higher-order EW corrections is taken into account as a variation in the shape of the \ptv distributions for $VH$ production. %
Acceptance uncertainties, evaluated according to STXS regions,  accounting for the migration and correlations between regions, are evaluated for the scale variations, PS/UE models and PDF+$\alphas$.
All the contributions, apart from the branching ratio uncertainties, are treated as fully correlated between the $H\to\bbbar$ and $H\to\ccbar$ processes.

For the STXS measurement, the signal uncertainties are separated into
two groups. The first group contains uncertainties in the
acceptance and shape of kinematic distributions, which alter the
signal modelling (theoretical modelling uncertainties). The second set
contains uncertainties in the prediction of the production cross-section for
each kinematic region (theoretical cross-section uncertainties).
Whilst theoretical modelling uncertainties enter the STXS
measurements, theoretical cross-section uncertainties only affect the
predictions with which they are compared, and are therefore not
included in the likelihood function.

%


%
%

\section{Statistical model}
\label{sec:stat}

The statistical procedure is based on fits using a likelihood function $\mathcal{L}(\mu, \vec{\theta})$ constructed as the product of
Poisson probability terms over the bins of the input distributions, with parameters of interest (POIs) $\vec{\mu}$ extracted by maximising the
likelihood.
The effects of systematic uncertainties enter the likelihood as nuisance parameters (NPs), $\vec{\theta}$. The
normalisations of the largest backgrounds, top-quark and $V$+jets processes in several analysis regions, are determined by the fit, so they are left unconstrained in the
likelihood.
Experimental, background and signal systematic uncertainties, discussed in \Sect{\ref{sec:background}}, are described by NPs that are constrained with
Gaussian or log-normal probability density functions.  The uncertainties due to the limited number of events in the simulated
samples used for the background predictions are included using the ``light'' Beeston--Barlow technique~\autocite{Barlow:1993dm}. As detailed in
\Refn{\cite{HIGG-2013-23}}, systematic variations that are subject to large statistical fluctuations are smoothed, and systematic
uncertainties that have a negligible impact on the final results are pruned away region-by-region (treating signal and control regions
separately). %

The global likelihood fit comprises 59 signal regions (27 targeting \hbb and 32 targeting \hcc), where BDT distributions are considered, %
and 97 control regions. The list of variables used in each CR of the resolved regime is shown in \Tab{\ref{tab:CRvars}}; in the boosted regime the mass of the large-$R$ jet is considered in the Top CRs.
The variables are selected by balancing fit complexity, data statistical uncertainty and the capability to control critical background modelling uncertainties.

The binning of the BDT distributions is determined following the algorithm defined in \Refn{\cite{HIGG-2016-29}} to obtain a smoother distribution for the background processes and finer binning in the regions with the largest signal contribution, whilst ensuring that the statistical uncertainty of the
simulated background is less than 20\% in each bin.  In addition, bins are required to have at least three expected events considering the sum of the total backgrounds and the SM expected signals.
The final number of bins for each distribution varies from 15 to 3 depending on the expected number of background events. The analysis
is run blinded meaning all selection criteria and the systematic model are
determined without looking at the data in any histogram bin with
significant expected signal or looking at signal yields from the
likelihood fit.

\begin{table}[tbhp]
\caption{A schematic of the fit variables used in the control regions.
In the signal regions, the BDT output is used as the fit variable.
The `Norm. Only' label indicates that only the event yield is used in the fits
and `---' indicates that the region is not used in the fits.}
\label{tab:CRvars}
\renewcommand{\arraystretch}{1.3}
\setlength\tabcolsep{5.2pt}
\centering
\begin{tabular}{
l | l | c | c | c | c | c | c
}
\toprule
Channel & Region & \bb & \ctn & \ctcl & \ctct & \bct & \cln \\
\midrule
\multirow{4}{*}{0-lepton}
& \hdr CR & \multicolumn{4}{c|}{Norm. Only} & \multicolumn{2}{c}{---} \\
\cmidrule{2-8}
& \topbc CR & \multicolumn{4}{c|}{---} & $m_{j_1 j_2}$ & --- \\
\cmidrule{2-8}
& \Vlf CR & \multicolumn{5}{c|}{---} & Norm. Only \\
\toprule
\toprule
\multirow{5}{*}{1-lepton}
& \ldr CR & BDT$_{\ldr\,\mathrm{CR}}$ & \multicolumn{5}{c}{---} \\
\cmidrule{2-8}
& \hdr CR & \multicolumn{2}{c|}{\ptv} & \multicolumn{2}{c|}{$m_{j_1 j_2}$} & \multicolumn{2}{c}{---} \\
\cmidrule{2-8}
& \topbc CR & \multicolumn{4}{c|}{---} & $m_{j_1 j_2}$ & --- \\
\cmidrule{2-8}
& \Vlf CR & \multicolumn{5}{c|}{---} & \ptv \\
\toprule
\toprule
\multirow{4}{*}{2-lepton}
& \hdr CR & \multicolumn{2}{c|}{\ptv} & \multicolumn{2}{c|}{$m_{j_1 j_2}$} & \multicolumn{2}{c}{---} \\
\cmidrule{2-8}
& \topem CR & --- &\multicolumn{3}{c|}{Norm. Only} & -- & --- \\
\cmidrule{2-8}
& \Vlf CR & \multicolumn{5}{c|}{---} & \ptv \\
\toprule
\end{tabular}
\end{table}

Multiple alternative sets of POIs are probed by the analysis.
First in the $VH$ fit, two signal strength parameters, $\mu_{VH}^{bb}$ and $\mu_{VH}^{cc}$, are considered as multiplication factors that scale the expected \vhbb~and \vhcc~predictions respectively. An upper limit at 95\% confidence level (CL) on the \vhcc\ signal strength is computed using a modified frequentist CLs method~\cite{Read:2002hq}, with the profile-likelihood ratio as the test statistic~\cite{Cowan:2010js}.
Second, a fit with 4-POI is performed to extract simultaneously the signal strength for the \wh~and \zh~productions in \hbb and \hcc ($\mu_{WH}^{bb}$, $\mu_{ZH}^{bb}$, $\mu_{WH}^{cc}$, $\mu_{ZH}^{cc}$).
Third, a 13-POI fit version measures the signal cross-section multiplied by the \hbb~ and $V\to$leptons branching fractions in the STXS categories defined in \Fig{\ref{fig:STXScat}}. For the categories with $\pTVt<\SI{75}{\GeV}$ no dedicated signal region exist and their contribution is fixed to the SM predictions including the full set of uncertainties described in
\Sect{\ref{sec:sigmod}}. For this fit configuration, an inclusive signal strength is used for the \vhcc~component. Fourth, an additional 10-POI fit is performed where no split is considered in the number of jets (reduced STXS schema).

Finally, alternative fits are performed to extract signal strengths
for the diboson processes as a cross-check. First a 2-POI fit is used to extract $\mu_{VZ}^{bb}$ and $\mu_{VZ}^{cc}$, then a
4-POI $VZ$ fit measures  the signal strength of the
$WZ$ and $ZZ$, with \zcc\ or \zbb. These fits use the BDT
$VZ$ output distributions instead of the $VH$ BDT. The SM Higgs boson
is included as a background process normalised to the predicted SM
cross-section and including the full set of uncertainties described in
\Sect{\ref{sec:sigmod}}.
%

%
%


%
%

\section{Results}
\label{sec:res}

The background predictions in all post-fit distributions and tables are obtained by normalising the
backgrounds and setting the nuisance parameters according to the values determined by the fit configuration with two inclusive signal strengths.

The post-fit normalisation factors of the unconstrained backgrounds in
the $VH$ fit are shown in \Tabrange{\ref{tab:WNF}}{\ref{tab:TNF}}. The
\Vjets normalisation factors have a similar behaviour for \Wjets and
\Zjets; the normalisation factors increase with the number of
additional jets and with \ptv. The same trends are observed in a
dedicated measurement of the \Zhf processes in
\Refn{\cite{ATLAS:2024tnr}} for comparable multijet merged MC samples
generated at NLO precision.  The normalisation factors for the top-quark
background processes show that the NLO \POWPY[8] generator
set-up overshoots the data with increasing \ptv, similarly to the
trend observed in a dedicated measurement of the top-quark transverse
momentum in \ttbar events~\cite{TOPQ-2019-23}.

\begin{figure}[tb!]
\flushright
\subfloat[]{
\includegraphics[width=0.32\textwidth]{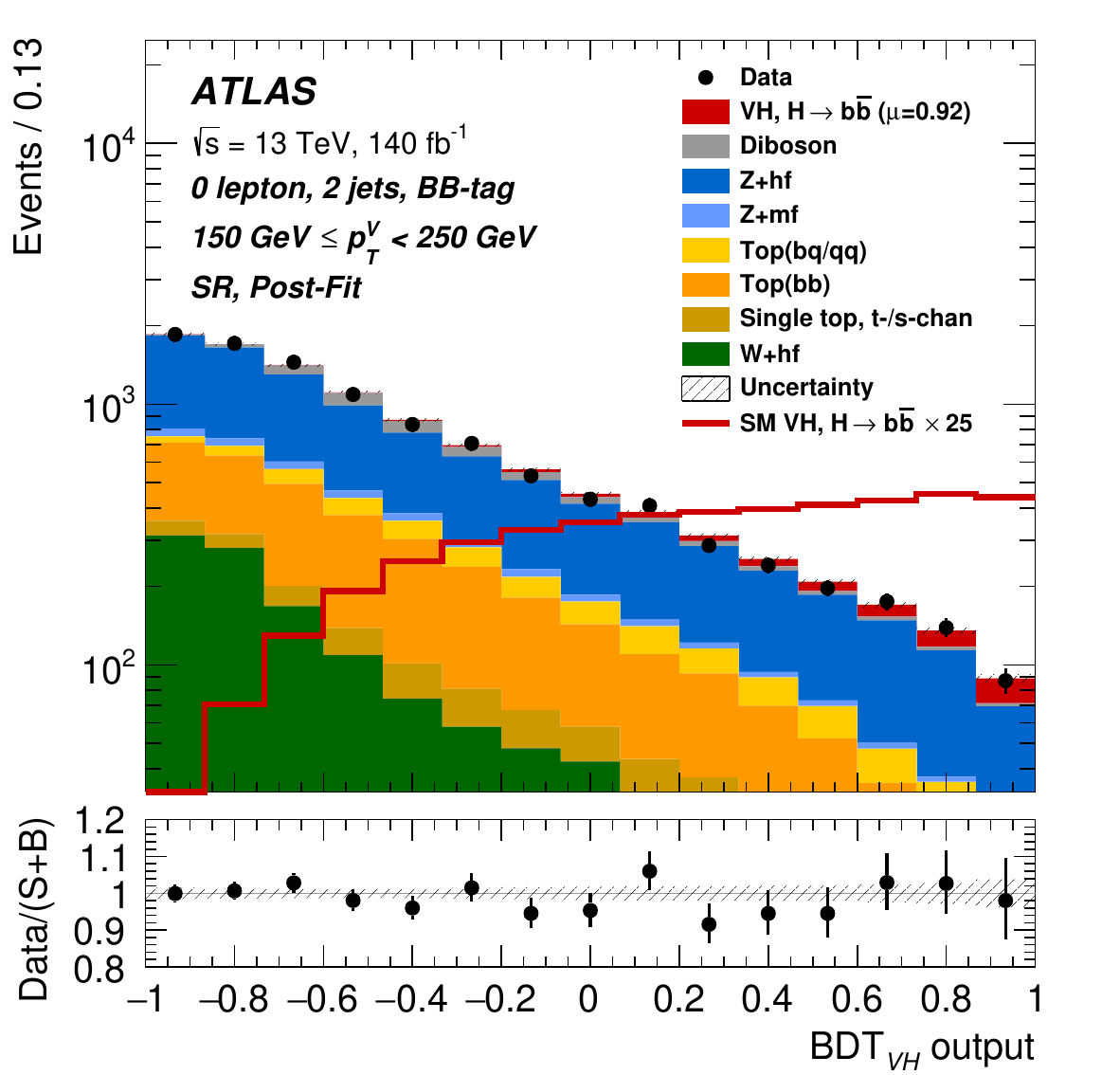}
\label{fig:BDT_1a}
}
\subfloat[]{
\includegraphics[width=0.32\textwidth]{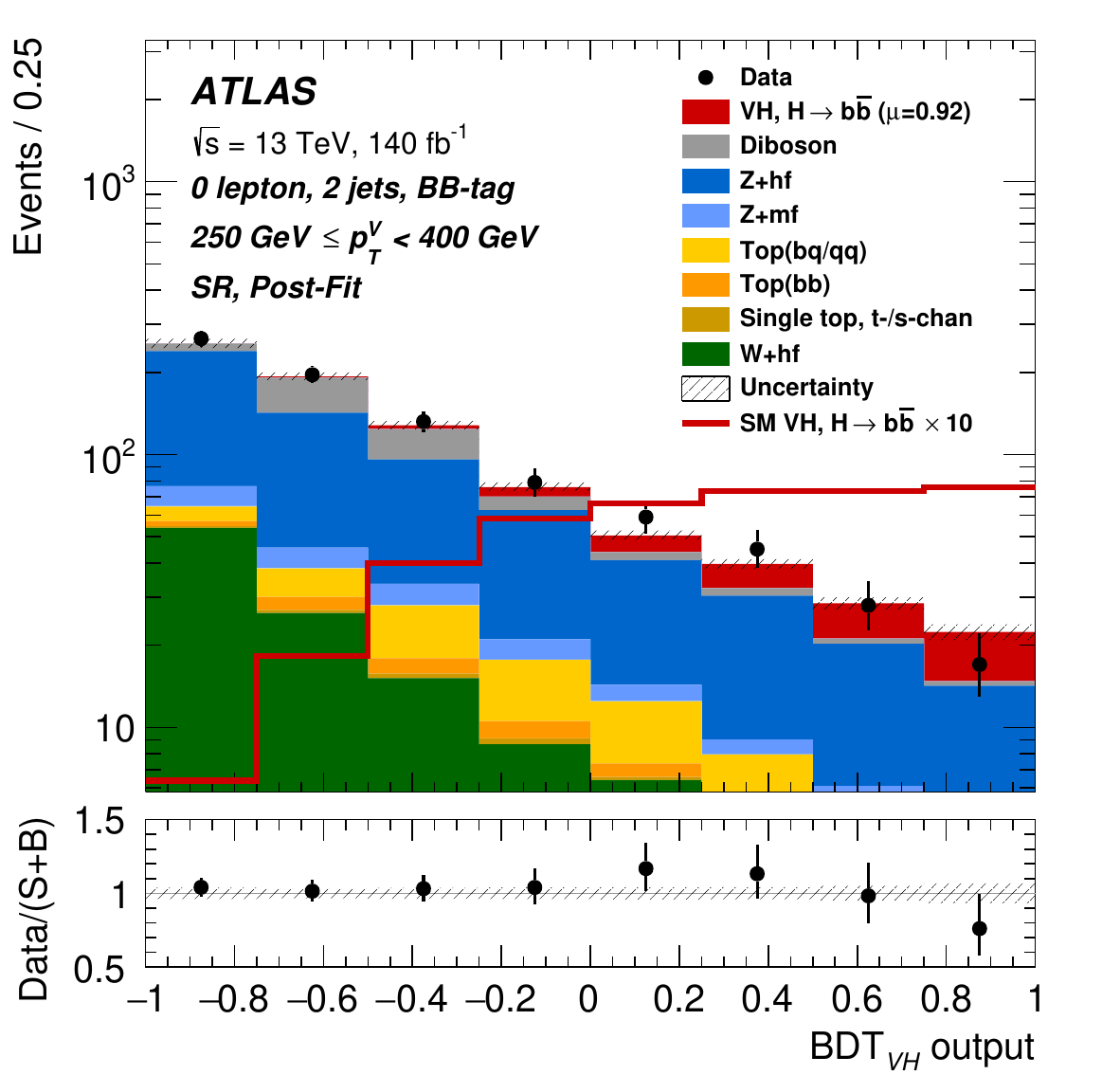}
\label{fig:BDT_1b}
}\\
\subfloat[]{
\includegraphics[width=0.32\textwidth]{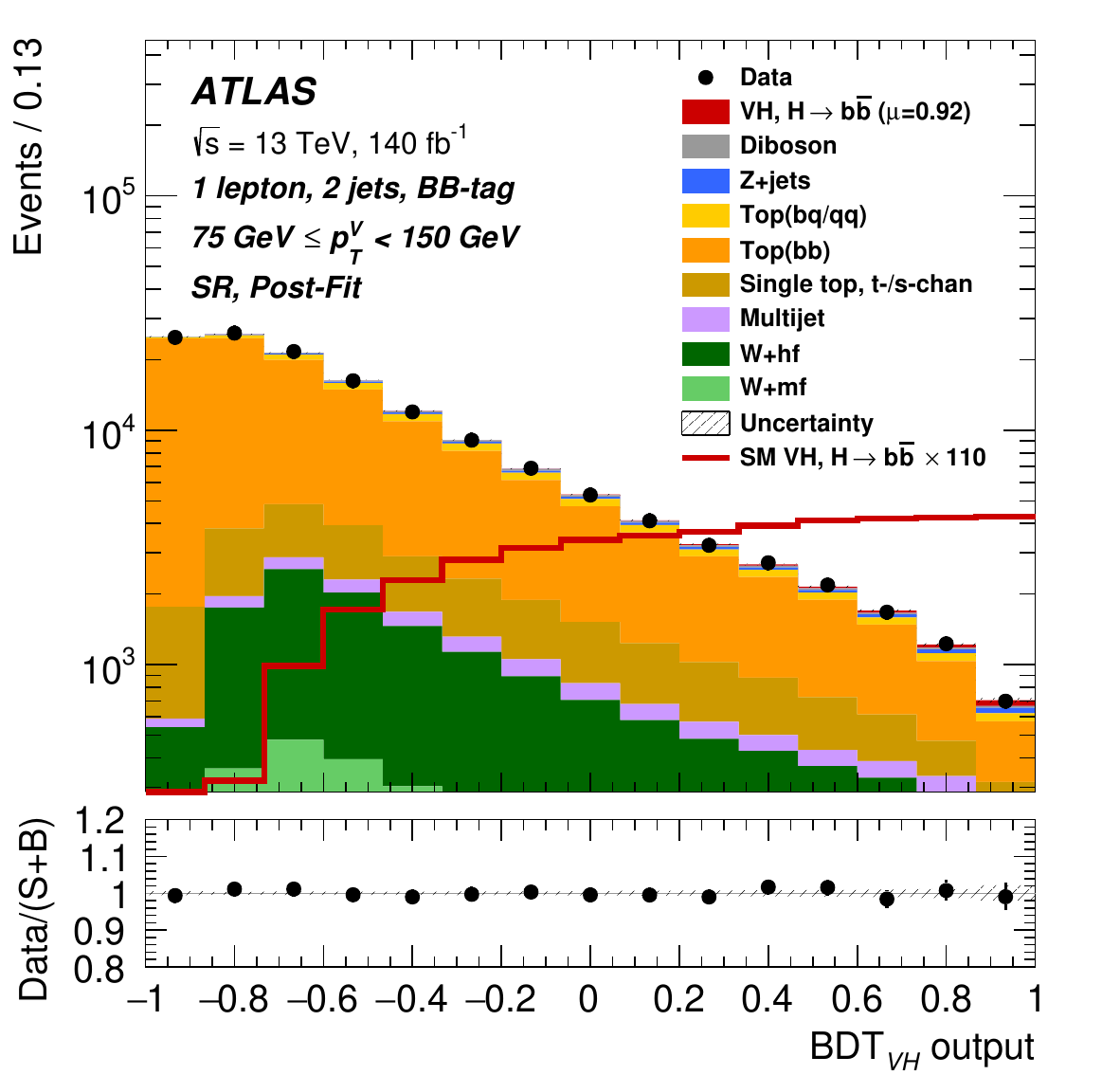}
\label{fig:BDT_1c}
}
\subfloat[]{
\includegraphics[width=0.32\textwidth]{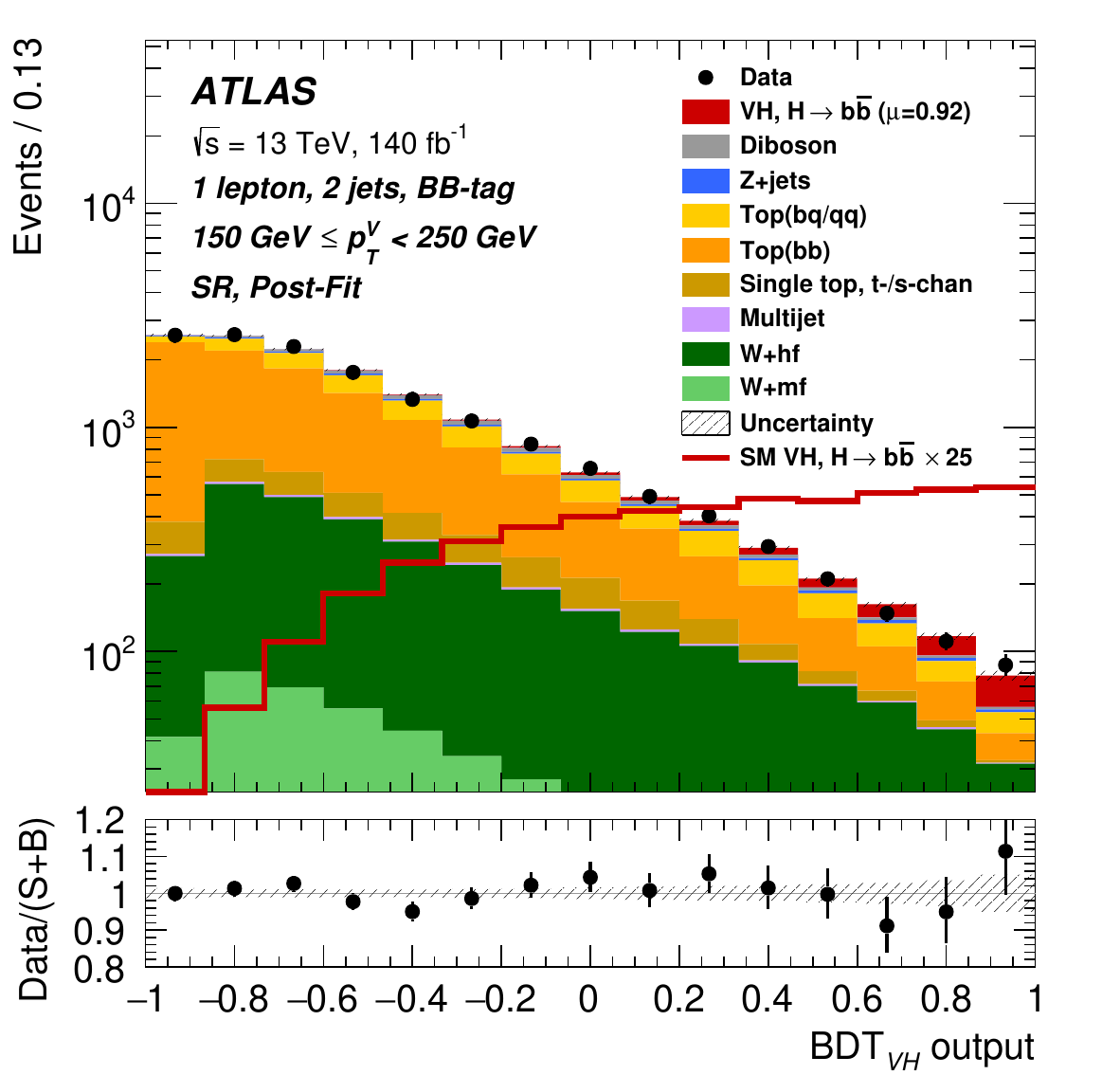}
\label{fig:BDT_1d}
}
\subfloat[]{
\includegraphics[width=0.32\textwidth]{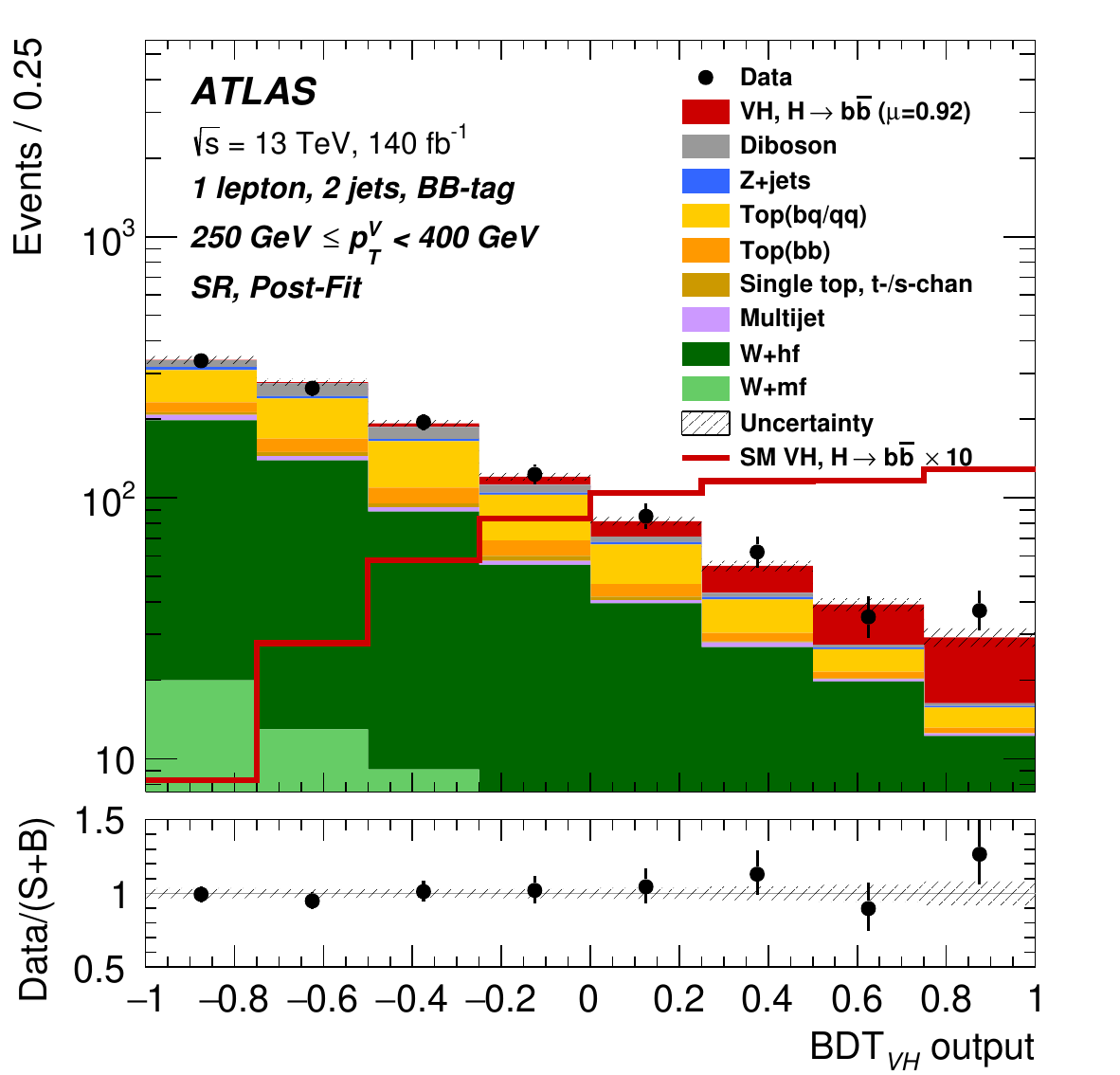}
\label{fig:BDT_1e}
}\\
\subfloat[]{
\includegraphics[width=0.32\textwidth]{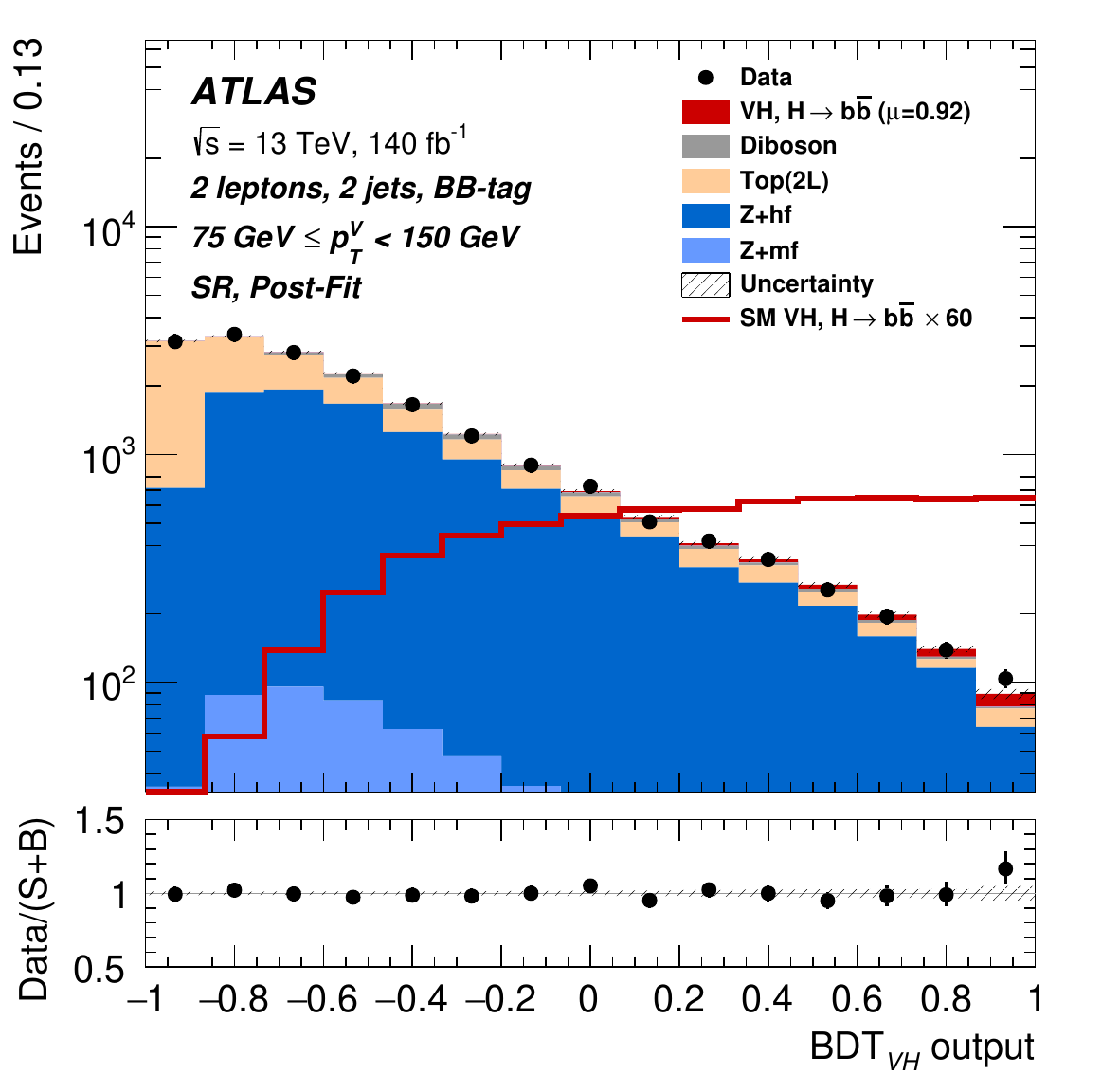}
\label{fig:BDT_1f}
}
\subfloat[]{
\includegraphics[width=0.32\textwidth]{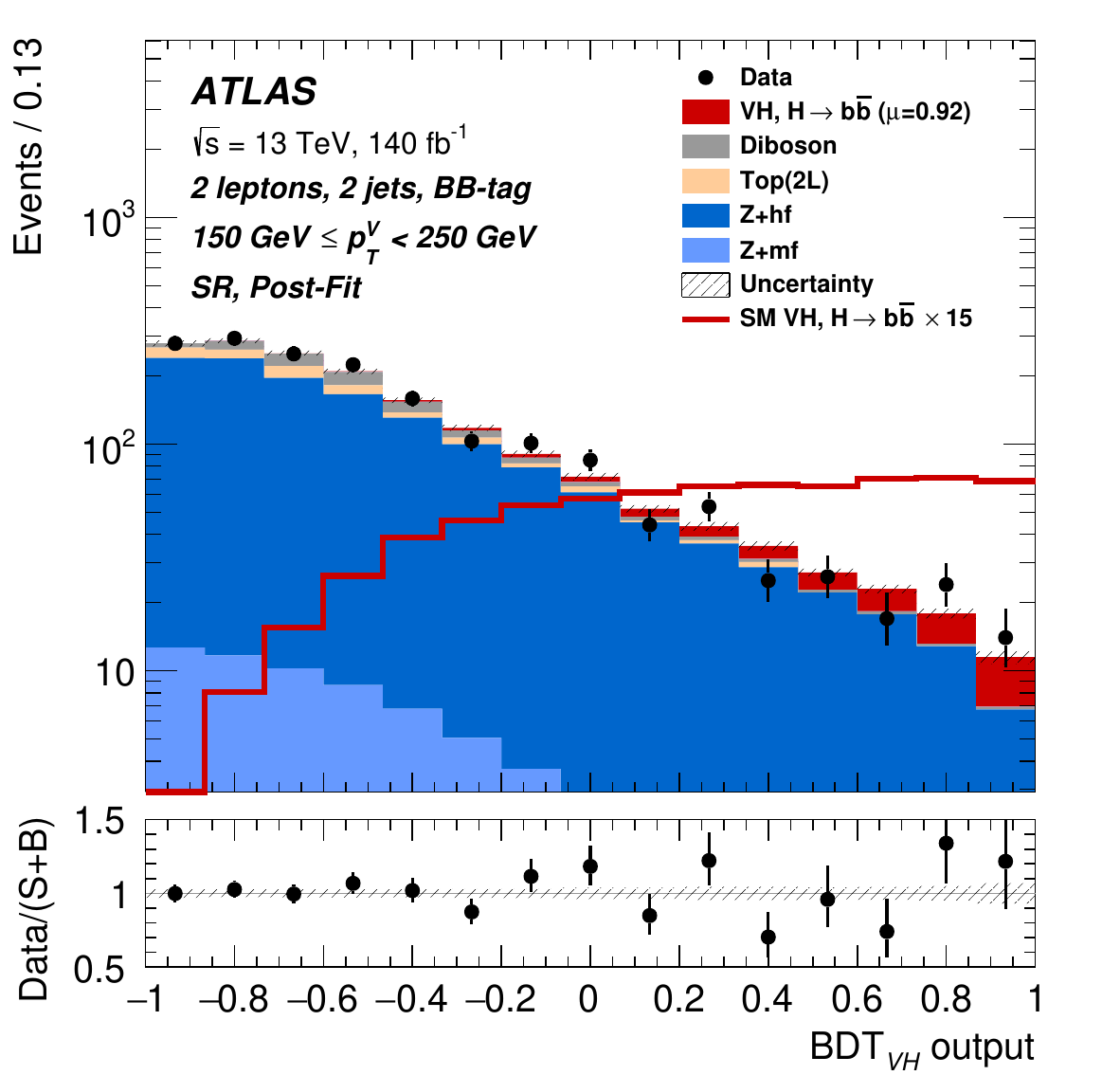}
\label{fig:BDT_1g}
}
\subfloat[]{
\includegraphics[width=0.32\textwidth]{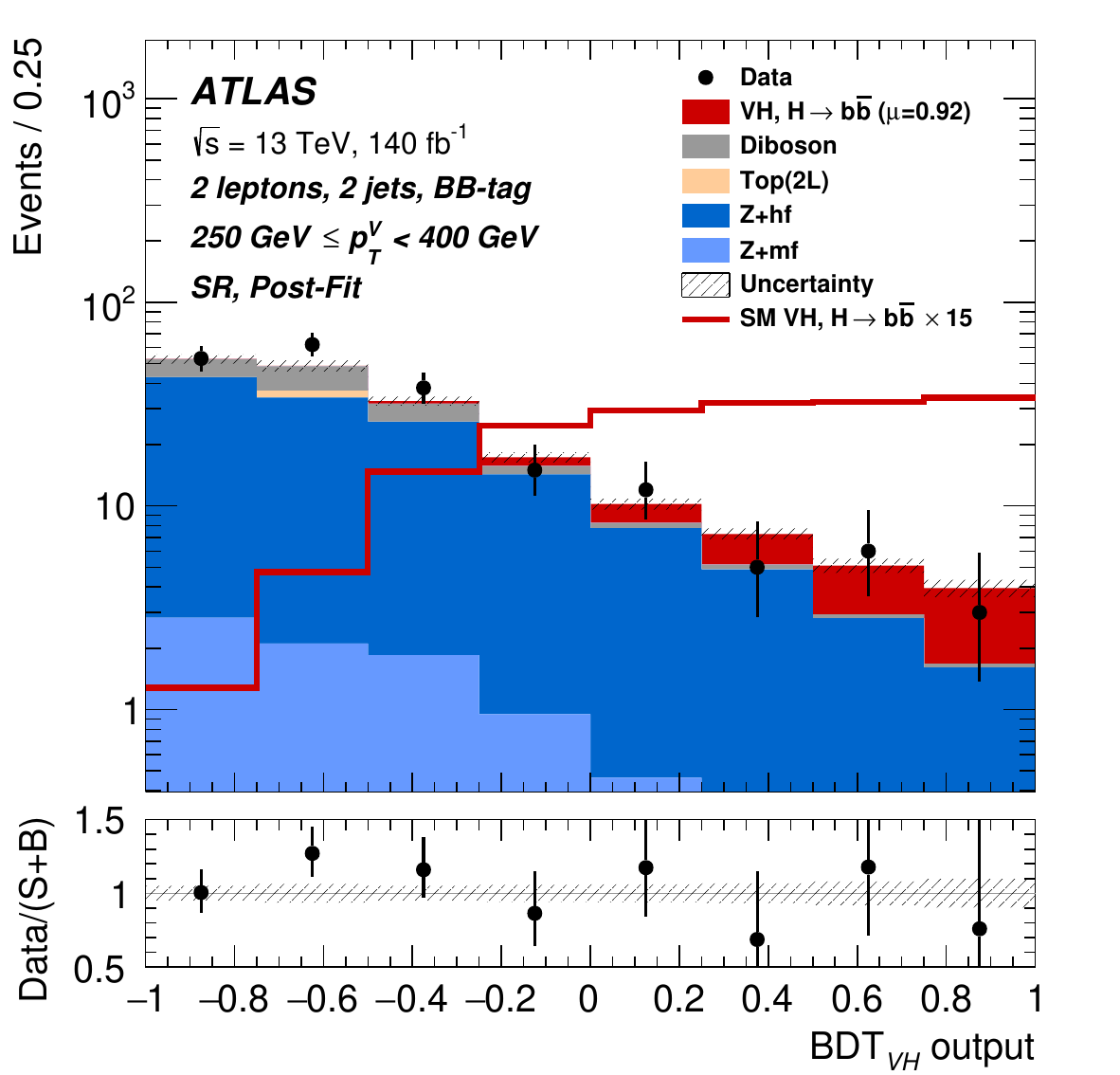}
\label{fig:BDT_1h}
}
\caption[]{The BDT$_{VH}$ post-fit distributions in the
\protect\subref{fig:BDT_1a} $\SI{150}{\GeV}<\ptv<\SI{250}{\GeV}$ and
\protect\subref{fig:BDT_1b} $\SI{250}{\GeV}<\ptv<\SI{400}{\GeV}$ signal regions of the \hbbCAT in the 0-lepton,
\protect\subref{fig:BDT_1c} $\SI{75}{\GeV}<\ptv<\SI{150}{\GeV} $,
\protect\subref{fig:BDT_1d} $\SI{150}{\GeV}<\ptv<\SI{250}{\GeV}$ and
\protect\subref{fig:BDT_1e} $\SI{250}{\GeV}<\ptv<\SI{400}{\GeV}$ in the 1-lepton and
\protect\subref{fig:BDT_1f} $\SI{75}{\GeV}<\ptv<\SI{150}{\GeV} $,
\protect\subref{fig:BDT_1g} $\SI{150}{\GeV}<\ptv<\SI{250}{\GeV}$ and
\protect\subref{fig:BDT_1h} $\SI{250}{\GeV}<\ptv<\SI{400}{\GeV}$ in the 2-lepton
channel for events with $2$ jets. The background contributions after the $VH$  fit are shown as filled histograms. The Higgs boson signal \vhbb\ is shown as a filled histogram on top of the fitted backgrounds normalised to the signal yield extracted from data ($\mu=\muvhbb$), and unstacked as an unfilled histogram, scaled by a value reported in the legend for better visualisation.
The size of the combined statistical and systematic uncertainties for the sum of the fitted signal and background (S+B) are indicated by the hatched band. The ratios of the data to the sum of the fitted signal and background are shown in the lower panel.
}
\label{fig:BDT_1}
\end{figure}

\begin{figure}[tb!]
\centering
\subfloat[]{
\includegraphics[width=0.32\textwidth]{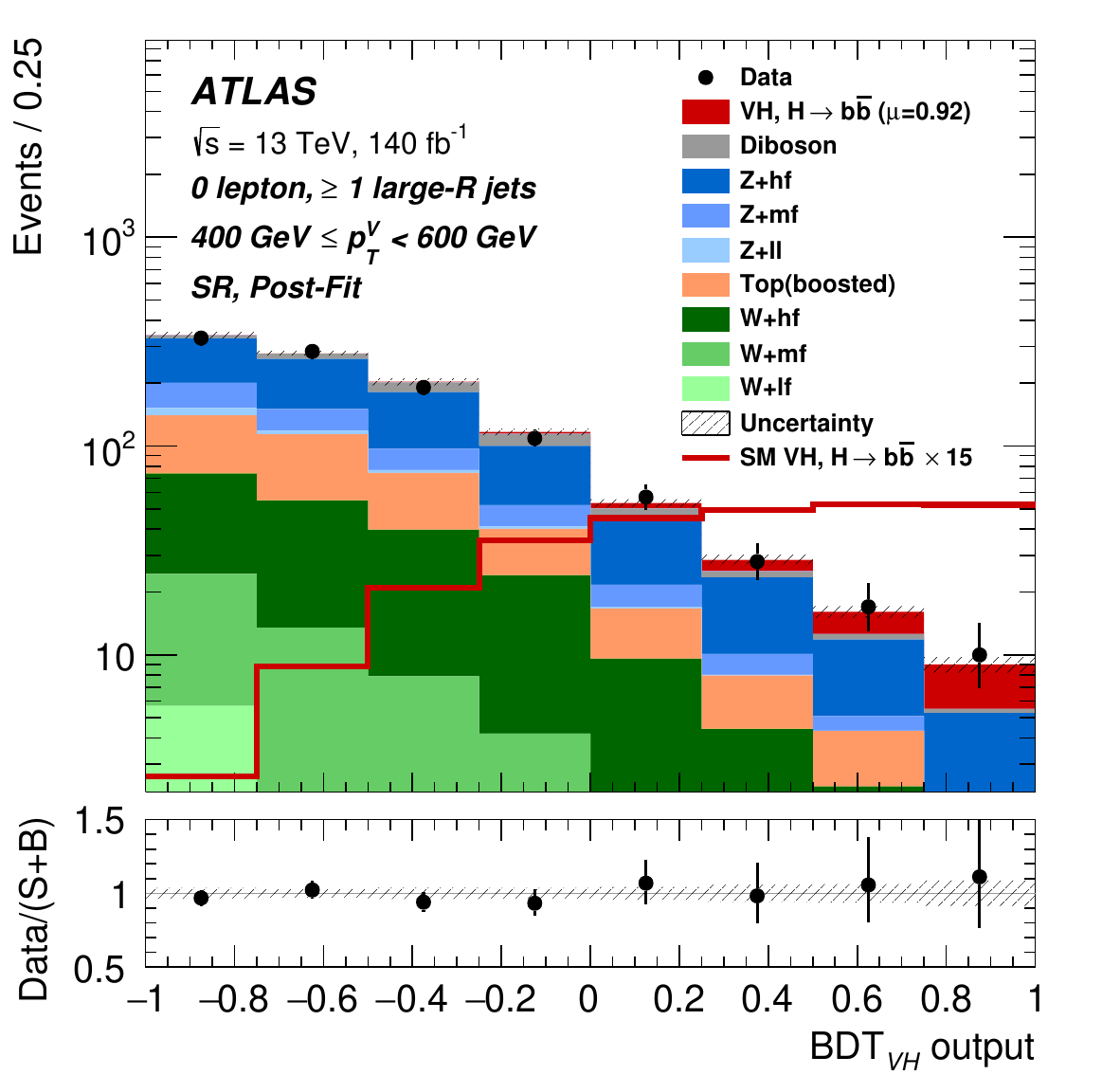}
\label{fig:BDT_3a}
}
\subfloat[]{
\includegraphics[width=0.32\textwidth]{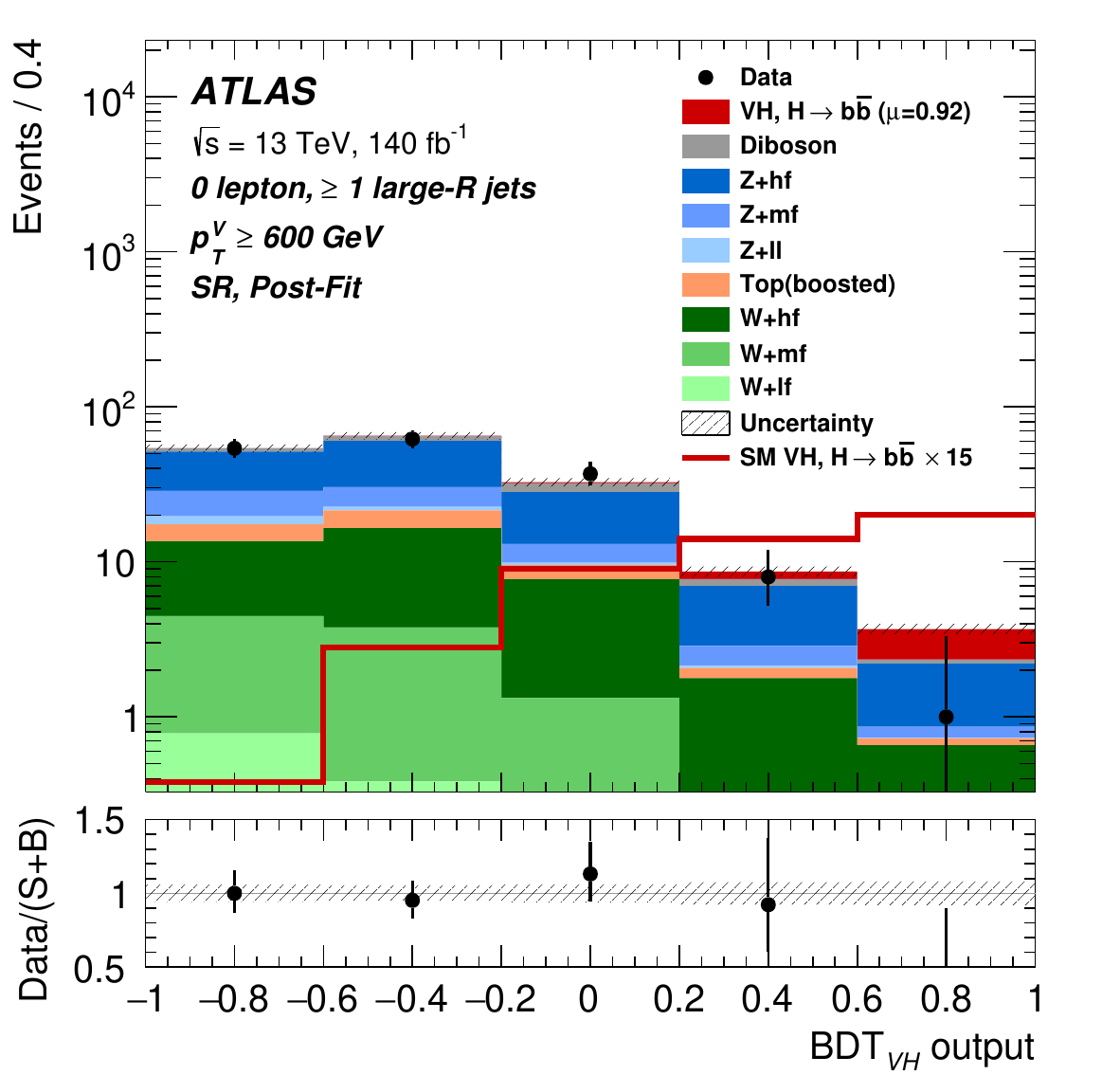}
\label{fig:BDT_3b}
}\\
\subfloat[]{
\includegraphics[width=0.32\textwidth]{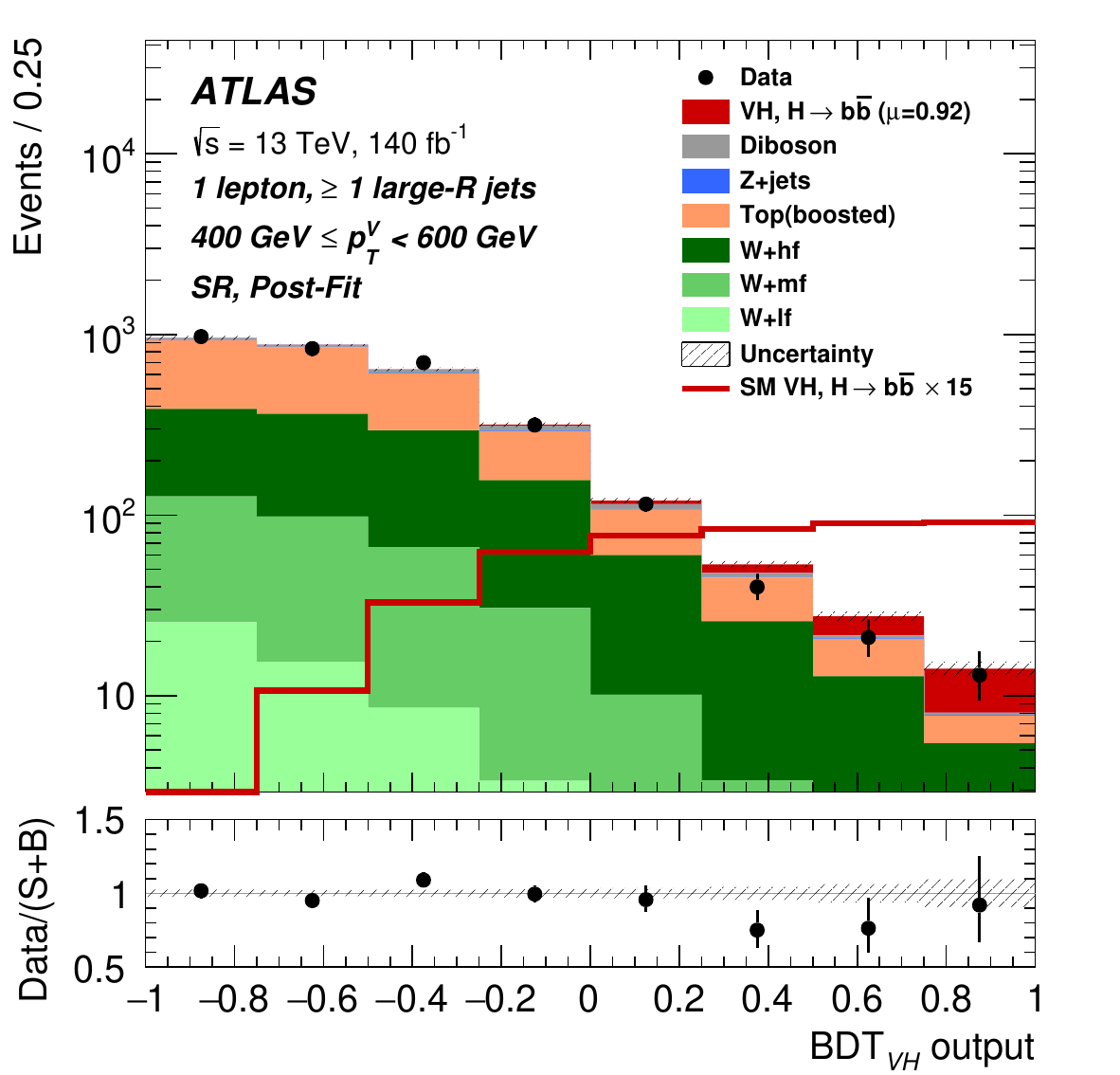}
\label{fig:BDT_3c}
}
\subfloat[]{
\includegraphics[width=0.32\textwidth]{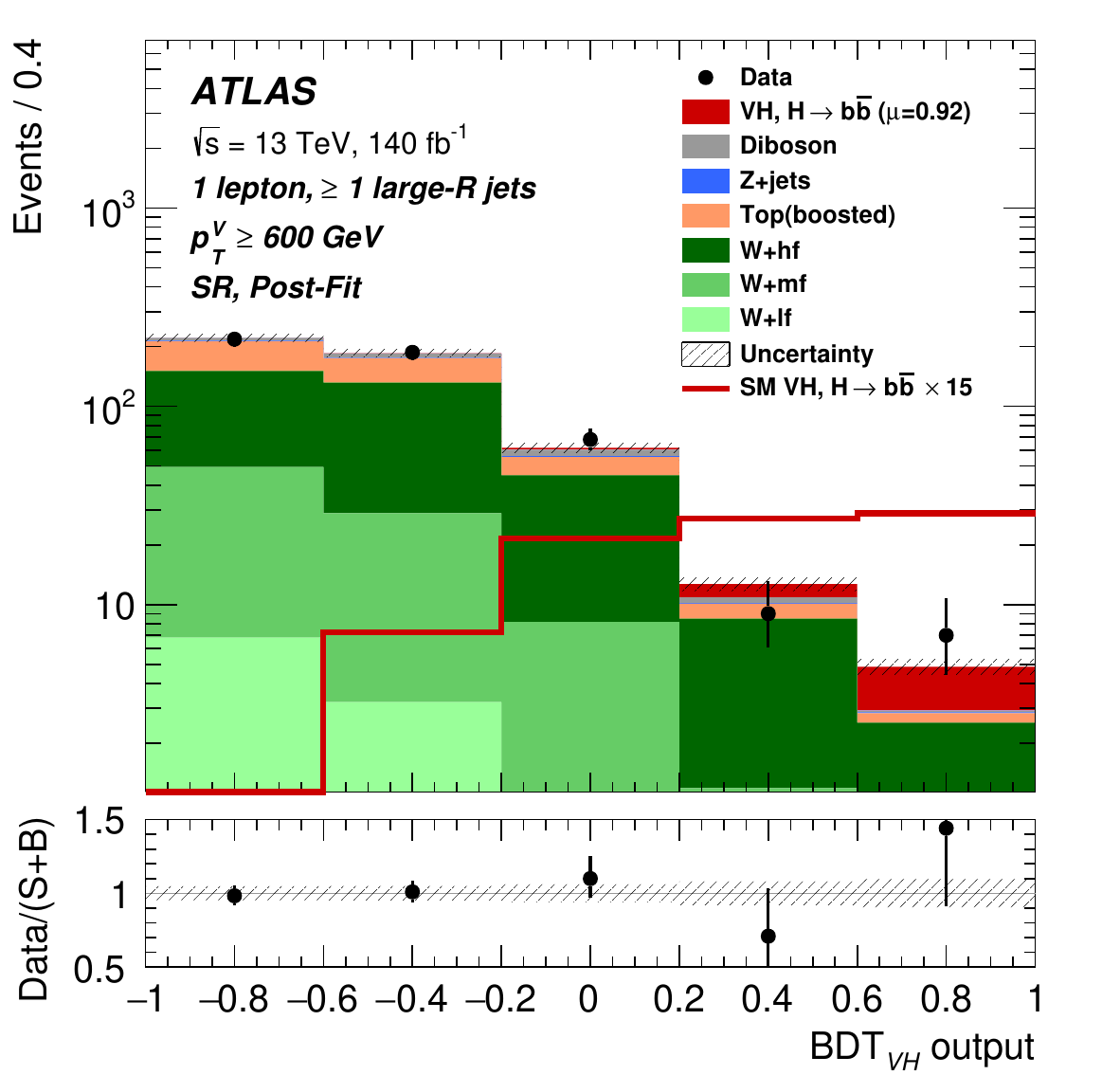}
\label{fig:BDT_3d}
}\\
\subfloat[]{
\includegraphics[width=0.32\textwidth]{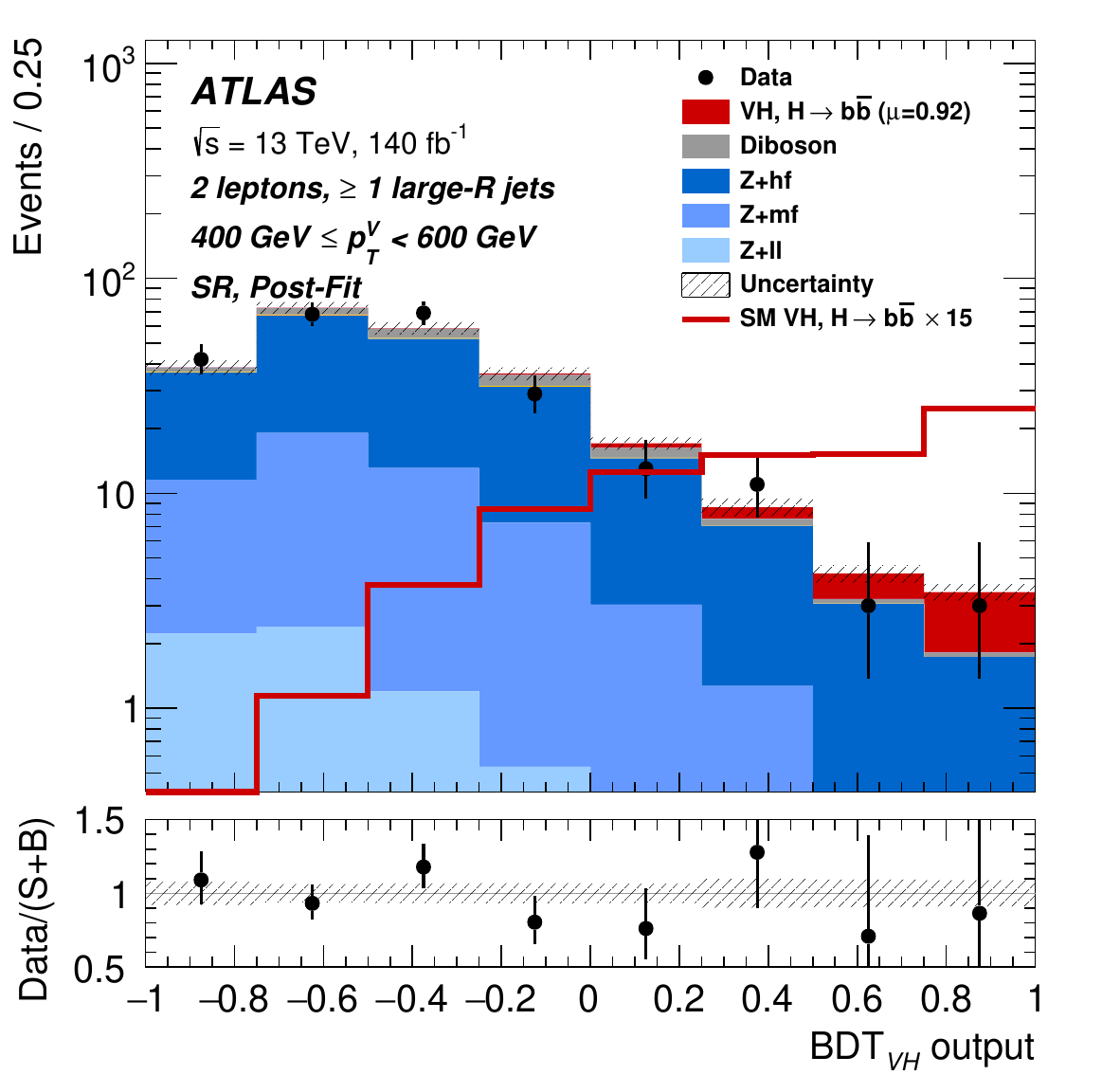}
\label{fig:BDT_3e}
}
\subfloat[]{
\includegraphics[width=0.32\textwidth]{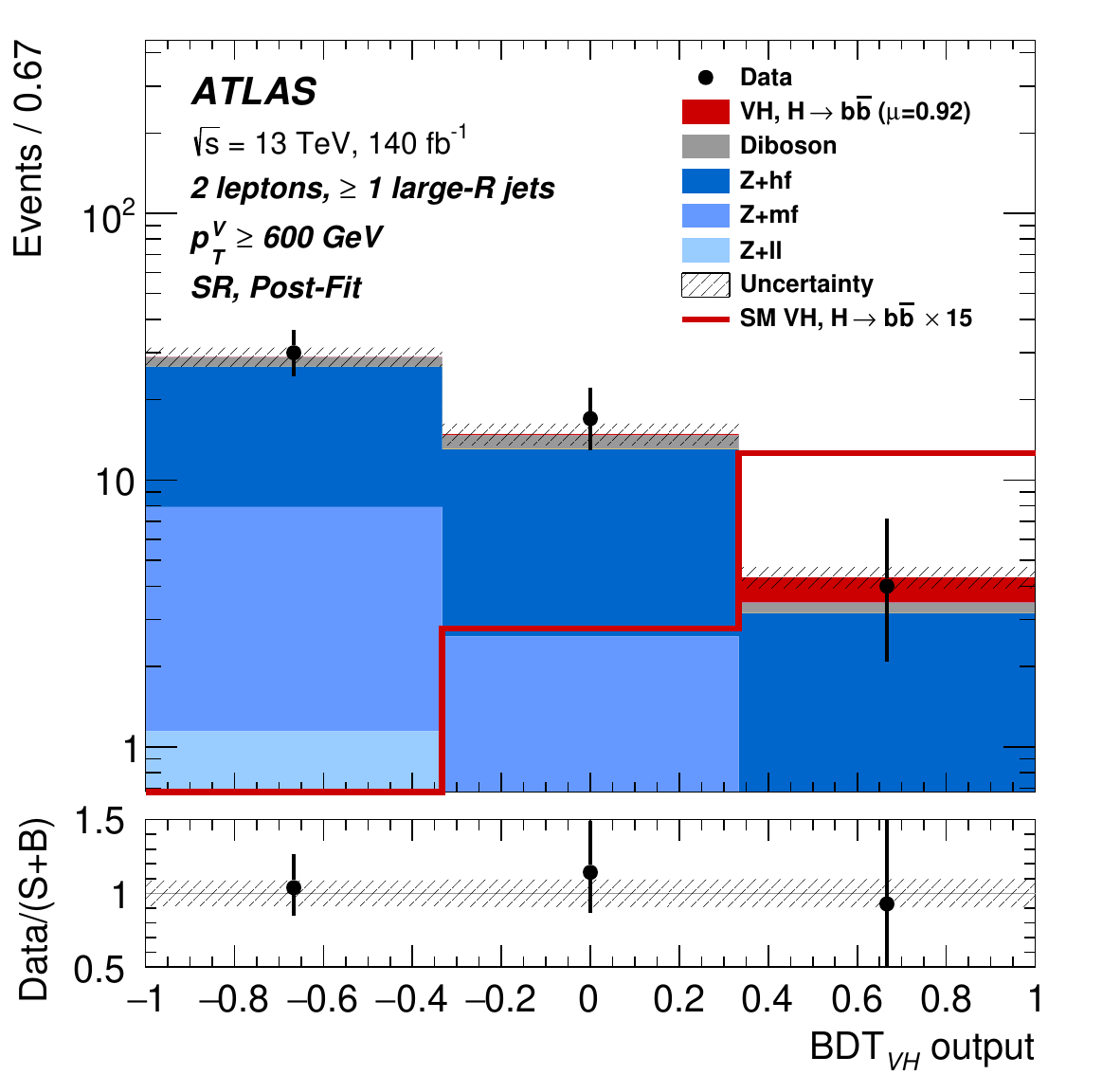}
\label{fig:BDT_3f}
}
\caption[]{The BDT$_{VH}$ post-fit distributions in the
\protect\subref{fig:BDT_3a} $\SI{400}{\GeV}<\ptv<\SI{600}{\GeV}$ and
\protect\subref{fig:BDT_3b} $\ptv>\SI{600}{\GeV}               $ signal regions of the boosted regime in the 0-lepton,
\protect\subref{fig:BDT_3c} $\SI{400}{\GeV}<\ptv<\SI{600}{\GeV}$ and
\protect\subref{fig:BDT_3d} $\ptv>\SI{600}{\GeV}               $ in the 1-lepton and
\protect\subref{fig:BDT_3e} $\SI{400}{\GeV}<\ptv<\SI{600}{\GeV}$ and
\protect\subref{fig:BDT_3f} $\ptv>\SI{600}{\GeV}               $ in the 2-lepton channels.
The background contributions after the $VH$ fit are shown as filled histograms.
The Higgs boson signal \vhbb\ is shown as a filled histogram on top of the fitted backgrounds normalised to the signal yield extracted from data ($\mu=\muvhbb$), and unstacked as an unfilled histogram, scaled by a value reported in the legend for better visualisation.
The size of the combined statistical and systematic uncertainties for the sum of the fitted signal and background (S+B) are indicated by the hatched band. The ratios of the data to the sum of the fitted signal and background are shown in the lower panel.
}
\label{fig:BDT_3}
\end{figure}

\Figrange{\ref{fig:BDT_1}}{\ref{fig:BDT_3}} show the BDT$_{VH}$ output distribution for a selection of the most sensitive signal regions for both the \hbb\ and \hcc\ signals.
The goodness-of-fit evaluated using a saturated model~\cite{Baker:1983tu,Saturated} yields a probability of more than 75\% for all fit configurations.

\begin{table}[tb!]
\caption{The normalisation factors applied to the $W$+jets background processes in the analysis as obtained from the $VH$ fit to data.
The errors represent the combined statistical and systematic uncertainties.}
\centering
\scalebox{0.95}{
\begin{tabular}{c | c || c | c | c }
\toprule
\ptv~interval & Number of jets & \Whf & \Wmf & \Wlf \\
\midrule
\multirow{2}{*}{ 75--150~\GeV }	&   2	  		& 1.09 $\pm$ 0.06 & 1.20 $\pm$ 0.03 & 1.03 $\pm$ 0.04  \\
&   $\geq$3	& 1.30 $\pm$ 0.07 & 1.16 $\pm$ 0.04 & 1.07 $\pm$ 0.05  \\
\hline
\multirow{2}{*}{ 150--250~\GeV  }   &   2	  		& 1.00 $\pm$ 0.05 & 1.31 $\pm$ 0.03 & 1.08 $\pm$ 0.03  \\
&   $\geq$3	& 1.28 $\pm$ 0.07 & 1.31 $\pm$ 0.04 & 1.07 $\pm$ 0.04  \\
\hline
\multirow{2}{*}{ 250--400~\GeV }   &   2	  		& 0.97 $\pm$ 0.08 & 1.35 $\pm$ 0.07 & 1.05 $\pm$ 0.03 \\
&   $\geq$3	& 1.46 $\pm$ 0.12 & 1.32 $\pm$ 0.07 & 1.10 $\pm$ 0.04 \\
\hline
400--600~\GeV					&    -			& \multicolumn{2}{  c |}{1.49 $\pm$ 0.25} 		  & --			       \\
\hline
$>600$~\GeV 					&    -			& \multicolumn{2}{  c |}{2.03 $\pm$ 0.25} 		  & --			       \\
\bottomrule
\end{tabular}
}
\label{tab:WNF}
\end{table}

\begin{table}[tb!]
\caption{The normalisation factors applied to the $Z$+jets background processes in the analysis as obtained from the $VH$ fit to data.
The $3/\geq3$ value is an additional correction factor applied to \Zhf events in categories with exactly three jets on top of the $\geq$3 normalisation factors. The errors represent the combined statistical and systematic uncertainties.}
\centering
\scalebox{0.95}{
\begin{tabular}{c | c || c | c | c }
\toprule
\ptv~interval & Number of jets & \Zhf & \Zmf & \Zlf \\
\midrule
\multirow{2}{*}{ 75--150~\GeV }		&   2	  		& 1.20 $\pm$ 0.04 & 1.04 $\pm$ 0.04 & 1.12 $\pm$ 0.03  \\
&   $\geq$3	& 1.49 $\pm$ 0.06 & 1.11 $\pm$ 0.05 & 1.12 $\pm$ 0.05  \\
&   3/$\geq$3	& 0.77 $\pm$ 0.03 & --	 & --  \\
\hline
\multirow{2}{*}{ 150--250~\GeV  }   	&   2	  		& 1.30 $\pm$ 0.04 & 1.08 $\pm$ 0.04 & 1.17 $\pm$ 0.02  \\
&   $\geq$3	& 1.59 $\pm$ 0.07 & 1.14$\pm$ 0.05  & 1.17 $\pm$ 0.04  \\
&   3/$\geq$3	& 0.80 $\pm$ 0.04 & -- 	 & --  \\
\hline
\multirow{2}{*}{ 250--400~\GeV }  	&   2	  		& 1.40 $\pm$ 0.07 & 1.31 $\pm$ 0.08 & 1.16 $\pm$ 0.03 \\
&   $\geq$3	& 1.78 $\pm$ 0.09 & 1.32 $\pm$ 0.07 & 1.20 $\pm$ 0.04 \\
&   3/$\geq$3	& 0.74 $\pm$ 0.04 & --	 & --  \\
\hline
$>$400    \GeV					&    -			&  \multicolumn{2}{  c |}{1.63 $\pm$ 0.13}   & --			       \\
\bottomrule
\end{tabular}
}
\label{tab:ZNF}
\end{table}

\begin{table}[tb!]
\caption{The normalisation factors applied to the top-quark background processes in the analysis as obtained from the $VH$ fit to data.
The errors represent the combined statistical and systematic uncertainties.}
\centering
\scalebox{0.95}{
\begin{tabular}{c | c || c | c | c }
\toprule
\ptv~interval & Number of jets & Top(bb) & Top(bq,qq) & Top 2L \\
\midrule
\multirow{2}{*}{ 75--150~\GeV }		&   2	  		& 1.02 $\pm$ 0.04 & 0.98 $\pm$ 0.05 & 1.05 $\pm$ 0.05  \\
&   3			& 0.97 $\pm$ 0.03 & 0.98 $\pm$ 0.03 & 0.98 $\pm$ 0.05  \\
\hline
\multirow{3}{*}{ 150--250~\GeV  }   	&   2	  		& 0.89 $\pm$ 0.05 & 0.83 $\pm$ 0.04 & 1.07 $\pm$ 0.16  \\
&   3			& 0.91 $\pm$ 0.03 & 0.86 $\pm$ 0.03 & \multirow{2}{*}{ 0.95 $\pm$ 0.14 }  \\
&   4	& 0.97 $\pm$ 0.02 & 0.95 $\pm$ 0.03 & 							  \\
\hline
\multirow{2}{*}{ 250--400~\GeV }   	&   2	  		& 0.78 $\pm$ 0.08 & 0.82 $\pm$ 0.05 &  \multirow{3}{*}{ 1.10 $\pm$ 0.50 } \\
&   3			& 0.83 $\pm$ 0.04 & 0.80 $\pm$ 0.03 & 							 \\
&   4	& 0.93 $\pm$ 0.05 & 0.86 $\pm$ 0.04 & 							  \\
\hline
400--600~\GeV					&    -			& \multicolumn{2}{  c |}{0.83 $\pm$ 0.05}   & --			       \\
\hline
$>$600    \GeV					&    -			& \multicolumn{2}{  c |}{0.69 $\pm$ 0.07}   & --			       \\
\bottomrule
\end{tabular}
}
\label{tab:TNF}
\end{table}

\begin{figure}[tb!]
\flushright
\subfloat[]{
\includegraphics[width=0.32\textwidth]{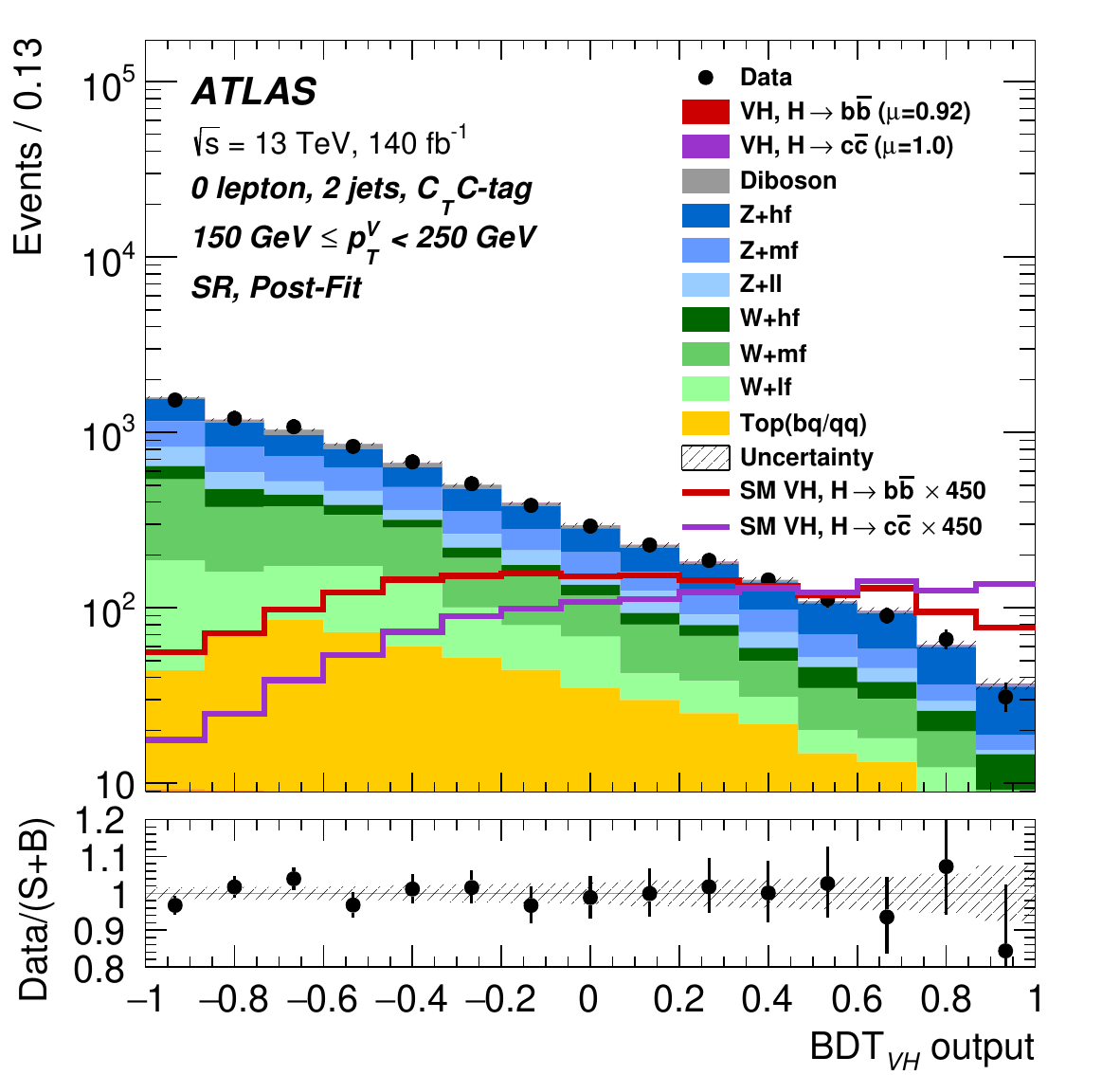}
\label{fig:BDT_2a}
}
\subfloat[]{
\includegraphics[width=0.32\textwidth]{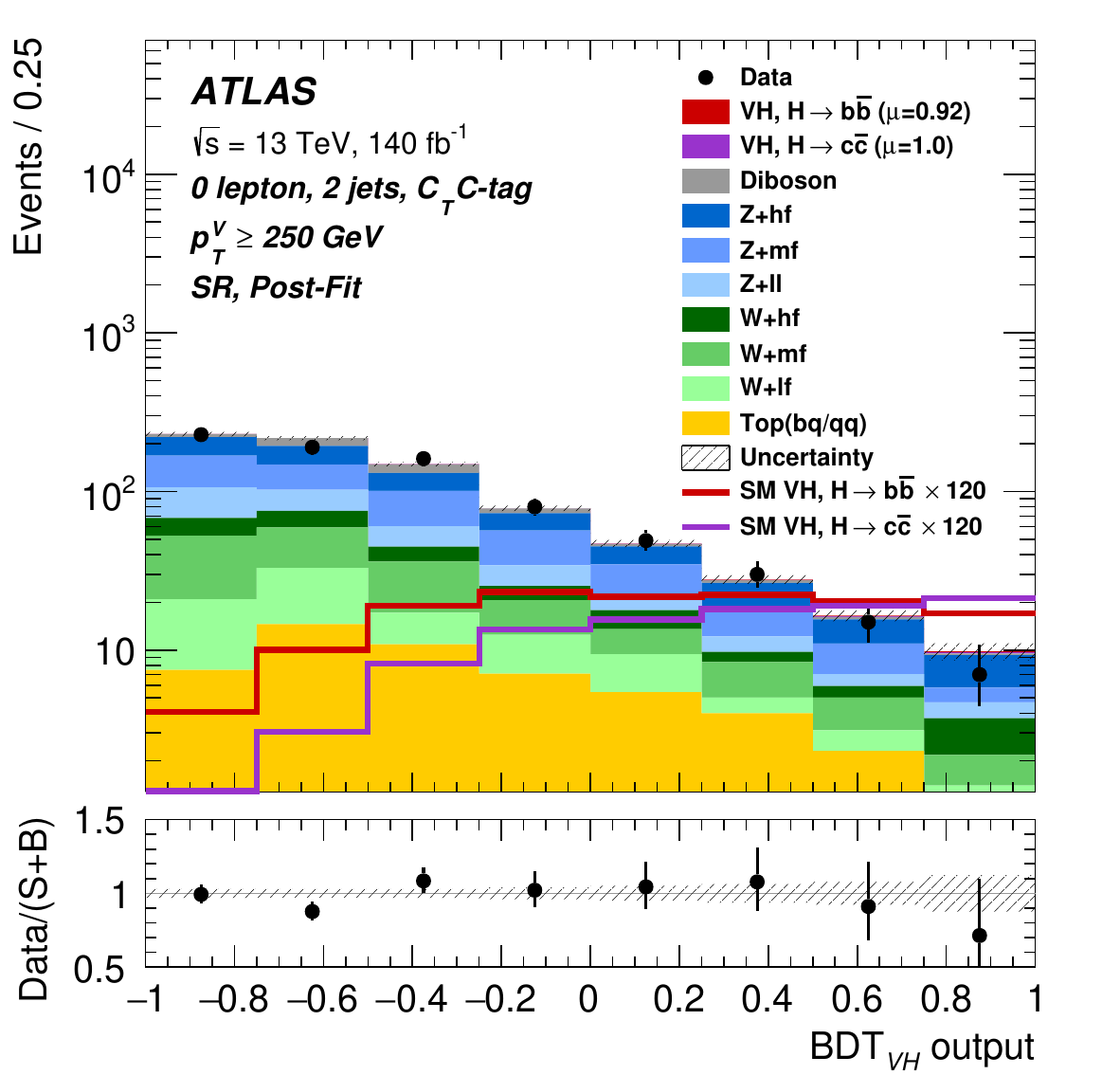}
\label{fig:BDT_2b}
}\\
\subfloat[]{
\includegraphics[width=0.32\textwidth]{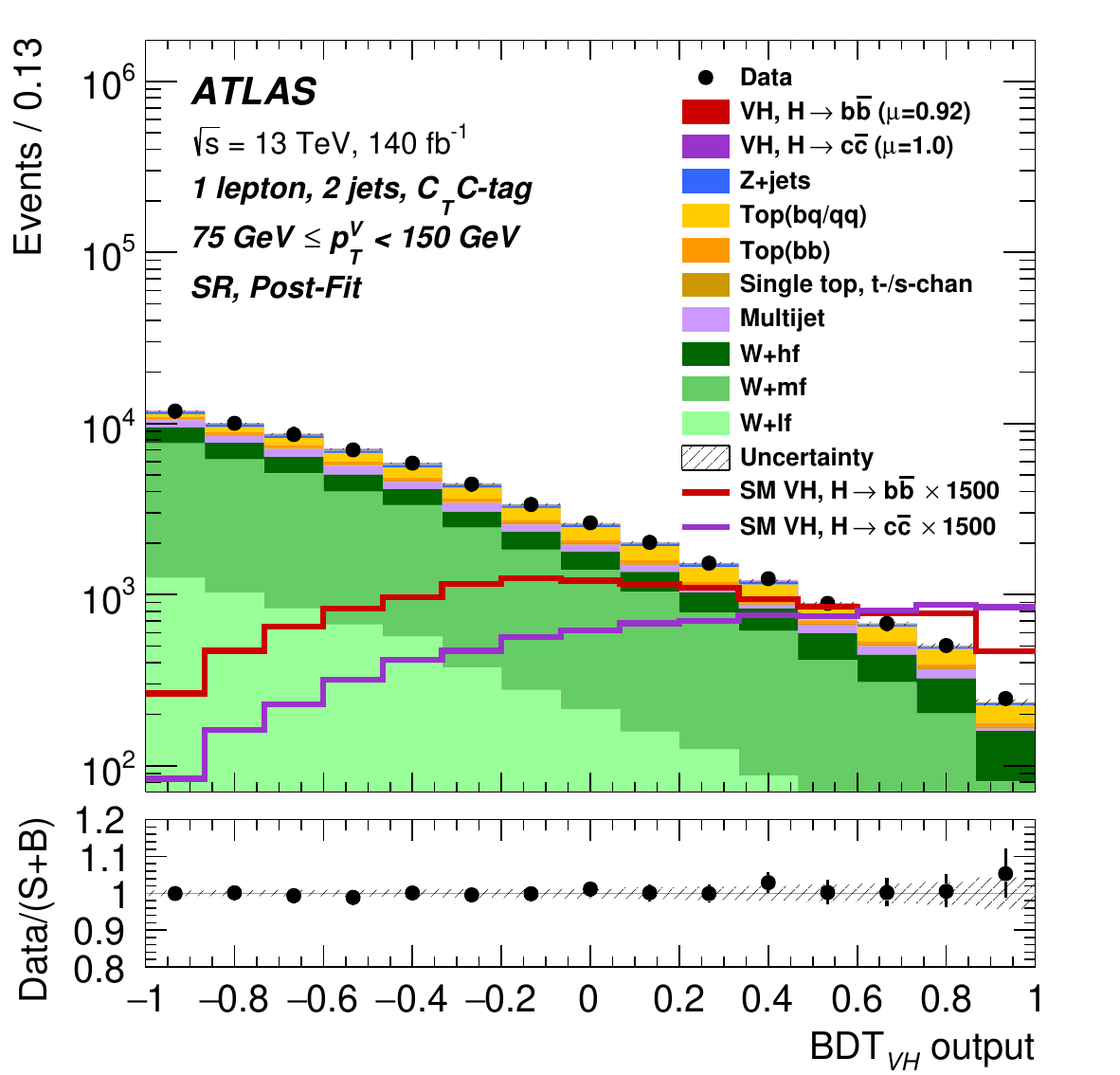}
\label{fig:BDT_2c}
}
\subfloat[]{
\includegraphics[width=0.32\textwidth]{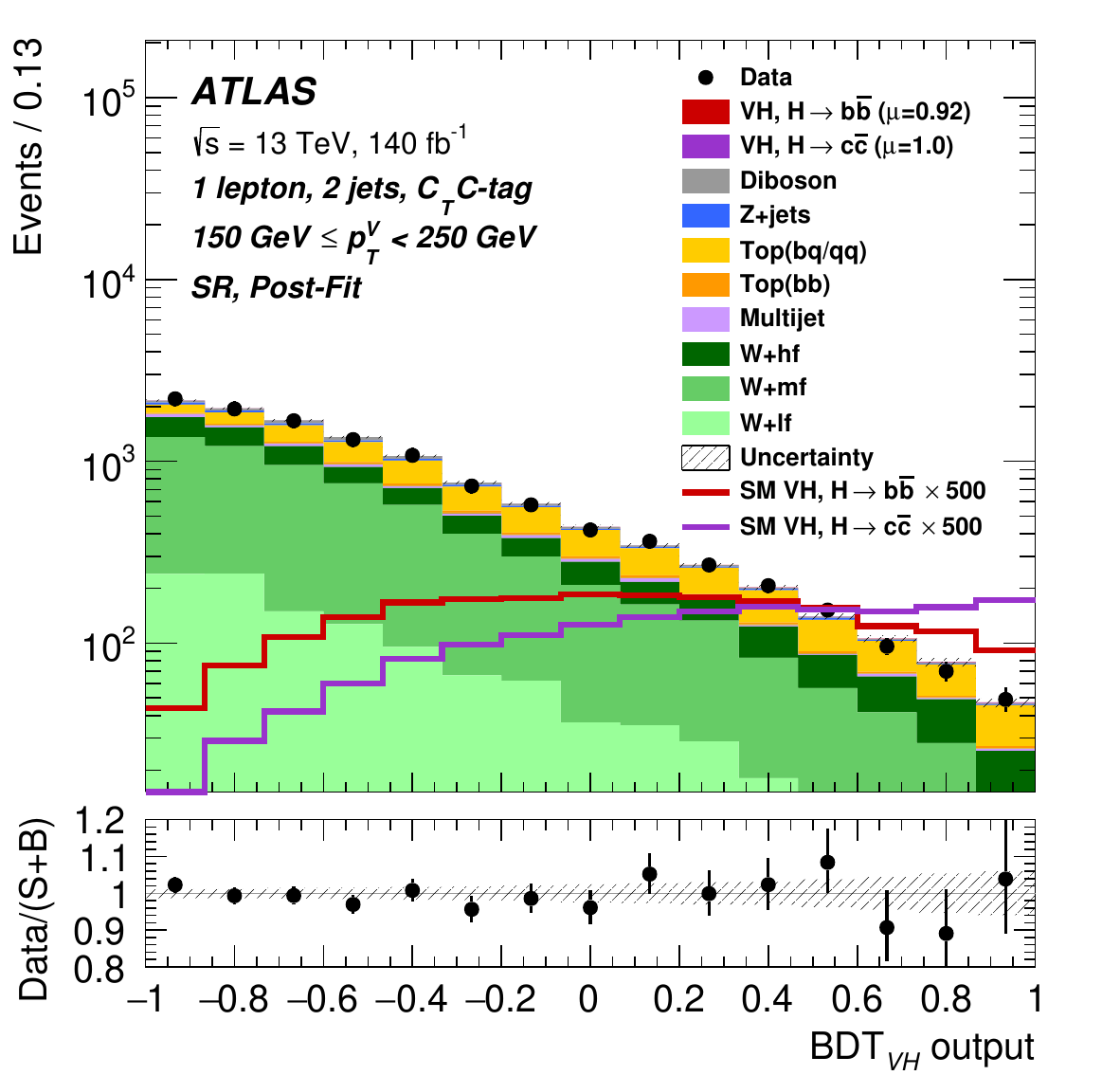}
\label{fig:BDT_2d}
}
\subfloat[]{
\includegraphics[width=0.32\textwidth]{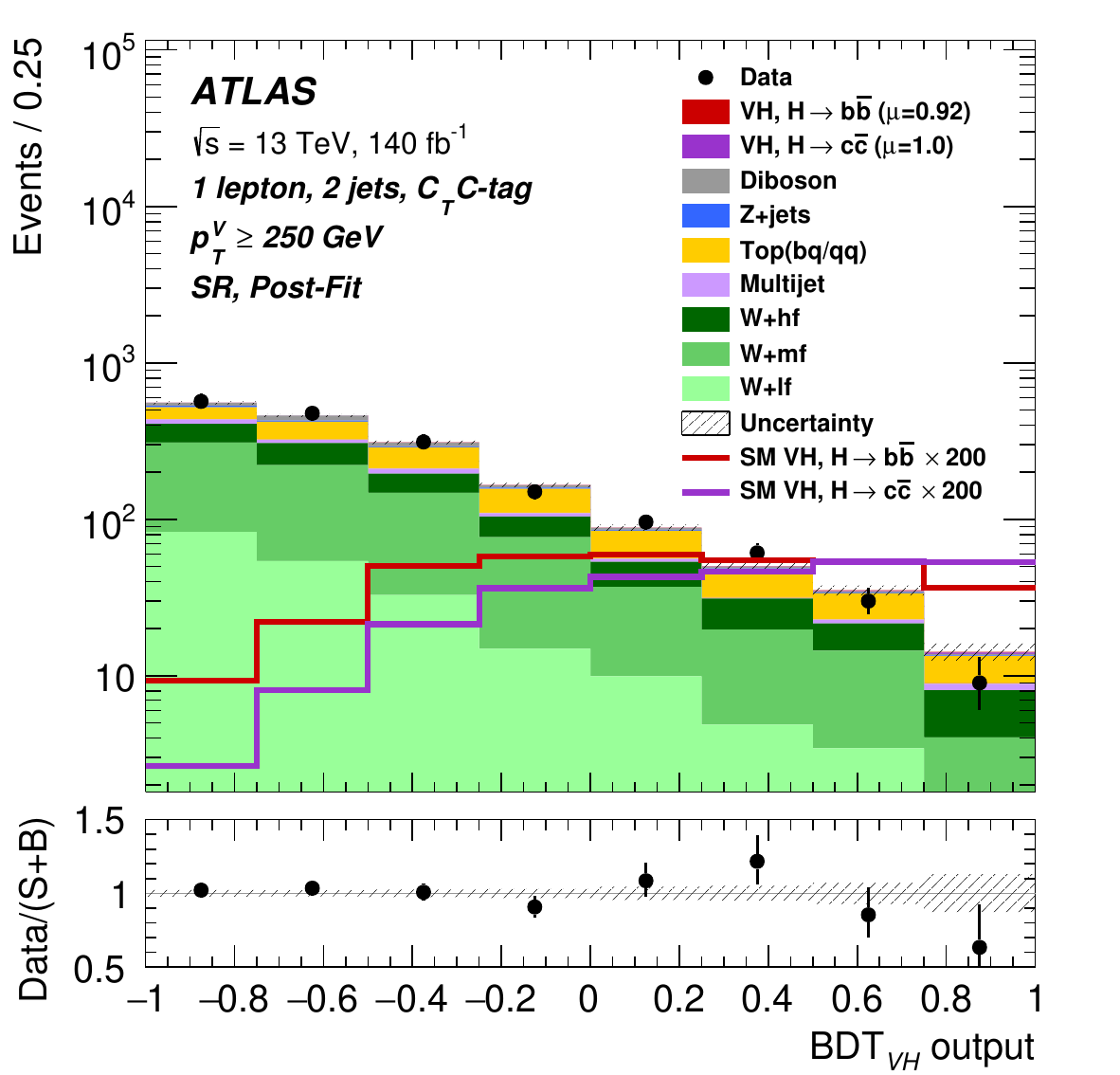}
\label{fig:BDT_2e}
}\\
\subfloat[]{
\includegraphics[width=0.32\textwidth]{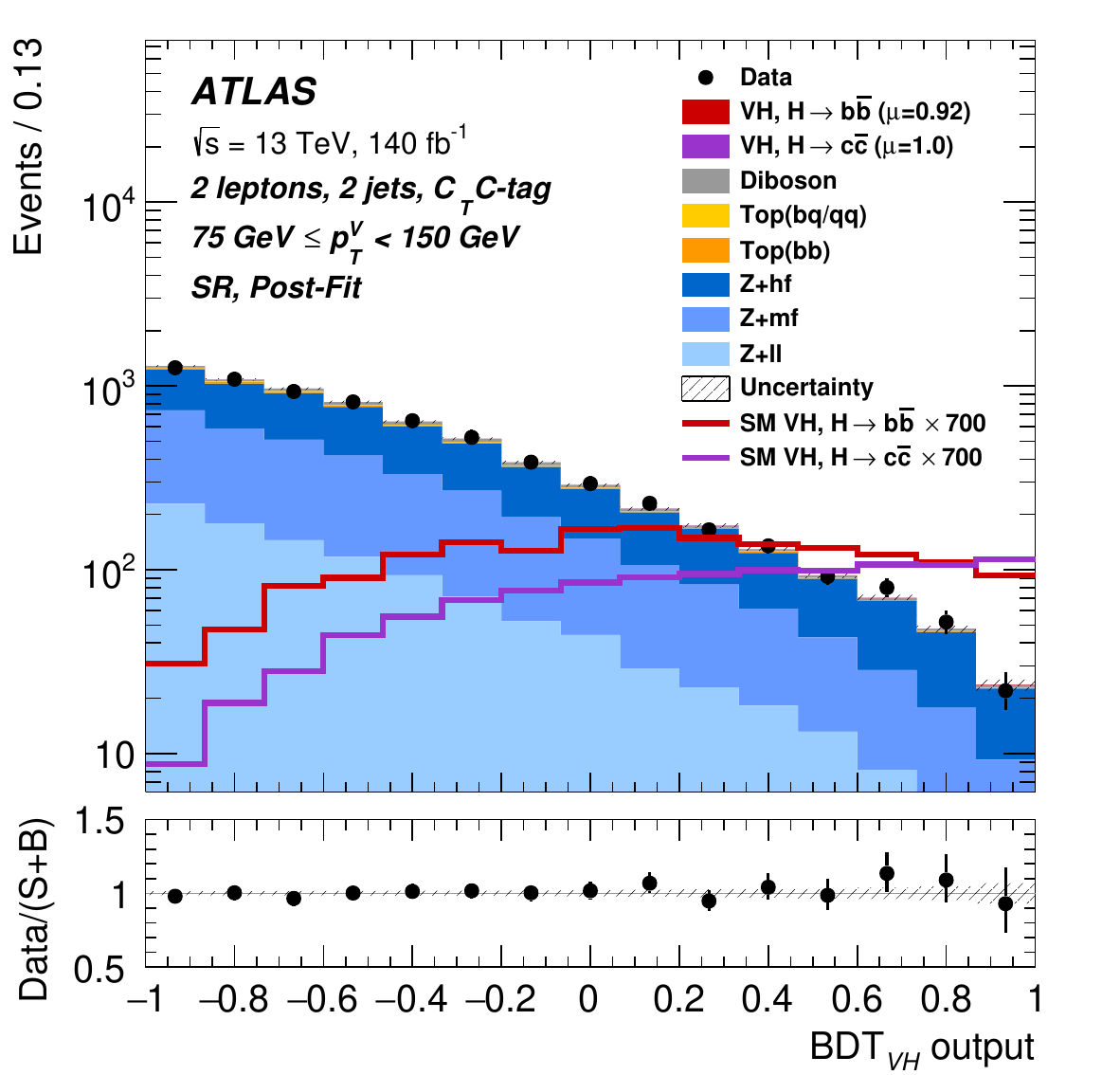}
\label{fig:BDT_2f}
}
\subfloat[]{
\includegraphics[width=0.32\textwidth]{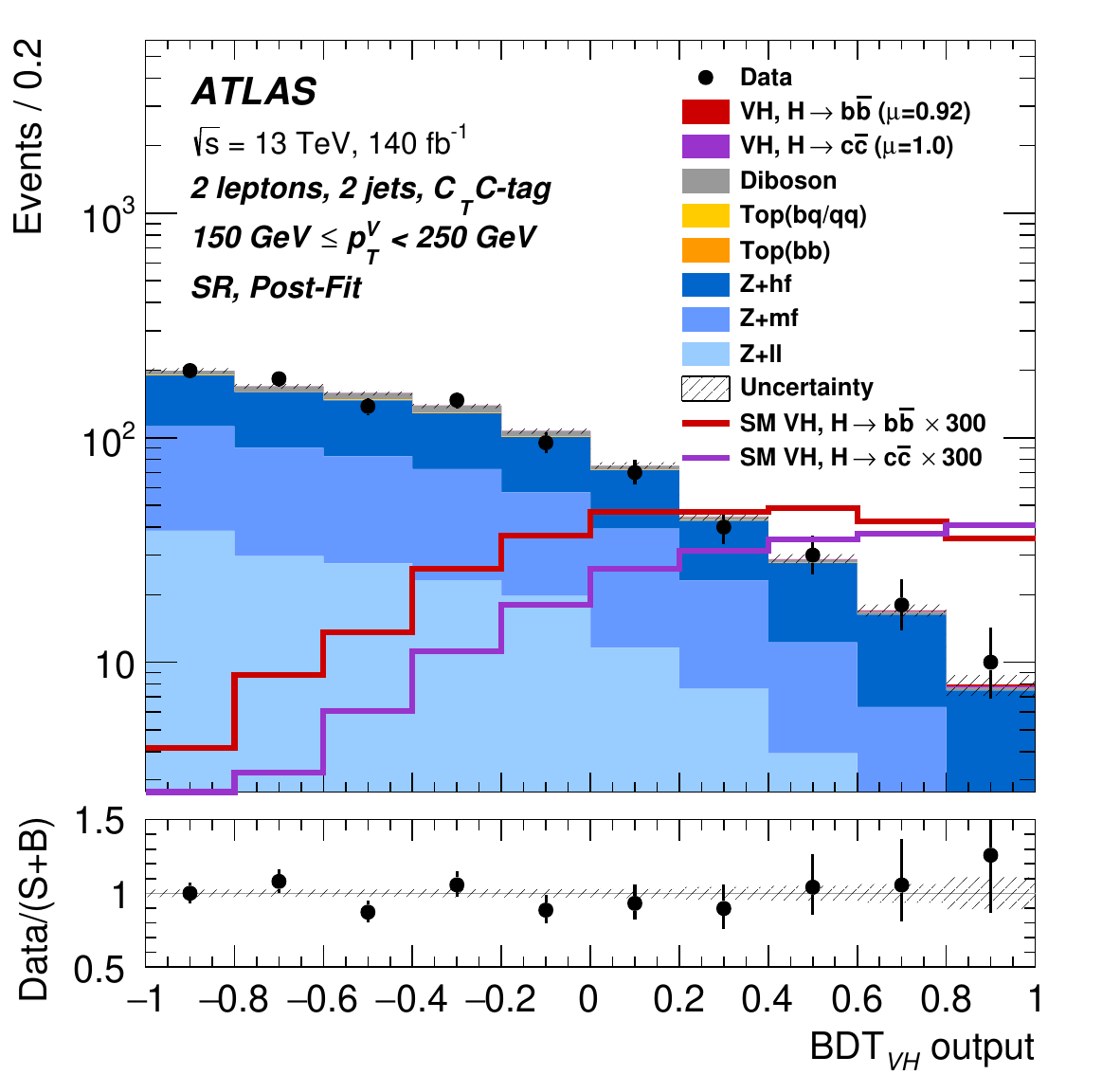}
\label{fig:BDT_2g}
}
\subfloat[]{
\includegraphics[width=0.32\textwidth]{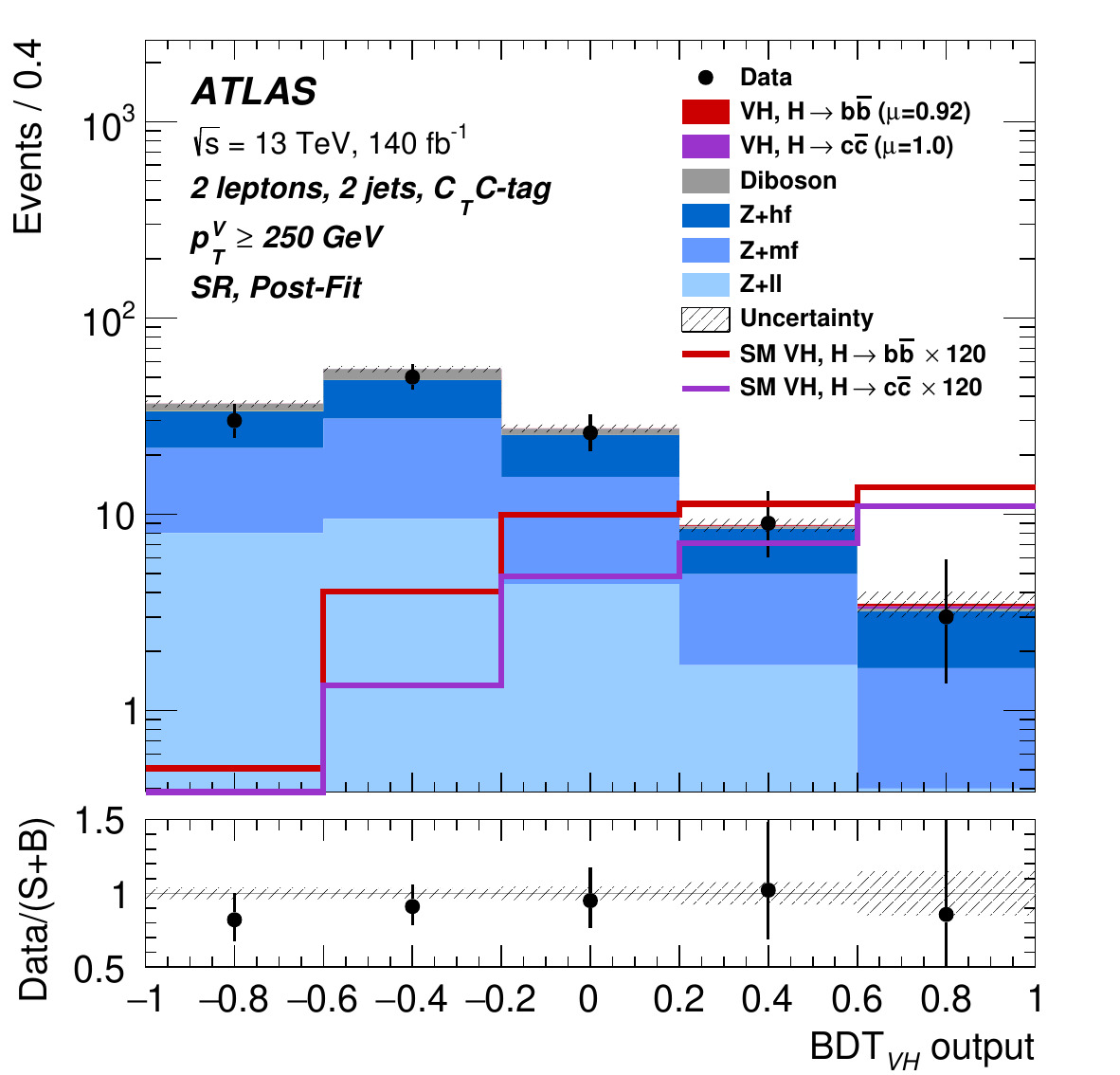}
\label{fig:BDT_2h}
}
\caption[]{The BDT$_{VH}$ post-fit distributions in the
\protect\subref{fig:BDT_2a} $\SI{150}{\GeV}<\ptv<\SI{250}{\GeV}$ and
\protect\subref{fig:BDT_2b} $\SI{250}{\GeV}<\ptv<\SI{400}{\GeV}$ signal regions of the Hcc \ctc category in the 0-lepton,
\protect\subref{fig:BDT_2c} $\SI{75}{\GeV}<\ptv<\SI{150}{\GeV} $,
\protect\subref{fig:BDT_2d} $\SI{150}{\GeV}<\ptv<\SI{250}{\GeV}$ and
\protect\subref{fig:BDT_2e} $\SI{250}{\GeV}<\ptv<\SI{400}{\GeV}$ in the 1-lepton and
\protect\subref{fig:BDT_2f} $\SI{75}{\GeV}<\ptv<\SI{150}{\GeV} $,
\protect\subref{fig:BDT_2g} $\SI{150}{\GeV}<\ptv<\SI{250}{\GeV}$ and
\protect\subref{fig:BDT_2h} $\SI{250}{\GeV}<\ptv<\SI{400}{\GeV}$ in the 2-lepton
channel for events with $2$ jets. The background contributions after the $VH$ fit are shown as filled histograms.
The \vhcc\  signal  and the contribution from \vhbb\  are shown unstacked as  unfilled histograms, scaled by the factor indicated in the legend.
The size of the combined statistical and systematic uncertainties for the sum of the fitted signal and background (S+B) are indicated by the hatched band. The ratios of the data to the sum of the fitted signal and background are shown in the lower panel.}
\label{fig:BDT_2}
\end{figure}

\FloatBarrier
\subsection{Diboson signal strength measurements}
\begin{figure}[tb!]
\centering
\subfloat[]{
\includegraphics[width=0.50\textwidth]{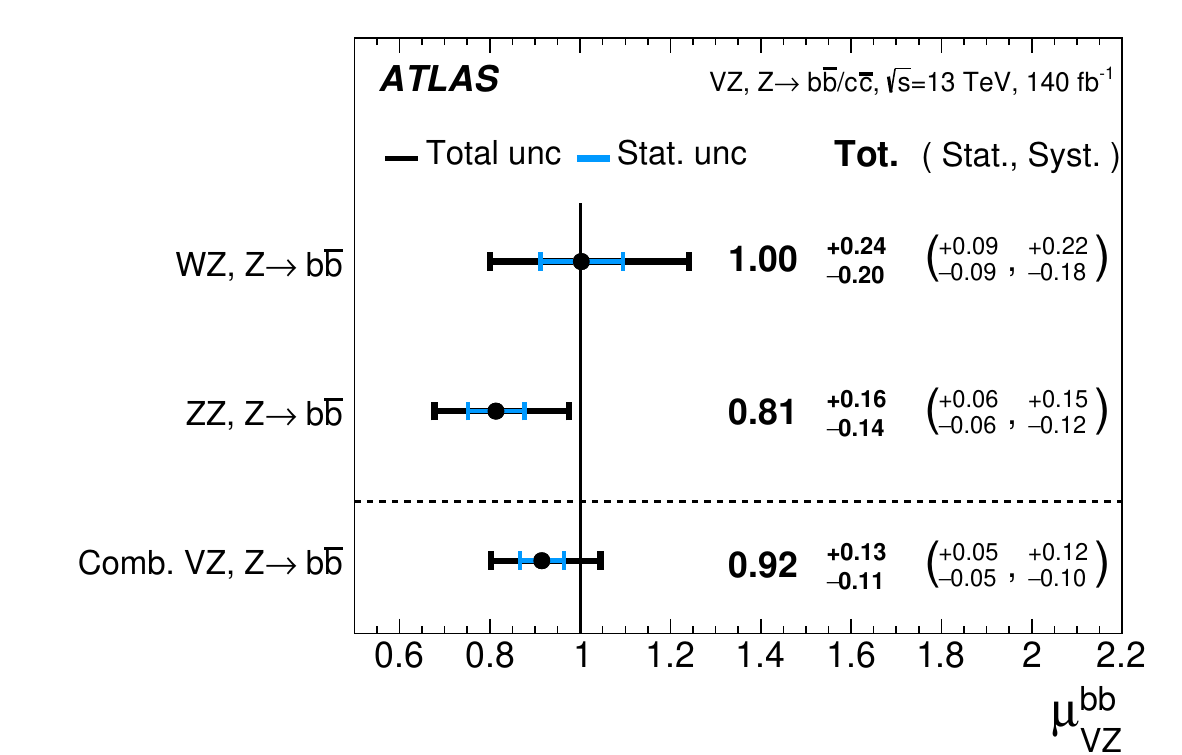}
\label{fig:diboson1a}
}
\subfloat[]{
\includegraphics[width=0.50\textwidth]{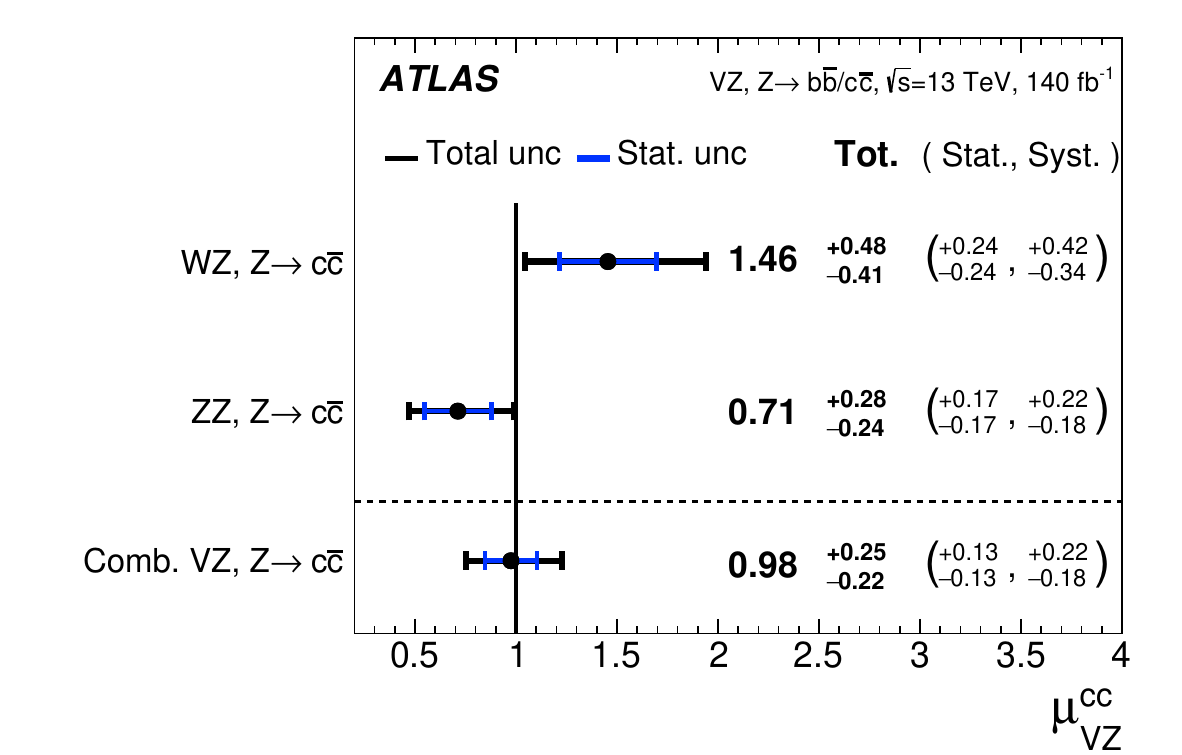}
\label{fig:diboson1b}
}
\caption{The fitted values of the $VZ$ signal strengths for the \protect\subref{fig:diboson1a} \zbb and \protect\subref{fig:diboson1b} \zcc processes.
The values for the $WZ$ and $ZZ$ signal strengths are obtained from a simultaneous fit with the signal strengths for each of the $WZ$ and $ZZ$ processes floating independently for the two decay modes. %
}
\label{fig:diboson1}
\end{figure}

Measurements of $VZ$ production, which are used to validate the Higgs boson analysis, yield  signal strengths of
\begin{align*}
\mu_{VZ}^{bb} & =  0.92^{+ 0.13}_{- 0.11} = 0.92\pm 0.05\:(\textrm{Stat.})^{+0.12}_{-0.10}\:(\textrm{Syst.}), \\
\mu_{VZ}^{cc} & =  0.98^{+ 0.25}_{- 0.22} = 0.98\pm 0.13\:(\textrm{Stat.})^{+0.22}_{-0.18}\:(\textrm{Syst.}),
\end{align*}
in good agreement with the SM predictions. The correlation
between the two measurements is +46\% primarily driven by signal
modelling uncertainties, which also represent the leading contribution
to the overall uncertainty. For the $VZ$, $\zcc$ process, the observed
(expected) significance over the background-only prediction is \vzccSigObs~
(\vzccSigExp) standard deviations. The $VZ$, $\zbb$ process is observed with a significance greater than 10 standard deviations.  A fit is also
performed with separate signal strengths for the $WZ$ and $ZZ$ processes in the two decay modes of the $Z$ boson; the results are shown in \Fig{\ref{fig:diboson1}}.
The $WZ$, $\zbb$ process is observed (expected) at \wzbbSigObs~(\wzbbSigExp) standard deviations, while the significance is greater than 10 standard deviations for $ZZ$, $\zbb$.
Evidence for both $WZ$ and $ZZ$ with $\zcc$ is found with an observed (expected) significance of
\wzccSigObs~(\wzccSigExp) and \zzccSigObs~(\zzccSigExp) standard deviations respectively.

\subsection{$VH$ signal strength measurements}

The measured signal strength for the \hbb\ and \hcc\ signals are:
\begin{align*}
\mu_{VH}^{bb} & = \muvhbb^{+ 0.16}_{- 0.15} =\muvhbb \pm 0.10\:(\textrm{Stat.})^{+0.13}_{-0.11}\:(\textrm{Syst.}), \\
\mu_{VH}^{cc} & = \muvhcc^{+5.4}_{- 5.2} =\muvhcc ^{+4.0}_{-3.9}\:(\textrm{Stat.})^{+3.7}_{-3.5}\:(\textrm{Syst.}).
\end{align*}

\begin{figure}[tb!]
\centering
\includegraphics[width=0.45\textwidth]{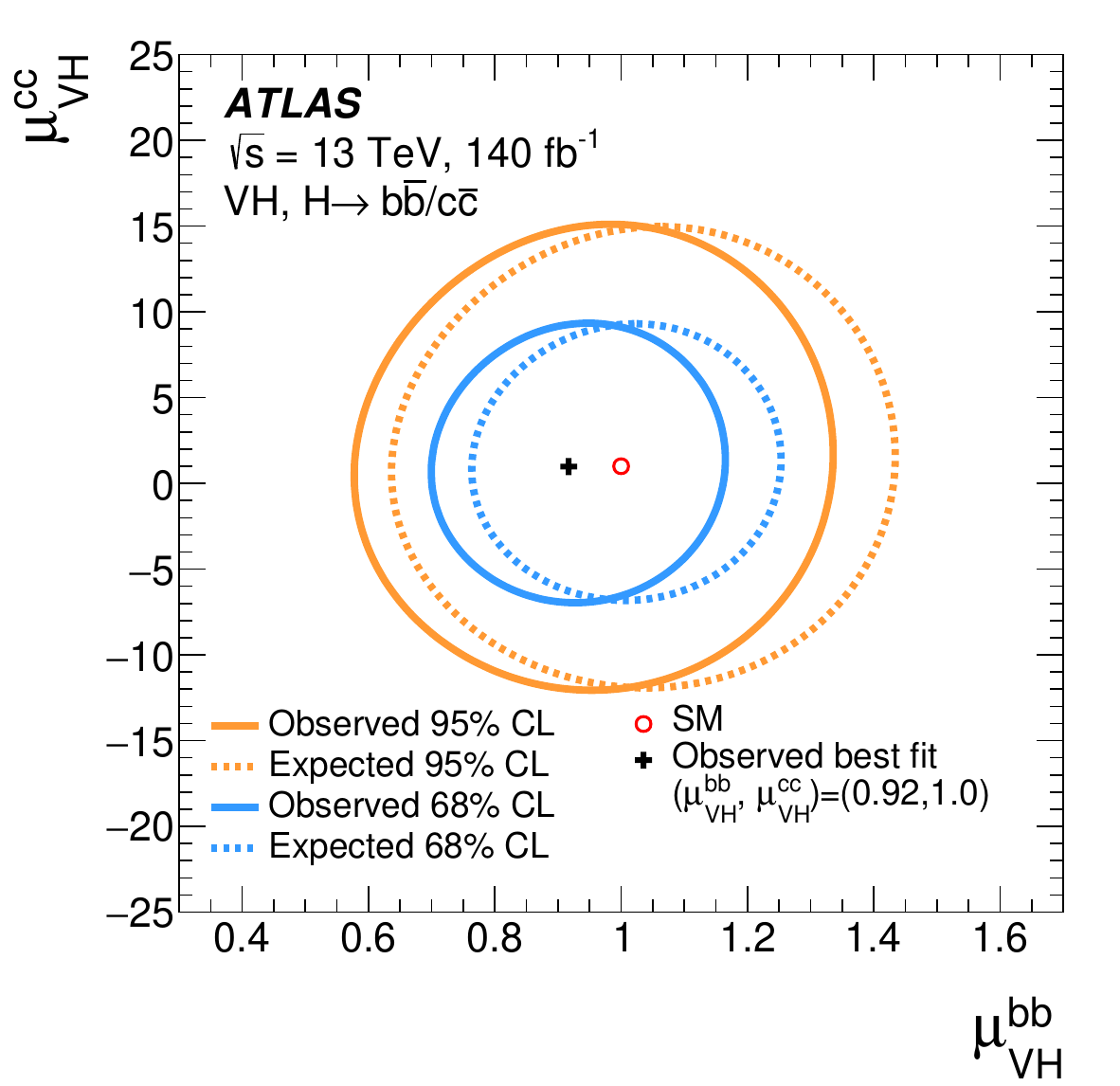}
\caption{The observed (solid lines) and expected (dashed lines) 68\% and 95\% CL contours of the  \vhbb\ and \vhcc\ signal strengths, along with their best-fit values.}
\label{fig:vh1}
\end{figure}

\begin{figure}[tb!]
\centering
\includegraphics[width=0.6\textwidth]{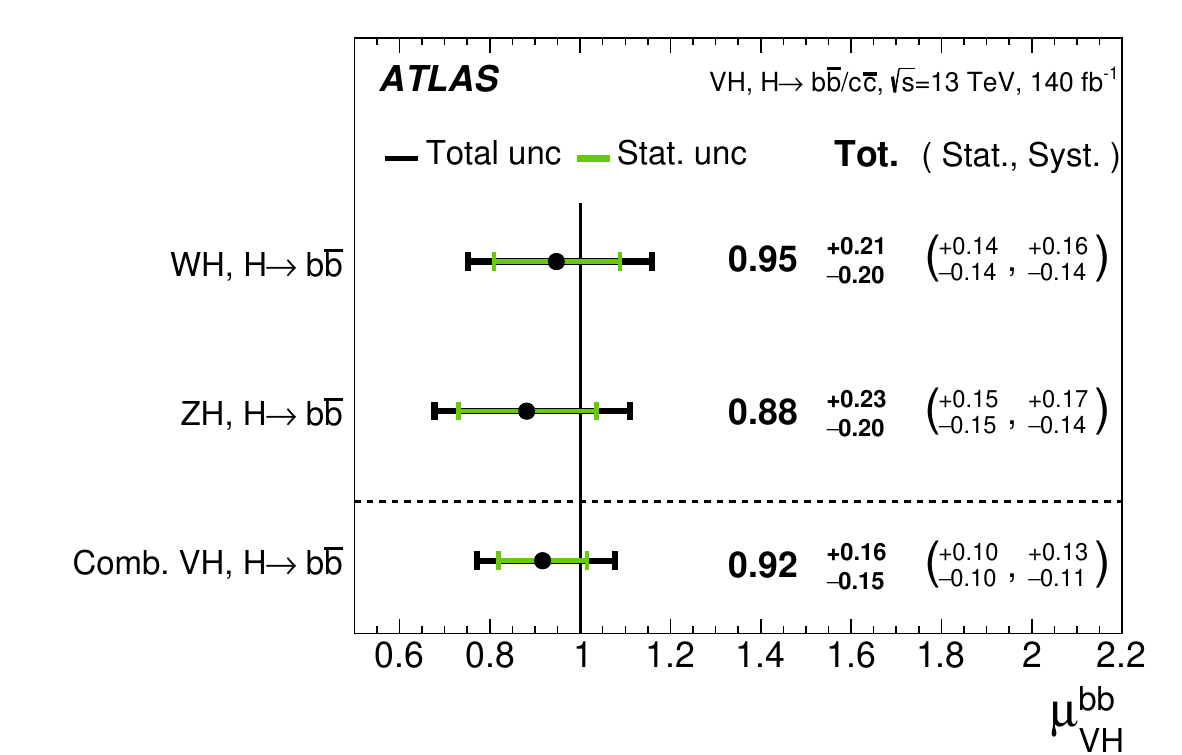}
\caption{The fitted values of the $WH$, \hbb\ and $ZH$, \hbb\ signal strengths, along with their combination.}
\label{fig:vh2}
\end{figure}

The results are shown in \Fig{\ref{fig:vh1}} together with the expected and observed 68\% and 95\% CL contours.
Both measurements show good agreement with the SM and their correlation is +5\%.

The measurement of $\mu_{VH}^{bb}$ has an observed (expected) significance of \vhbbSigObs~(\vhbbSigExp) standard deviations.
The measured \hbb\ signal strengths of the $WH$ and $ZH$ processes separately are shown in
\Fig{\ref{fig:vh2}}, in agreement with the SM predictions. The $ZH$, \hbb\ measurement has an observed (expected) significance of \zhbbSigObs~(\zhbbSigExp) standard deviations,
while the observed (expected) significance for $WH$, \hbb\  is \whbbSigObs~(\whbbSigExp) standard deviations. This is the first observation of the $WH$, \hbb\ process.

The effects of systematic uncertainties in the measurement of the \vhbb, \whbb, \zhbb and \vhcc signal strengths
are shown in~\Tab{\ref{tab:break}}. The impact of a set of systematic uncertainties is defined as the difference
in quadrature between the uncertainty in $\mu$ computed when all NPs are fitted and that when the NPs in
the set are fixed to their best-fit values. The total statistical uncertainty
is defined as the uncertainty on $\mu$
when all the NPs are fixed to their best-fit values. The total systematic uncertainty is defined as the
difference in quadrature between the total uncertainty in $\mu$ and the total statistical uncertainty.

\begin{table}[tb!]
\caption{The breakdown of contributions to the uncertainty in the fitted value of the signal strengths ($\sigma_\mu$)  for \vhbb, \whbb, \zhbb and \vhcc. The sum in quadrature of uncertainties from different sources may differ from the total due to correlations. In cases where the upward and downward systematic variations have different values, the mean of the absolute values is shown. }
\centering
\begin{tabular}{ l | l  c | c c || c}
\toprule
\multicolumn{2}{l}{\multirow{2}{*}{Source of uncertainty}} & \multicolumn{4}{c}{$\sigma_\mu$}  \\
\cline{3-6}
\multicolumn{2}{l}{} &  $\vhbb$ \rule{0pt}{2.5ex} & $\whbb$ \rule{0pt}{2.5ex} &  $\zhbb$ \rule{0pt}{2.5ex} &  $\vhcc$ \rule{0pt}{2.5ex} \\
\midrule
\multicolumn{2}{l}{Total} 							&   0.153  &   0.204  &  0.216  & 5.31 \\
\multicolumn{2}{l}{Statistical} 						&   0.097  &   0.139  &  0.153  & 3.94 \\
\multicolumn{2}{l}{Systematic} 						&   0.118  &   0.149  &  0.153  & 3.57 \\
\midrule
\multicolumn{4}{l}{Statistical uncertainties}\\
\midrule
\multicolumn{2}{l}{Data statistical} 					&   0.090  &    0.129  &    0.139 & 3.67 \\
\multicolumn{2}{l}{\ttbar $e\mu$ control region} 			&   0.009  &    0.014 	&    0.027 	& 0.08 \\
\multicolumn{2}{l}{Background floating normalisations} 	&   0.034  &    0.049  &    0.042	&1.24  \\
\multicolumn{2}{l}{Other $VH$ floating normalisation} 	&   0.007  &    0.018  &    0.014 	& 0.33 \\
\midrule
\multicolumn{2}{l}{Simulation samples size} 			& 0.023 	& 0.033 	&  0.030 	& 1.62  \\
\midrule
\multicolumn{4}{l}{Experimental uncertainties}\\
\midrule
\multicolumn{2}{l}{Jets} 							&  0.027 	&   0.035 	&    0.030 & 1.02 \\
\multicolumn{2}{l}{\met} 							&  0.010	&   0.005  	&    0.021 & 0.23 \\
\multicolumn{2}{l}{Leptons} 						&  0.003 	&   0.002  	&    0.010 & 0.25 \\
\multirow{3}{*}{Flavour tagging}
& $b$-jets 					&  0.020	&  0.018  	&    0.026 & 0.29   \\
& $c$-jets 					&  0.013 	&  0.017 	&    0.012 & 0.73   \\
& light-flavour jets 				&  0.005 	&  0.008 	&    0.008 & 0.66   \\
\multicolumn{2}{l}{\Pileup} 						&  0.008 	&  0.017 	&    0.002 & 0.23   \\
\multicolumn{2}{l}{Luminosity} 						&  0.006 	&  0.007 	&    0.006 & 0.08  \\
\midrule
\multicolumn{4}{l}{Theoretical and modelling uncertainties}\\
\midrule
\multicolumn{2}{l}{Signal} 							&  0.076	& 0.074	&   0.101 	& 0.72  \\
\multicolumn{2}{l}{$Z$~+~jets} 						& 0.042	& 0.018	&  0.081	& 1.77  \\
\multicolumn{2}{l}{$W$~+~jets} 					& 0.054 	& 0.087 	&  0.026	&  1.42 \\
\multicolumn{2}{l}{\ttbar and $Wt$} 					& 0.018 	& 0.033 	&  0.018 	& 1.02  \\
\multicolumn{2}{l}{Single top-quark ($s$-, $t$-ch.)} 		& 0.010  	& 0.018 	&  0.002 	& 0.16 \\
\multicolumn{2}{l}{Diboson} 						& 0.033	& 0.039 	&  0.049 	& 0.52 \\
\multicolumn{2}{l}{Multijet} 						& 0.005	& 0.010	&  0.005 	& 0.55 \\
\bottomrule
\end{tabular}

\label{tab:break}
\end{table}

For the $WH$ and $ZH$ signal strength measurements the total statistical and systematic uncertainties are similar in
size. The background modelling, in particular $W$+jets and $Z$+jets production, and signal modelling represent the
largest contribution to the total systematic uncertainty while detector-related systematic uncertainties, primarily jet and flavour tagging,
have a subdominant effect.

An improvement of 23\% (10\%) is observed in the total uncertainty of the $WH$ ($ZH$) measurement compared with the previous \vhbb result in \Refn{\cite{HIGG-2018-51}};
the $WH$ channel, in particular, profits from the 25\% increased $c$-jet rejection of the \DLoner~algorithm for a similar $b$-jet efficiency, improved treatment of \tauhad candidates,
addition of the $\SI{75}{\GeV} < \ptw < \SI{150}{\GeV}$ region, use of BDT$_{\ldr\,\mathrm{CR}}$, and the dedicated CRs to constrain the
\topbq~background. Additional reasons for the improved performance are more precise flavour tagging and jet calibrations, larger signal-background separation of re-optimised BDTs, improved boosted \hbb reconstruction and identification, and up to a 50\% increase in the effective size of the MC simulated samples primarily in the $V$+jets background.

\begin{figure}[tb!]
\centering
\includegraphics[width=0.6\textwidth]{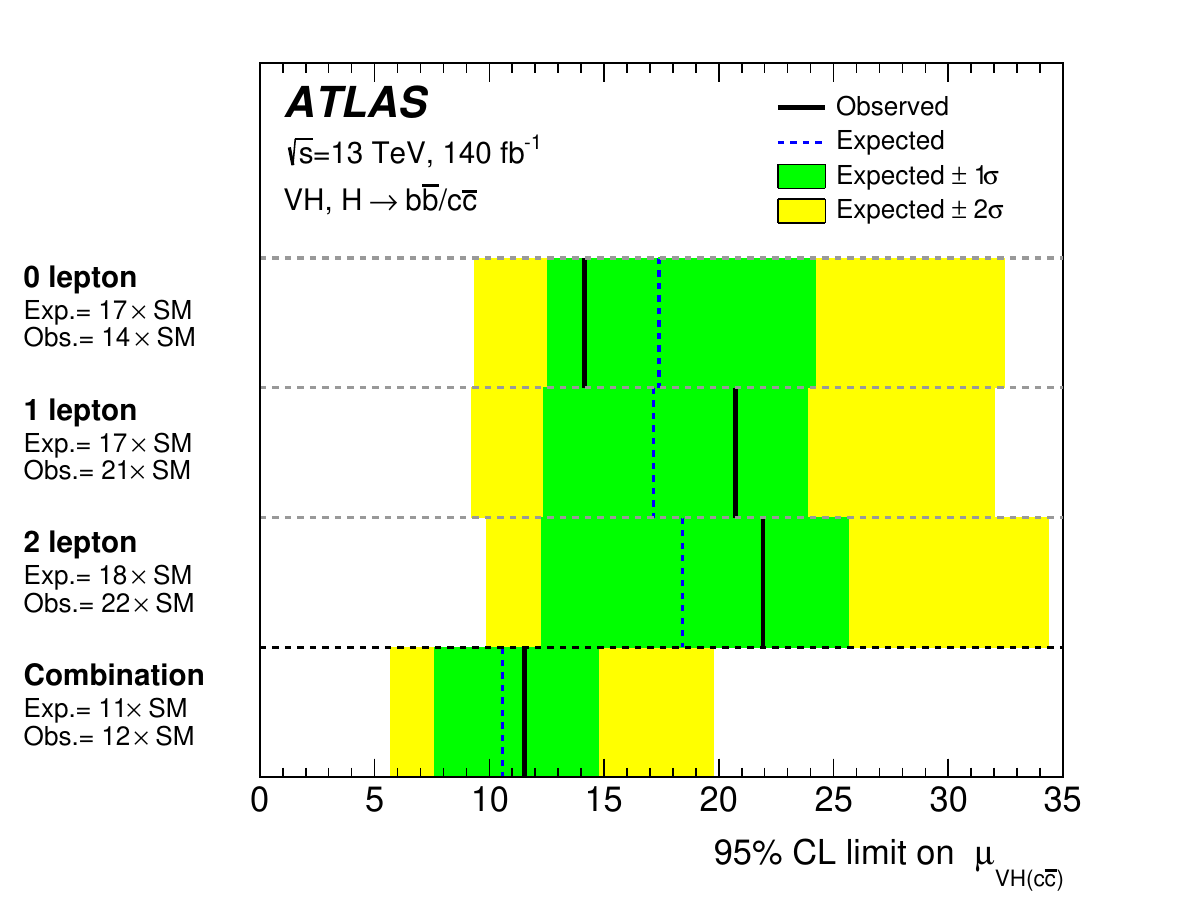}
\caption{The observed and expected 95\% CL upper limits on the \vhcc\ signal strength in each lepton channel and for the combined fit.
The single-channel limits are obtained in a fit in which each channel has a separate \vhcc\  signal strength.
A single parameter of interest is used in both fits for the \vhbb signal strength.
} \label{fig:vh3}
\end{figure}

The $\mu_{VH}^{cc}$ result corresponds to an observed upper limit of \hccLimObs~times the SM predictions
at 95\% CL, while a limit of \hccLimExp~is expected in the case of no \hcc process.
The upper limits at 95\% CL on $VH$, \hcc\  for each individual channel and the combinations  %
are shown in \Fig{\ref{fig:vh3}}.
The impacts of statistical and systematic uncertainties in the observed signal strength for $\vhcc$ are at a similar level.
The modelling of the $V$+jets background, jet and flavour tagging related calibration and the impact of the finite size of the MC samples %
have comparable effects.
The expected sensitivity has improved by roughly a factor of three compared with  the previous iteration of the analysis~\cite{HIGG-2021-12}.
Several interconnected factors contribute to this. The improved flavour tagging algorithms reduced the contamination of non $c$-jet background
by roughly 40\% while maintaining the same signal efficiency; at the same time the usage of an additional looser working point allows  the signal acceptance to be increased,
while maintaining the background contamination under control. The combined effect of the new working points and event classification is estimated as a 25\% improvement on the previous result. A more efficient generator set-up for \Vjets events,
bringing up to a factor of five times more effective statistics, and the more advanced technique to parameterise the tagging probabilities
has significantly reduced the uncertainty due to the finite size of the MC simulated samples.
Furthermore, an additional 40\% improvement in the expected sensitivity is obtained by deploying a multivariate discriminant over using the dijet invariant mass as fit variable.\footnote{In the current configuration, the BDT is further using flavour tagging information and has the capability to separate \ctct and \ctcl events.}
Finally, fitting the normalisation of the \Vhf components in Hbb and Hcc categories simultaneously reduces the related modelling uncertainties by up to a factor of two,
improving the expected sensitivity by 6\%.

\begin{figure}[tb!]
\centering
\subfloat[]{
\includegraphics[width=0.50\textwidth]{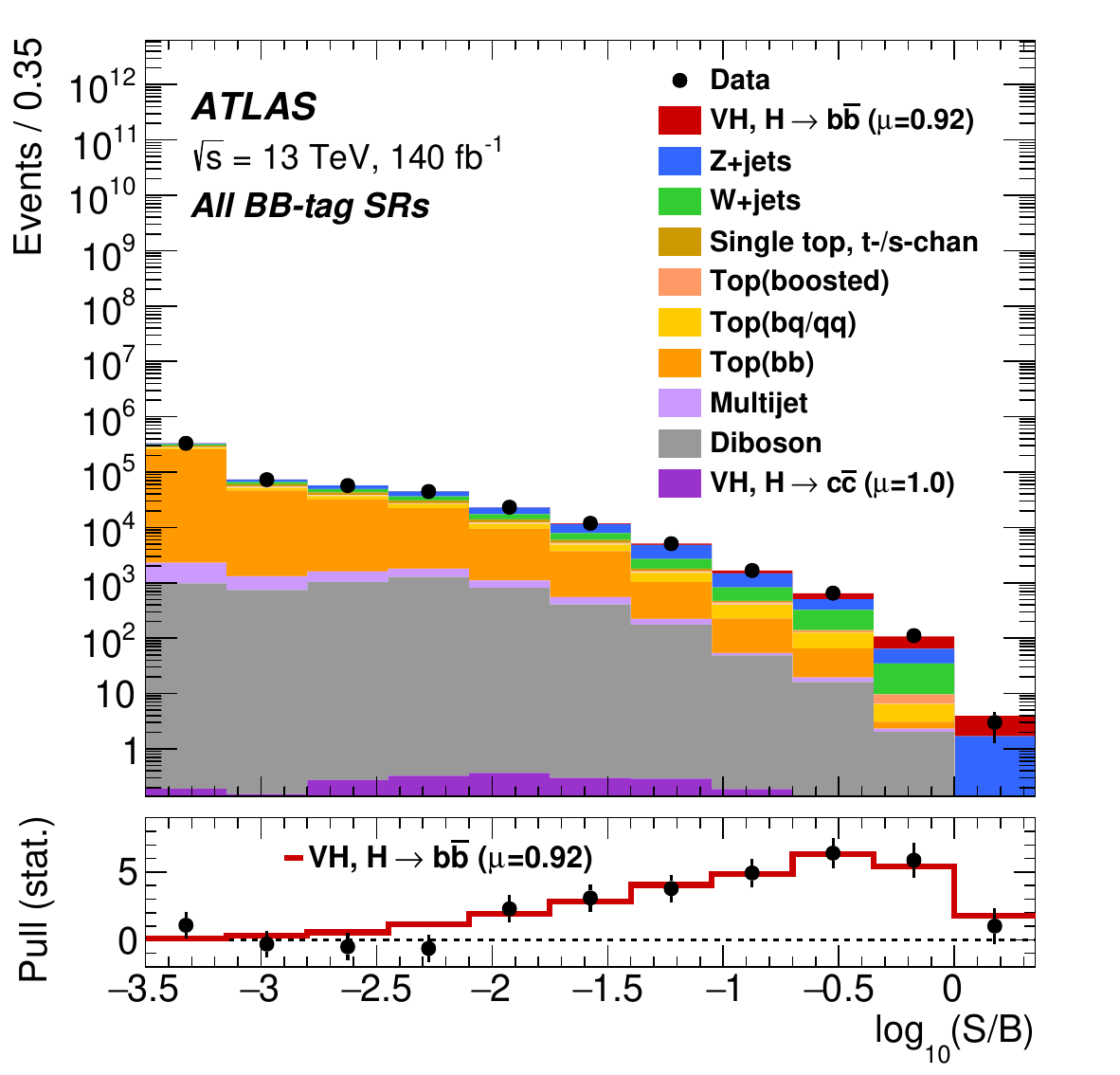}
\label{fig:weighted1}
}
\subfloat[]{
\includegraphics[width=0.50\textwidth]{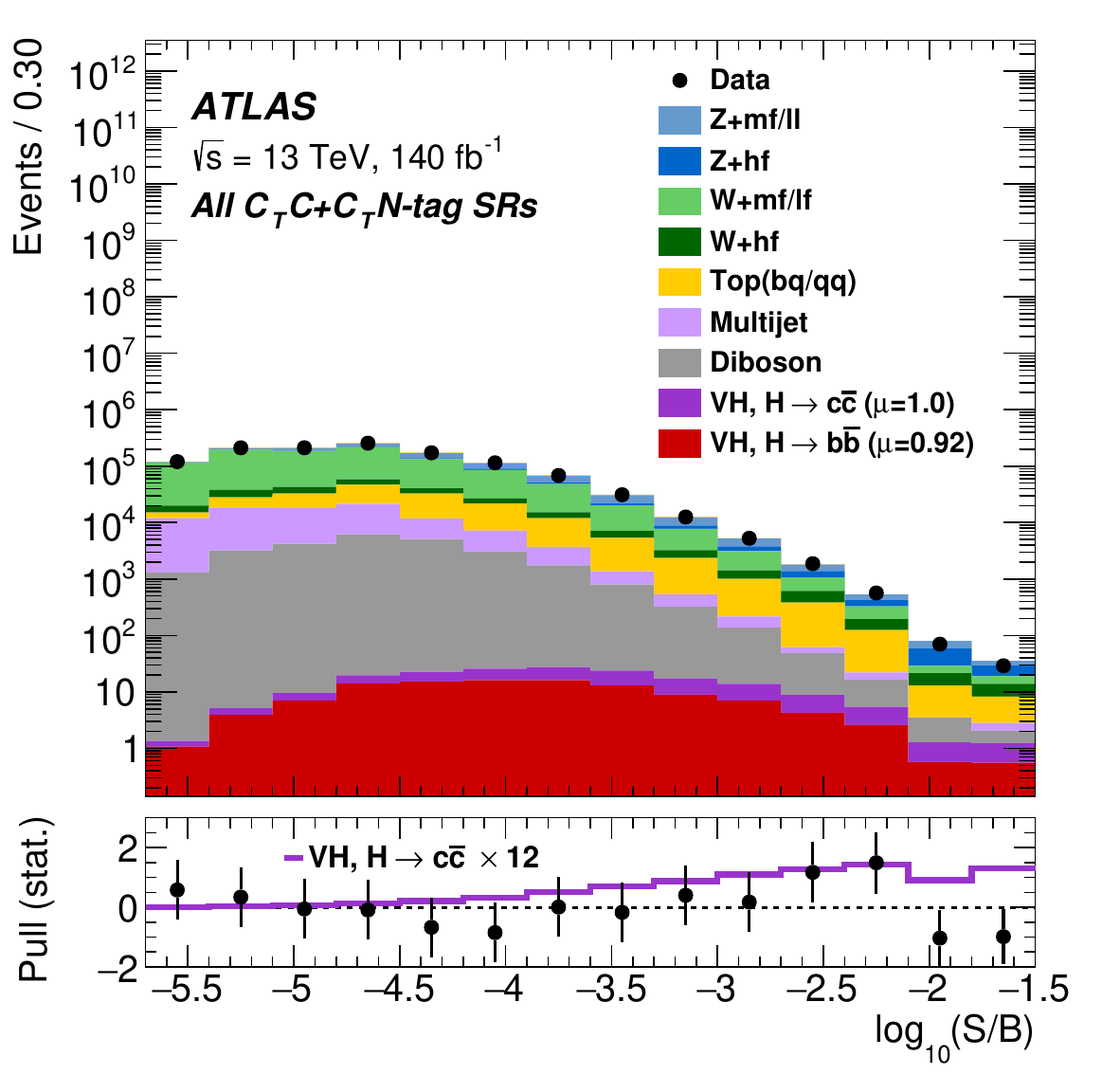}
\label{fig:weighted2}
}
\caption{The event yields as a function of $\log_{10} (\mathrm{S/B})$ for data, background and a Higgs boson signal for the nominal $VH$ fit.
Final-discriminant bins in the \protect\subref{fig:weighted1} Hbb and \protect\subref{fig:weighted2} Hcc regions are combined into bins of $\log_{10} (\mathrm{S/B})$ with $S$ being the fitted $VH$, \hbb\ and \hcc\ contribution respectively and $B$ the fitted background yields.
The Higgs boson contributions in the top pad are shown after rescaling the SM cross-section according to the value of the signal strength extracted from data:  0.92 for  \hbb and 1.0 for \hcc.
In the lower panel, the pull of the data relative to the background (the statistical significance of the difference between data and fitted background) is shown with statistical uncertainties only.
The full line indicates the pull expected from the sum of fitted signal and background relative to the fitted background; in the \hcc case, the fitted signal is  scaled to the value of the observed upper limit.}
\label{fig:weighted}
\end{figure}

\Fig{\ref{fig:weighted}} shows the data, background (B) and signal (S) yields, where the bins of the BDT output in SRs are
combined into bins of $\log_{10} (\mathrm{S/B})$. Distributions are shown separately for the Hbb and Hcc categories. In each category,
the definition of S is adapted to the targeted Higgs boson decay mode, while events from the other decay mode are counted as B.

\subsection{STXS result}

The measured values of the product of the \vh~cross-sections and the \hbb\ ~and $V\to$leptons branching fractions,
together with the SM predictions in the selected STXS regions, are summarised in \Tab{\ref{tab:STXS}}.
The cross-sections are consistent with the SM expectations within one standard deviation except in the
$\SI{400}{\GeV} < \ptwt < \SI{600}{\GeV}$ and $\ptzt > \SI{600}{\GeV}$ categories, where the observed cross-sections
are lower than the SM predictions by more than one standard deviation, but less than two standard deviations.
Overall, the $p$-value for compatibility of the measurements with the SM predictions is 90\%.

\begin{figure}[tb!]
\centering
\includegraphics[width=0.75\textwidth]{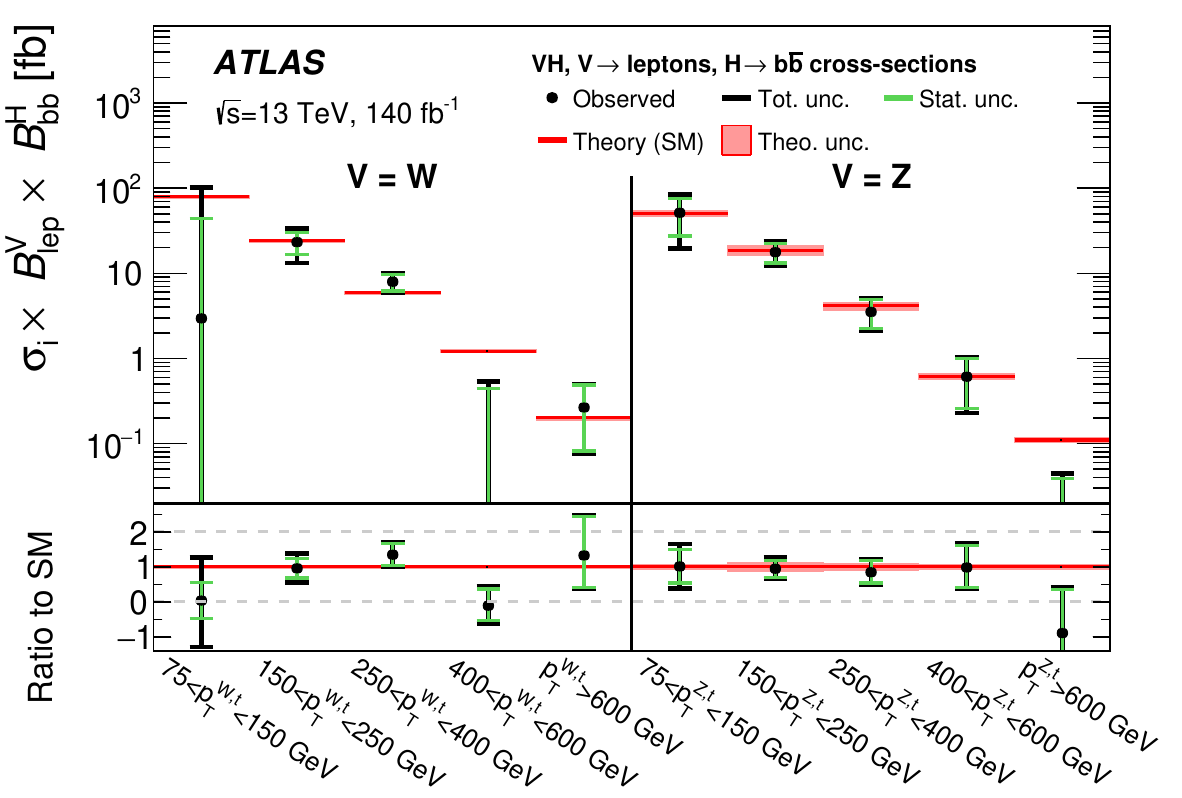}
\caption{The measured $VH$ cross-sections times the $V\to$~leptons and $H\to b\bar{b}$ branching fractions in the reduced STXS scheme with no split in the number of jets. }
\label{fig:stxs2}
\end{figure}

\begin{figure}[tb!]
\centering
\includegraphics[width=0.75\textwidth]{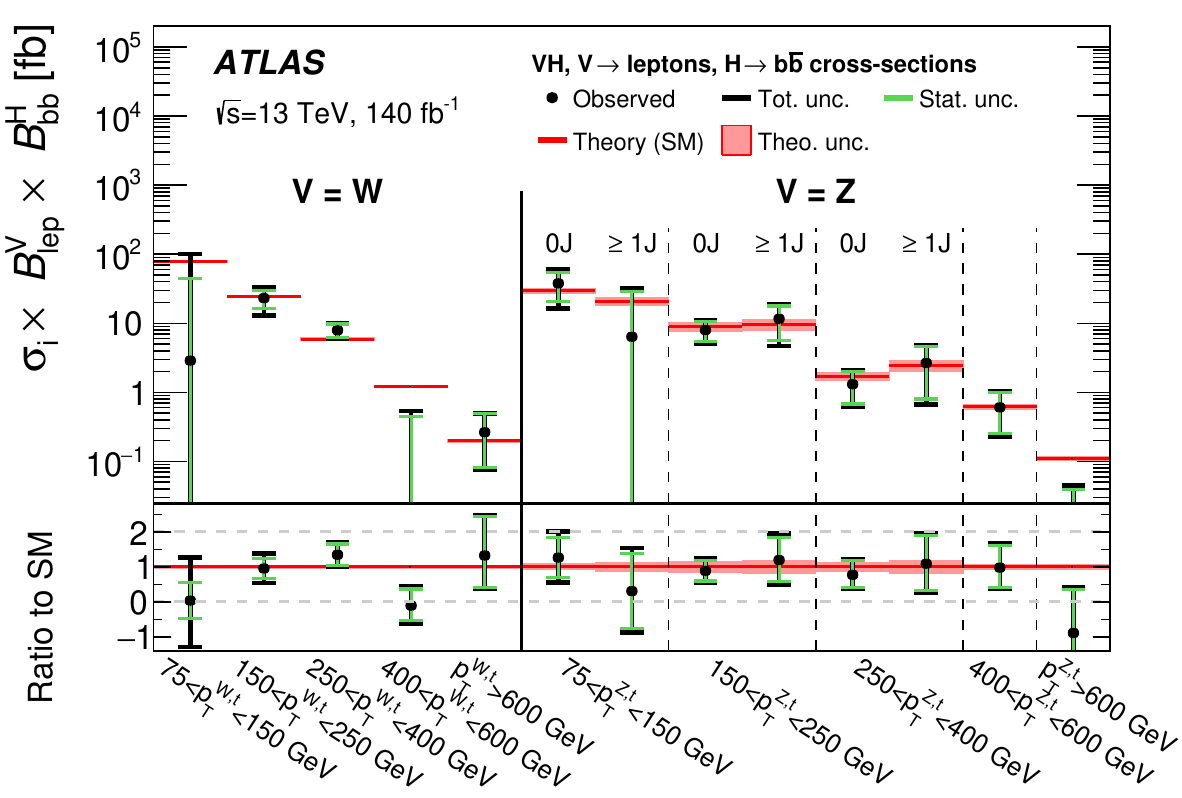}
\caption{The measured $VH$ cross-sections times the $V\to$~leptons and $H\to b\bar{b}$ branching fractions  in the extended STXS scheme. }
\label{fig:stxs1}
\end{figure}

\Fig{\ref{fig:stxs2}} shows  the result for a configuration with no categorisation as a function of additional truth jets. In the $WH$ channel, the relative uncertainty in the cross-sections is the smallest for the
$\SI{250}{\GeV} < \ptwt < \SI{400}{\GeV}$ category at 25\%. In the $ZH$ channel the smallest relative error of 33\% is in the $\SI{150}{\GeV} < \ptwt < \SI{250}{\GeV}$ category.
The statistical precision and the smaller systematic uncertainties in the $ZH$ channel allows a further split of the measurements
into the $\Njet=0$ and $\Njet \geq 1$ categories for $\SI{75}{\GeV} < \ptzt < \SI{400}{\GeV}$ as shown in  \Fig{\ref{fig:stxs1}}. The $\Njet=0$  categories have a relative uncertainty roughly a factor two smaller than the corresponding $\Njet \geq 1$ categories in the same $\pTVt$ interval.
The  categories with $\ptzt> \SI{400}{\GeV}$ remain inclusive.

In all cases, the uncertainties are dominated by the data statistical uncertainty, which also includes the contribution from the floating normalisation factors on the leading background contributions and %
the effect of signal migrations between categories,%
except in the
$\SI{75}{\GeV} < \ptwt < \SI{150}{\GeV}$ and $\SI{150}{\GeV} < \ptwt < \SI{250}{\GeV}$ categories, where the systematic uncertainties are the largest contributions to the total uncertainty. Compared to the previous result in \Refn{\cite{ATLAS-CONF-2021-051}}, a finer binning with a split at $\ptvt = \SI{400}{\GeV}$ was chosen. This was possible due to the 50\% improvement in the expected sensitivity of the boosted Hbb category, driven by the use of a multivariate discriminant over the dijet invariant mass as the fit variable and the improved performance of the flavour tagging algorithms.

The largest sources of systematic uncertainties are the modelling of \Zhf production, which primarily affects the $ZH$ categories, and the  modelling of the \Whf production and top-quark background that primarily affect the $WH$ categories. %
Uncertainties related to the jet calibration, both small- and large-$R$ jets, and $b$-tagging are the leading detector-related effects, but their contribution is small compared with the modelling uncertainties.
The relative impact of the limited size of the MC samples is the third largest component, and is similar across all the regions.
For the \zh~measurements, the signal uncertainties also have a significant contribution due to the limited
precision of the theoretical calculations of the gluon-initiated process, for which the lowest order is a box or triangle diagram.

\begin{table}[tb!]
\caption{The best-fit values and uncertainties for the measured \vh cross-sections times the $V\to$~leptons and the \hbb\ branching
fraction, in the extended 1.2 STXS scheme.  For the $ZH$ process, inclusive results in truth jet multiplicity are also reported.
Such results are obtained from an alternative fit configuration described in Section~\ref{sec:stat}; the cross-section results for other categories are unchanged
in this alternative configuration within the precision considered. The SM predictions for each region, computed using the inclusive cross-section
calculations and the simulated event samples are also shown. The symmetrised contributions to the total measurement uncertainty
from statistical (Stat.\ unc.) or systematic uncertainties (Syst.\ unc.) related to the signal prediction (Th.\ sig.),
background prediction (Th.\ bkg.), and the experimental performance (Exp.) are given separately. The total systematic
uncertainty, equal to the difference in quadrature between the total uncertainty and the statistical uncertainty, differs
from the sum in quadrature of the Th.\ sig., Th.\ bkg., and Exp.\ systematic uncertainties due to correlations. All
leptonic decays of the $V$ bosons (including those to $\tau$-leptons, $\ell$= $e$, $\mu$, $\tau$) are considered.}
\centering
\resizebox{\textwidth}{!}{
\begin{tabular}{ccc |  rcl |  rcr | r  r  r  r }
\toprule
\multicolumn{3}{c}{STXS region}                                      & \multicolumn{3}{c}{SM prediction}  & \multicolumn{3}{c|}{Measurement} & Stat. unc.    & \multicolumn{3}{c}{Syst. unc. [fb]}                     \\
\cline{11-13}
Process & $p_{\textrm{T}}^{V\textrm{, t}}$ interval & \Njet &\multicolumn{3}{c}{[fb]}           & \multicolumn{3}{c|}{[fb]}   & [fb]     & Th.\ sig.\    & Th.\ bkg.\            & Exp.\           \\
\midrule
\multirow{5}{*}{ $W(\ell\nu)H$ } &  75--150~\GeV 		& $\geq$0  	&	79 .2   & $\pm$ &  2.8 		& 3 		& $\pm$  & 102 			& 41  	& 13    &  89  & 36 \\
&  150--250~\GeV 		& $\geq$0  	&     	24.3 	  & $\pm$ &  1.0 		& 23	 	& $\pm$  &   10				&  7   	& 2      &   7   &   3 \\
&  250--400~\GeV 		& $\geq$0  	&	5.90 	  & $\pm$ & 0.25		& 7.9 	&$\pm$   &  2.1				&  1.8  	& 0.5    &    0.8 &  0.3 \\
&  400--600~\GeV 		& $\geq$0	  	&	1.03 	  & $\pm$ & 0.05 		& -0.11 	& $\pm$  &  0.54			&  0.46  	& 0.05    &    0.25 &  0.09 \\
&  $>600$~\GeV 		& $\geq$0		&	0.20   & $\pm$ & 0.01 		& 0.26 	& $\pm$   & 0.21                        &  0.20  	& 0.02    &    0.04  & 0.03  \\
\hline
\multirow{11}{*}{ $Z(\ell\ell/\nu\nu)H$}  &   \multirow{3}{*}{75--150~\GeV} 	& $\geq$0	        & 50.7       & $\pm$ &   3.9		& 51          & $\pm$   & 32 	 & 24  & 5   & 18   & 11 \\   %
&                                                       & =0 		& 29.9  	& $\pm$ &   2.5 	& 38     	 & $\pm$  	& 22	          & 17  & 3  & 12   & 6 \\
&   							& $\geq$1		& 20.7 	& $\pm$ &   2.6		&   6          & $\pm$  	& 25  	  & 22  & 4  &   9   & 8 \\
\cmidrule{2-13}
&  \multirow{3}{*}{150--250~\GeV} 	& $\geq$0         & 18.7      & $\pm$ &   3.5  	&  17.7      & $\pm$   & 5.8	   &  4.6   &  1.7   & 3.0    & 1.0 \\    %
&							& =0  		& 9.0. 	& $\pm$ &   1.3		&   8.0   	& $\pm$ 	& 3.1           &  2.7    & 0.6 & 1.4   & 0.5 \\
&   							& $\geq$1		& 9.7 	& $\pm$ &   1.9		&  11.6     	& $\pm$ 	& 7.1           &  6.1   &  1.0  & 3.2   & 1.4  \\
\cmidrule{2-13}
&  \multirow{3}{*}{250--400~\GeV} 	& $\geq$0		& 4.15 	 & $\pm$ &   0.45	&   3.5      & $\pm$   & 1.5  	   &  1.3   & 0.4   & 0.5    & 0.2 \\    %
&							& =0  		& 1.70 	& $\pm$ &   0.22	&   1.31    	& $\pm$ 	& 0.72          & 0.66    & 0.14  & 0.25    & 0.10 \\
&  							& $\geq$1		& 2.45 	& $\pm$ &   0.45	&   2.7      	& $\pm$  	& 2.1		  & 1.9   & 0.3  & 0.7  & 0.3 \\
\cmidrule{2-13}
&  400--600~\GeV 				& $\geq$0		& 0.62 	& $\pm$  &  0.05	&   0.61 	& $\pm$  	& 0.40 	  & 0.37  &  0.05  & 0.12  & 0.08 \\
\cmidrule{2-13}
&  $>600$~\GeV 				& $\geq$0		& 0.11 	& $\pm$  &  0.01	& -0.10	& $\pm$  	& 0.12         & 0.12   & 0.01  & 0.03 & 0.01 \\

\bottomrule
\end{tabular}
}
\label{tab:STXS}
\end{table}

\section{Interpretation in the $\kappa$-framework }
The best-fit values of the $VH$ signal strength are interpreted in the context of the $\kappa$-framework~\cite{Heinemeyer:2013tqa,deFlorian:2016spz} by reparameterising the $\mu_{VH}^{bb}$ and $\mu_{VH}^{cc}$
in the likelihood function in terms of the Higgs-bottom and Higgs-charm multiplicative coupling modifiers, $\kappa_b$ and $\kappa_c$, assuming that they affect only  the Higgs boson decays.\footnote{The effect of anomalous $\kappa$ values on the $ggZH$ production process is found to be negligible within the range of parameters probed by this analysis.
}
Including effects in both the partial and full width, considering only SM decays and setting all other couplings to their SM predictions, the parameterisation is:
\begin{eqnarray}
\mu_{VH}^{bb} = \frac{\kappa_b^2}{1+B^{\mathrm{SM}}_{Hbb}(\kappa_b^2-1)+B^{\mathrm{SM}}_{Hcc}(\kappa_c^2-1)} \label{eq:kb}, \\
\mu_{VH}^{cc} = \frac{\kappa_c^2}{1+B^{\mathrm{SM}}_{Hbb}(\kappa_b^2-1)+B^{\mathrm{SM}}_{Hcc}(\kappa_c^2-1)}  \label{eq:kc},
\end{eqnarray}
where $B^\mathrm{SM}_{Hbb}$ and $B^\mathrm{SM}_{Hcc}$ are the \hbb\ and \hcc\ branching fraction predictions in the SM.

\begin{figure}[tb!]
\centering
\includegraphics[width=0.6\textwidth]{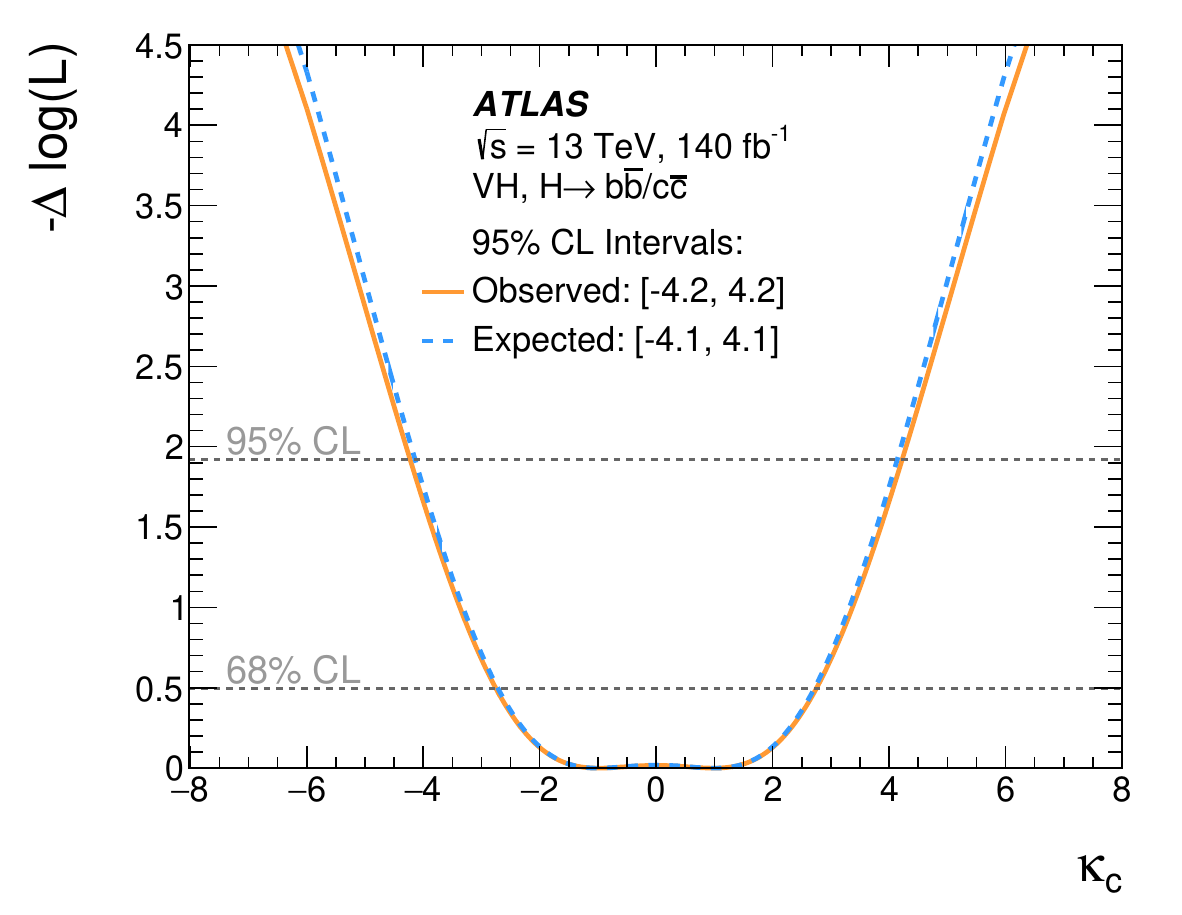}
\caption{The observed  (solid) and expected (dashed) values of the negative profile log-likelihood ratio as a function of $\kappa_c$, is obtained in a fit with
$\kappa_b$ fixed to unity. The signal strength $\mu_{VH}^{bb}$ is not parameterised but taken as a single parameter of interest in the fit.
} %
\label{fig:kc}
\end{figure}

First, the direct $\kappa_c$ constraint from the \vhcc\ process is extracted by setting $\kappa_b=1$ in \Eqn{\ref{eq:kc}} and not parameterising $\mu_{VH}^{bb}$.
Constraints on $\kappa_c$ are set using the profile-likelihood ratio test statistic and are shown in \Fig{\ref{fig:kc}}. %
The result from the full fit achieves an observed (expected) constraint of $|\kappa_c|<\kappacObs$ ($|\kappa_c| < 4.1$) at 95\% CL.
An equivalent approach for $\kappa_b$ yields an observed (expected) 95\% CL interval of $0.65< |\kappa_b| < 1.37$  ($0.72< |\kappa_b| < 1.62$).

\begin{figure}[tb!]
\centering
\subfloat[]{
\includegraphics[width=0.50\textwidth]{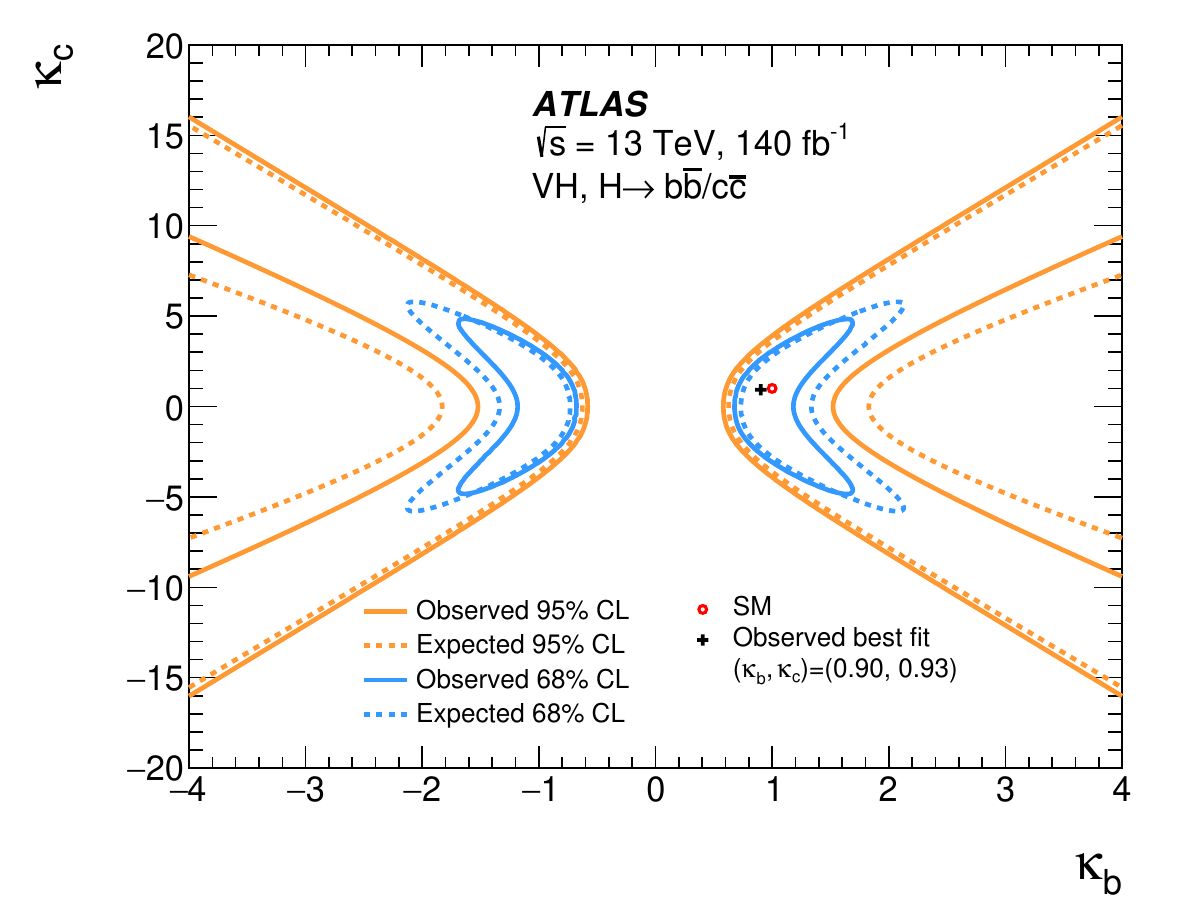}
\label{fig:kappabkappac}
}
\subfloat[]{
\includegraphics[width=0.50\textwidth]{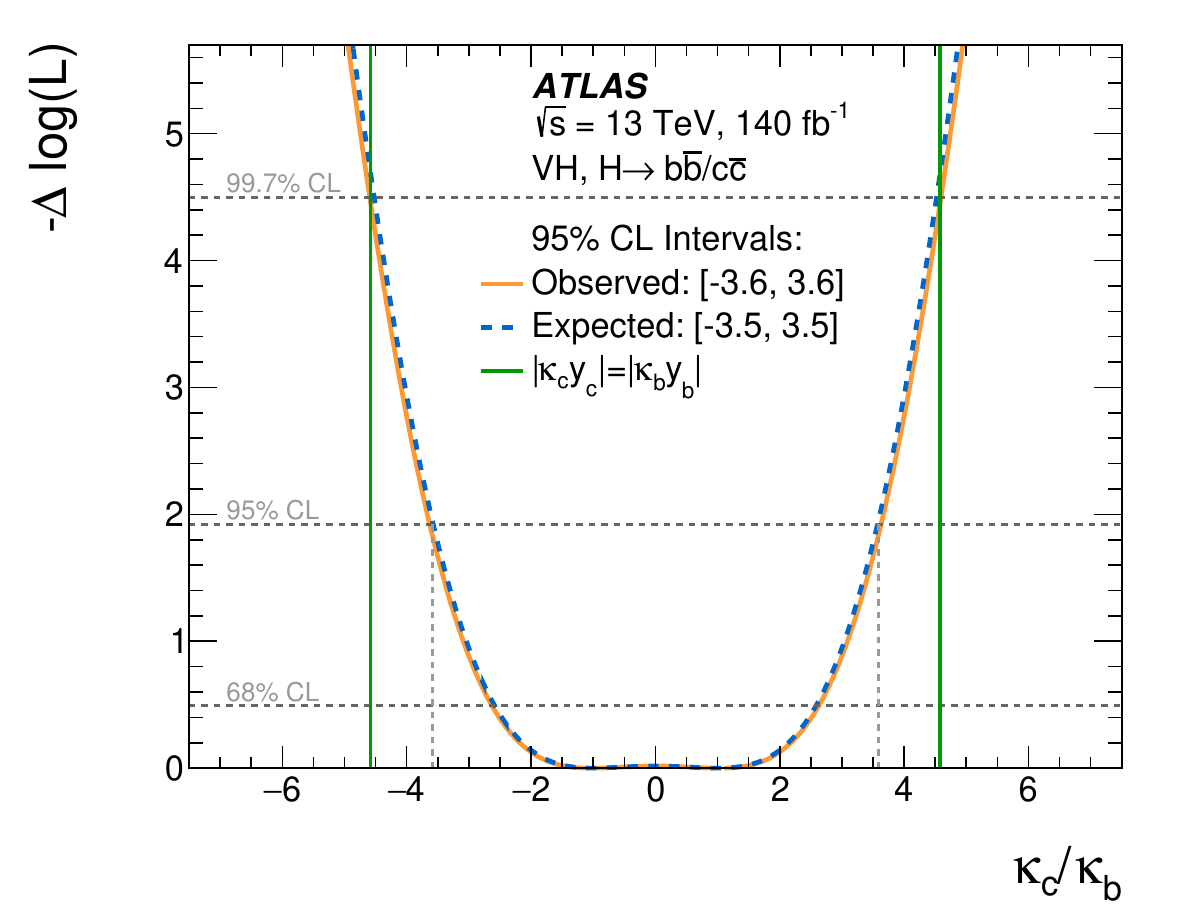}
\label{fig:kapparatio}
}
\caption{\protect\subref{fig:kappabkappac} The observed (solid) and expected (dashed) constraints on $\kappa_b$ and $\kappa_c$ at 68\% CL and 95\% CL confidence levels. \protect\subref{fig:kapparatio} The
observed and expected  values of the combined  negative profile log-likelihood ratio as a function of $\kappa_c/\kappa_b$ where  $\mu_{VH}^{bb}$ is a free parameter. The solid vertical lines correspond to the values of $|\kappa_c/\kappa_b|$ for which the Higgs-charm and Higgs-bottom couplings are equal.
}
\label{fig:kcratio}
\end{figure}

Second, a simultaneous determination of $\kappa_b$ and $\kappa_c$ is performed using \Eqns{\ref{eq:kb}}{\ref{eq:kc}}.
The contours at 68\% CL and 95\% CL extracted from a likelihood scan are reported in \Fig{\ref{fig:kappabkappac}}. For most values of $\kappa_b$, a value of $\kappa_c$ is allowed at 95\% CL
that compensates for the effect of $\kappa_b$ via the width of the Higgs boson and vice versa. The best-fit value is ($\kappa_b$, $\kappa_c$) = (0.90, 0.93)%
, no 95\% CL contours can be set on each parameter independent of the other in this model. These constraints complement those from measurements of the Higgs boson $\pt$ spectrum~\cite{HIGG-2022-04}.

An alternative parameterisation is performed targeting the ratio $|\kappa_c/\kappa_b|$, which can be performed without assumptions on the Higgs boson's width.
The signal strength $\mu_{VH}^{bb}$ is a profiled free parameter in the fit absorbing both the effect of potentially anomalous $\kappa_b$ and width values; this assumes that the width is still negligible relative to the experimental resolution, which is more than 1000 times worse than the SM intrinsic width of the Higgs boson.
The $VH, $\hcc signal strength is parameterised as $\mu_{VH}^{cc}$= $(\kappa_c/\kappa_b)^2 \mu_{VH}^{bb}$.
The results are shown in~\Fig{\ref{fig:kapparatio}} reporting the profile likelihood scan as a function of $|\kappa_c/\kappa_b|$. The observed (expected) upper limit on $|\kappa_c/\kappa_b|$ is  \kappaRatioObs~(\kappaRatioExp) at 95\% CL. Both upper limits are smaller than the ratio of the $b$- and $c$-quark masses, $4.578\pm0.008$~\cite{Bazavov:2018}, evaluated at a renormalisation scale equal to \SI{125}{\GeV}, confirming that the coupling of the Higgs boson to charm-quarks is weaker than the coupling of the Higgs boson to bottom-quarks.


%
%
%
%
%
%

\FloatBarrier

\section{Conclusion}
\label{sec:conclusion}
%

The study of the Higgs boson decay into a $b\bar{b}$ or $c\bar{c}$ pair, when produced in association
with a $W$ or $Z$ boson, is presented using data collected by the ATLAS
experiment in proton--proton collisions from Run~2 of the LHC\@. The data
correspond to an integrated luminosity of 140 fb$^{-1}$  collected at a centre-of-mass energy of $\sqrt{s} = $13~\TeV.

Measurements of the diboson processes $WZ$ and $ZZ$ with $\zbb$ and $\zcc$ decays are used
as a validation of the analysis methodology. All four processes are
observed with significances larger than 3 standard deviations. The  $WZ$, $\zbb$ process is measured with a
significance of \wzbbSigObs~standard deviations, while the $ZZ$, $\zbb$ process has a significance of more than 10 standard deviations.
For the first time, the $VZ$, $\zcc$ process is observed with a significance greater than five standard deviations by the ATLAS Collaboration.
No significant deviations are observed from the SM predictions.

For a Higgs boson with a mass of  $125$~\GeV\ produced in association with
either a $W$ or $Z$ boson, the signal strengths relative to the SM
prediction in the \hbb\ and \hcc\ decay channels are measured to be
$\mu_{VH}^{bb} = \muvhbb^{+ 0.16}_{- 0.14}$  and
$\mu_{VH}^{cc} = \muvhcc^{+5.4}_{- 5.2} $ respectively.
The measurement of $\mu_{VH}^{bb}$ has an observed (expected) significance of \vhbbSigObs~(\vhbbSigExp) standard deviations, while the
$\mu_{VH}^{cc}$ results correspond to an observed upper limit at 95\% confidence level of \hccLimObs~times the SM prediction, with an expected limit of  \hccLimExp~times the SM prediction in the case of no \hcc process. The uncertainties in the signal strengths have improved by 15\% in the
\hbb\ channel and roughly a factor of three in the \hcc\ channel compared with the previous ATLAS results. %
Measurements made separately for the $WH$ or $ZH$ processes are in agreement with the SM predictions.

Differential cross-sections of $WH$ and $ZH$ production with  \hbb\ decays are made
as a function of the vector boson transverse momentum  in kinematic fiducial volumes within the simplified template cross-section framework.
For the first time at ATLAS, the $ZH$ measurements are further split into events with 0 and 1 or more jets in addition to those associated with the Higgs boson decay
and the granularity of the $VH$ cross-section measurement was extended including dedicated signal extraction for $\ptvt > \SI{600}{\GeV}$.
All measurements are  in agreement with the SM predictions.

The results are also used to set constraints on a charm Yukawa
coupling modifier; $|\kappa_c|<\kappacObs$~at  95\% confidence level. The
$H \to b\bar{b}$ and $H \to c\bar{c}$ data constrain the absolute value of the ratio of the coupling modifiers of the Higgs boson to $c$- and $b$-quarks ($|\kappa_c / \kappa_b|$) to less than \kappaRatioObs~at  95\% confidence
level, confirming that the coupling of the Higgs boson to charm-quarks is weaker than the coupling of the Higgs boson to bottom quarks.

%
%


%
\section*{Acknowledgements}
%

%
%

%
%

This paper is dedicated to the memory of Prof. Usha Mallik, whose contributions, alongside her PhD students and postdocs, helped shape this research, and whose dedication and support continue to inspire.

We thank CERN for the very successful operation of the LHC and its injectors, as well as the support staff at
CERN and at our institutions worldwide without whom ATLAS could not be operated efficiently.

The crucial computing support from all WLCG partners is acknowledged gratefully, in particular from CERN, the ATLAS Tier-1 facilities at TRIUMF/SFU (Canada), NDGF (Denmark, Norway, Sweden), CC-IN2P3 (France), KIT/GridKA (Germany), INFN-CNAF (Italy), NL-T1 (Netherlands), PIC (Spain), RAL (UK) and BNL (USA), the Tier-2 facilities worldwide and large non-WLCG resource providers. Major contributors of computing resources are listed in Ref.~\cite{ATL-SOFT-PUB-2025-001}.

We gratefully acknowledge the support of ANPCyT, Argentina; YerPhI, Armenia; ARC, Australia; BMWFW and FWF, Austria; ANAS, Azerbaijan; CNPq and FAPESP, Brazil; NSERC, NRC and CFI, Canada; CERN; ANID, Chile; CAS, MOST and NSFC, China; Minciencias, Colombia; MEYS CR, Czech Republic; DNRF and DNSRC, Denmark; IN2P3-CNRS and CEA-DRF/IRFU, France; SRNSFG, Georgia; BMBF, HGF and MPG, Germany; GSRI, Greece; RGC and Hong Kong SAR, China; ISF and Benoziyo Center, Israel; INFN, Italy; MEXT and JSPS, Japan; CNRST, Morocco; NWO, Netherlands; RCN, Norway; MNiSW, Poland; FCT, Portugal; MNE/IFA, Romania; MSTDI, Serbia; MSSR, Slovakia; ARIS and MVZI, Slovenia; DSI/NRF, South Africa; MICIU/AEI, Spain; SRC and Wallenberg Foundation, Sweden; SERI, SNSF and Cantons of Bern and Geneva, Switzerland; NSTC, Taipei; TENMAK, T\"urkiye; STFC/UKRI, United Kingdom; DOE and NSF, United States of America.

Individual groups and members have received support from BCKDF, CANARIE, CRC and DRAC, Canada; CERN-CZ, FORTE and PRIMUS, Czech Republic; COST, ERC, ERDF, Horizon 2020, ICSC-NextGenerationEU and Marie Sk{\l}odowska-Curie Actions, European Union; Investissements d'Avenir Labex, Investissements d'Avenir Idex and ANR, France; DFG and AvH Foundation, Germany; Herakleitos, Thales and Aristeia programmes co-financed by EU-ESF and the Greek NSRF, Greece; BSF-NSF and MINERVA, Israel; NCN and NAWA, Poland; La Caixa Banking Foundation, CERCA Programme Generalitat de Catalunya and PROMETEO and GenT Programmes Generalitat Valenciana, Spain; G\"{o}ran Gustafssons Stiftelse, Sweden; The Royal Society and Leverhulme Trust, United Kingdom.

In addition, individual members wish to acknowledge support from Armenia: Yerevan Physics Institute (FAPERJ); CERN: European Organization for Nuclear Research (CERN PJAS); Chile: Agencia Nacional de Investigaci\'on y Desarrollo (FONDECYT 1230812, FONDECYT 1230987, FONDECYT 1240864); China: Chinese Ministry of Science and Technology (MOST-2023YFA1605700, MOST-2023YFA1609300), National Natural Science Foundation of China (NSFC - 12175119, NSFC 12275265, NSFC-12075060); Czech Republic: Czech Science Foundation (GACR - 24-11373S), Ministry of Education Youth and Sports (FORTE CZ.02.01.01/00/22\_008/0004632), PRIMUS Research Programme (PRIMUS/21/SCI/017); EU: H2020 European Research Council (ERC - 101002463); European Union: European Research Council (ERC - 948254, ERC 101089007), Horizon 2020 Framework Programme (MUCCA - CHIST-ERA-19-XAI-00), European Union, Future Artificial Intelligence Research (FAIR-NextGenerationEU PE00000013), Italian Center for High Performance Computing, Big Data and Quantum Computing (ICSC, NextGenerationEU); France: Agence Nationale de la Recherche (ANR-20-CE31-0013, ANR-21-CE31-0013, ANR-21-CE31-0022, ANR-22-EDIR-0002), Investissements d'Avenir Labex (ANR-11-LABX-0012); Germany: Baden-Württemberg Stiftung (BW Stiftung-Postdoc Eliteprogramme), Deutsche Forschungsgemeinschaft (DFG - 469666862, DFG - CR 312/5-2); Italy: Istituto Nazionale di Fisica Nucleare (ICSC, NextGenerationEU), Ministero dell'Università e della Ricerca (PRIN - 20223N7F8K - PNRR M4.C2.1.1); Japan: Japan Society for the Promotion of Science (JSPS KAKENHI JP22H01227, JSPS KAKENHI JP22H04944, JSPS KAKENHI JP22KK0227, JSPS KAKENHI JP23KK0245); Netherlands: Netherlands Organisation for Scientific Research (NWO Veni 2020 - VI.Veni.202.179); Norway: Research Council of Norway (RCN-314472); Poland: Ministry of Science and Higher Education (IDUB AGH, POB8, D4 no 9722), Polish National Agency for Academic Exchange (PPN/PPO/2020/1/00002/U/00001), Polish National Science Centre (NCN 2021/42/E/ST2/00350, NCN OPUS nr 2022/47/B/ST2/03059, NCN UMO-2019/34/E/ST2/00393, NCN \& H2020 MSCA 945339, UMO-2020/37/B/ST2/01043, UMO-2021/40/C/ST2/00187, UMO-2022/47/O/ST2/00148, UMO-2023/49/B/ST2/04085, UMO-2023/51/B/ST2/00920); Slovenia: Slovenian Research Agency (ARIS grant J1-3010); Spain: Generalitat Valenciana (Artemisa, FEDER, IDIFEDER/2018/048), Ministry of Science and Innovation (MCIN \& NextGenEU PCI2022-135018-2, MICIN \& FEDER PID2021-125273NB, RYC2019-028510-I, RYC2020-030254-I, RYC2021-031273-I, RYC2022-038164-I), PROMETEO and GenT Programmes Generalitat Valenciana (CIDEGENT/2019/027); Sweden: Carl Trygger Foundation (Carl Trygger Foundation CTS 22:2312), Swedish Research Council (Swedish Research Council 2023-04654, VR 2018-00482, VR 2022-03845, VR 2022-04683, VR 2023-03403, VR grant 2021-03651), Knut and Alice Wallenberg Foundation (KAW 2018.0157, KAW 2018.0458, KAW 2019.0447, KAW 2022.0358); Switzerland: Swiss National Science Foundation (SNSF - PCEFP2\_194658); United Kingdom: Leverhulme Trust (Leverhulme Trust RPG-2020-004), Royal Society (NIF-R1-231091); United States of America: U.S. Department of Energy (ECA DE-AC02-76SF00515), Neubauer Family Foundation.

%
%


%
%
%
%
%
%
%

%
%

%

%
%

%
%
%
\printbibliography
\clearpage
 
\begin{flushleft}
\hypersetup{urlcolor=black}
{\Large The ATLAS Collaboration}

\bigskip

\AtlasOrcid[0000-0002-6665-4934]{G.~Aad}$^\textrm{\scriptsize 104}$,
\AtlasOrcid[0000-0001-7616-1554]{E.~Aakvaag}$^\textrm{\scriptsize 17}$,
\AtlasOrcid[0000-0002-5888-2734]{B.~Abbott}$^\textrm{\scriptsize 123}$,
\AtlasOrcid[0000-0002-0287-5869]{S.~Abdelhameed}$^\textrm{\scriptsize 119a}$,
\AtlasOrcid[0000-0002-1002-1652]{K.~Abeling}$^\textrm{\scriptsize 56}$,
\AtlasOrcid[0000-0001-5763-2760]{N.J.~Abicht}$^\textrm{\scriptsize 50}$,
\AtlasOrcid[0000-0002-8496-9294]{S.H.~Abidi}$^\textrm{\scriptsize 30}$,
\AtlasOrcid[0009-0003-6578-220X]{M.~Aboelela}$^\textrm{\scriptsize 45}$,
\AtlasOrcid[0000-0002-9987-2292]{A.~Aboulhorma}$^\textrm{\scriptsize 36e}$,
\AtlasOrcid[0000-0001-5329-6640]{H.~Abramowicz}$^\textrm{\scriptsize 155}$
% ...
\bigskip
\\

% ...
$^{17}$Department for Physics and Technology, University of Bergen, Bergen; Norway.\\
% ...
$^{30}$Physics Department, Brookhaven National Laboratory, Upton NY; United States of America.\\
% ...
$^{36}$$^{(a)}$Facult\'e des Sciences Ain Chock, Universit\'e Hassan II de Casablanca;$^{(b)}$Facult\'{e} des Sciences, Universit\'{e} Ibn-Tofail, K\'{e}nitra;$^{(c)}$Facult\'e des Sciences Semlalia, Universit\'e Cadi Ayyad, LPHEA-Marrakech;$^{(d)}$LPMR, Facult\'e des Sciences, Universit\'e Mohamed Premier, Oujda;$^{(e)}$Facult\'e des sciences, Universit\'e Mohammed V, Rabat;$^{(f)}$Institute of Applied Physics, Mohammed VI Polytechnic University, Ben Guerir; Morocco.\\
% ...
$^{45}$Physics Department, Southern Methodist University, Dallas TX; United States of America.\\
% ...
$^{50}$Fakult\"{a}t Physik , Technische Universit{\"a}t Dortmund, Dortmund; Germany.\\
% ...
$^{56}$II. Physikalisches Institut, Georg-August-Universit\"{a}t G\"ottingen, G\"ottingen; Germany.\\
% ...
$^{104}$CPPM, Aix-Marseille Universit\'e, CNRS/IN2P3, Marseille; France.\\
% ...
$^{119}$$^{(a)}$New York University Abu Dhabi, Abu Dhabi;$^{(b)}$United Arab Emirates University, Al Ain; United Arab Emirates.\\
% ...
$^{123}$Homer L. Dodge Department of Physics and Astronomy, University of Oklahoma, Norman OK; United States of America.\\
% ...
$^{155}$Raymond and Beverly Sackler School of Physics and Astronomy, Tel Aviv University, Tel Aviv; Israel.\\
% ...
$^{a}$ Also Affiliated with an institute covered by a cooperation agreement with CERN.\\
$^{b}$ Also at An-Najah National University, Nablus; Palestine.\\
$^{c}$ Also at Borough of Manhattan Community College, City University of New York, New York NY; United States of America.\\
$^{d}$ Also at Center for High Energy Physics, Peking University; China.\\
$^{e}$ Also at Center for Interdisciplinary Research and Innovation (CIRI-AUTH), Thessaloniki; Greece.\\
$^{f}$ Also at CERN, Geneva; Switzerland.\\
% ...

\end{flushleft}


%

%
%

%

%
%
%
%

\end{document}